\documentclass{article}
\usepackage[utf8]{inputenc}
\usepackage[framemethod=TikZ]{mdframed}
\usepackage{tikz}
\usetikzlibrary{shapes,snakes}
\usetikzlibrary {positioning}
\RequirePackage[OT1]{fontenc}
\RequirePackage{amsthm}
\RequirePackage[cmex10]{amsmath}
\usepackage{natbib}

\numberwithin{equation}{section}
\theoremstyle{plain}
\newtheorem{assumption}[]{ASSUMPTION}

\newtheorem{assumptionalt}{Assumption}[assumption]

\newenvironment{assumptionp}[1]{
  \renewcommand\theassumptionalt{#1}
  \assumptionalt
}{\endassumptionalt}

\usepackage{tabularx}
\usepackage{adjustbox}
\usepackage{booktabs}

\usepackage{array}
\newcolumntype{$}{>{\global\let\currentrowstyle\relax}}
\newcolumntype{^}{>{\currentrowstyle}}

\newcommand{\ci}{\perp\!\!\!\perp}

\usepackage{paralist}
\usepackage[section]{placeins}
\usepackage{amsfonts}
\usepackage{booktabs}
\usepackage{siunitx}
\usepackage{tabularx}
\usepackage{mathptmx}
\usepackage[11pt]{moresize}
\usepackage{algorithm}
\usepackage{amsbsy}
\usepackage{amsmath}
\usepackage{amsmath,amssymb,amsthm,bm}
\usepackage{amsthm}
\usepackage{cmap}
\usepackage{cases}
\usepackage{xr}
\usepackage{algorithm}

\usepackage[english]{babel}
\usepackage{enumitem}
\usepackage{epsfig}
\usepackage{eucal}
\usepackage[font=small]{caption}
\usepackage[T1]{fontenc}
\usepackage{graphicx}
\usepackage[margin=1.35in]{geometry}
\usepackage{mathptmx}
\usepackage{mathtools}
\usepackage{natbib}
\usepackage{nccmath}
\usepackage[noend,]{algorithmic}
\usepackage{setspace}
\usepackage{sgame}
\usepackage{subcaption}
\usepackage{tikz}
\usepackage{url}
\usepackage{pdflscape}
\usepackage{afterpage}
\usepackage{quoting, lipsum}

\newcommand{\E}{\mathbb{E}}
\newcommand{\G}{\mathbb{G}}

\newcount\Comments
\theoremstyle{plain}

\newtheorem{theorem}{Theorem}[]

\newcounter{example}[section]
\newenvironment{example}[1][]{\refstepcounter{example}\par\medskip
   \noindent \textbf{Example~\theexample. #1} \rmfamily}{\medskip}

\theoremstyle{remark}

\newtheorem{definition}[]{Definition}

\newtheorem{remark}[]{Remark}

\newtheorem{lemma}[]{Lemma}

\renewcommand{\theremark}{\arabic{section}.\arabic{remark}}

\renewcommand{\thedefinition}{\arabic{section}.\arabic{definition}}

\newcommand{\ba}{\begin{array}}
\newcommand{\ea}{\end{array}}

\makeatletter
\DeclareRobustCommand{\varamalg}{%
  \mathbin{\mathpalette\var@malg\perp}%
}

\newcommand\var@malg[2]{%
  \rlap{$\m@th#1#2$}\mkern6mu{#1#2}%
}
\makeatother

\newcommand{\bs}{\begin{align}\begin{split}\nonumber}
\newcommand{\bsnumber}{\begin{align}\begin{split}}
\newcommand{\es}{\end{split}\end{align}}

\newcolumntype{Y}{>{\centering\arraybackslash}X}
\usepackage[T1]{fontenc}
\usepackage{etoolbox}

\makeatletter
\patchcmd{\maketitle}
 {\def\@makefnmark}
 {\def\@makefnmark{}\def\useless@macro}
 {}{}
\makeatother
\usepackage{amsmath}
\makeatletter
\renewcommand*\env@matrix[1][*\c@MaxMatrixCols c]{%
  \hskip -\arraycolsep
  \let\@ifnextchar\new@ifnextchar
  \array{#1}}
\makeatother

\usepackage{pdflscape}
\usepackage{afterpage}
\usepackage{capt-of}

\newtheorem{examplealt}{Example}[example]

\newenvironment{examplep}[1]{
  \renewcommand\theexamplealt{#1}
 \examplealt
}{\endexamplealt}

\begin{document}
\linespread{1.5}

\title {Generalized Lee Bounds}

\author{
	Vira Semenova\thanks{First version: August, 31, 2020. This version: February 24, 2023. Email: semenovavira@gmail.com. I am indebted to Victor Chernozhukov, Michael Jansson, Patrick Kline,  Anna Mikusheva,   Whitney Newey and Demian Pouzo for their patience and encouragement.  I am grateful to my discussants Jorg Stoye (Chamberlain seminar), Xiaoxia Shi (Cowles Econometrics Conference),  Isaiah Andrews (NBER Summer Institute (Labor Studies)) and Xiaohong Chen (Women in Applied Micro Conference),  whose comments have substantially improved the paper relative to the first version. I am also thankful to Alberto Abadie, Chris Ackerman, Stephane Bonhomme, Kirill Borusyak, Sydnee Caldwell, Matias  Cattaneo, Colin Cameron, Xiaohong Chen, Denis Chetverikov, Ben Deaner, Mert Demirer, Bulat Gafarov, Dalia Ghanem, Jerry Hausman, Keisuke Hirano, Peter Hull, Tetsuya Kaji, Michal Kolesar, Ivana Komunjer, Kevin Li, Elena Manresa, Eric Mbakop, David McKenzie, Rachael Meager, Francesca Molinari, Ismael Mourifie, Denis Nekipelov, Alexandre Poirier, Geert Ridder, Jonathan Roth, Oles Shtanko, Cory Smith, Sophie Sun, Takuya Ura, Roman Zarate, Edward Vytlacil and participants in numerous seminars and conferences for helpful comments. Mengsi Gao has provided superb research assistance.}  }

\date{}
\maketitle

\begin{abstract}
Lee (2009) is a common approach to bound the average causal effect in the presence of selection bias, assuming the treatment effect on selection has the same sign for all subjects. This paper generalizes Lee bounds to allow the sign of this effect to be identified by pretreatment covariates, relaxing the standard (unconditional) monotonicity to its conditional analog. Asymptotic theory for generalized Lee bounds is proposed in low-dimensional smooth and high-dimensional sparse designs. The paper also generalizes Lee bounds to accommodate multiple outcomes. Focusing on JobCorps job training program, I first show that unconditional monotonicity is unlikely to hold, and then demonstrate the use of covariates to tighten the bounds.
\end{abstract}

\newpage

\section{Introduction}

Randomized controlled trials are often complicated by endogenous sample selection and non-response. This problem occurs when treatment affects the researcher's ability to observe an outcome (a selection effect) in addition to the outcome itself (the causal effect of interest). For example, being randomized into a job training program affects both an individual's wage and employment status. Since wages exist only for employed individuals, treatment-control wage difference is contaminated by selection bias. A common way to proceed is
to bound the average causal effect from above and below, focusing on subjects whose outcomes are observed regardless of treatment receipt (the always-observed principal strata, \cite{FrangakisRubin} or the always-takers, \cite{LeeBound}).

Seminal work by \cite{LeeBound} proposes nonparametric bounds assuming the selection effect is non-negative for all subjects (monotonicity). For example, if JobCorps cannot deter employment, basic Lee lower bound is  the treatment-control difference in wages, where the top
wages in the treated group are trimmed until treated and control employment rates are equal. Furthermore, \cite{LeeBound}  shows that the covariate density-weighted conditional  Lee bound  is  weakly tighter than the basic bound that does not involve any covariates. However,  only a handful of discrete covariates can be utilized to tighten the bound, since each covariate cell is required to have a positive number of treated and control outcomes.  

This paper generalizes Lee bounds under \textit{conditional} monotonicity, which allows the sign of selection effect to be determined by pre-treatment covariates. Now, baseline covariates now have two roles: (1) to define the subspaces of positive and negative selection response and (2) to tighten the bound on each subspace. For the first one, using the full covariate vector  makes the conditional monotonicity assumption the least restrictive. For the second one,    \textit{any} covariate subvector -- even an empty one -- would suffice to define a valid bound. This paper studies the sharp -- the tightest possible -- generalized  bound, where all covariates are used for (1) and (2).

I represent the sharp bound as a ratio of two semiparametric moments whose nuisance functions are  the conditional probability of selection, the conditional quantile, and the propensity score (i.e., the conditional probability of treatment).  If the nonparametric functions are smooth functions of covariates, they can be estimated by logistic series regression of \cite{HIR2003} and quantile series of \cite{CherChetQuant}, respectively.  Alternatively, if these functions have a sparse representation with respect to some transformations of covariates, one could employ their $\ell_1$-penalized analogs proposed in \cite{orthogStructural} and  \cite{belloni:11}, \cite{Program}. To make the second stage moments insensitive to the first-stage estimation error, I derive Neyman-orthogonal (\cite{Neyman:1959}, \cite{Neyman:1979}, \cite{AiChen2003}, \cite{Newey1994}, \cite{chernozhukov2016double}, \cite{LRSP}) moment functions for the numerator and the denominator. Combining Neyman-orthogonality and sample splitting,  I derive a root-$N$ consistent and asymptotically normal estimator of the sharp bounds in the presence of (1) data-driven covariate selection, (2) possible misclassification into the regions of positive ($\mathcal{X}_{\text{help}}$) and negative ($\mathcal{X}_{\text{hurt}}$) selection response and (3) (any)  point mass on the region where the selection response is exactly zero (i.e., the boundary between $\mathcal{X}_{\text{help}}$ and $\mathcal{X}_{\text{hurt}}$). This proposal neither requires the covariates to be discrete nor the propensity score to be known, and considerably expands the scope for using Lee bounds in practice.

As an empirical application, I revisit Lee's JobCorps study as in \cite{LeeBound} using the data from \cite{Schochet}. The paper's first major empirical finding is to that the unconditional monotonicity of selection (i.e., employment) is unlikely to hold. After imposing conditional monotonicity  (and accounting for the differential JobCorps effect on employment), I find that the average JobCorps effect on the always-takers' week 90 wages ranges between $-5\%$ and $1\%$.  Finally, I provide evidence of mean reversion of the expected log wage for the always-takers in the control status. This mean reversion corroborates \cite{Ashenfelter} pattern and shows that earnings would have recovered even without JobCorps training. Therefore, evaluating JobCorps would have been very difficult without a randomized experiment, as one would need to explicitly model mean reversion in the potential wage in the control status.

Appendix B extends Lee's trimming approach to the case of  multiple outcomes. A naive approach to construct an identified set is to take a Cartesian product of scalar bounds for each component of the causal parameter. However, since outcomes may be correlated,  such a set may contain points that do not correspond to a data generating process. I characterize the sharp identified set for the causal parameter as well as its support function (\cite{BM,BMM}). Furthermore, I establish debiased  uniform over the boundary inference on the support function based on first-stage regularized estimators. The use of this theory is demonstrated for the parameters involving multi-dimensional outcomes, such as standardized treatment effect and wage growth.

\paragraph{Literature review. } The conditional monotonicity assumption has been previously proposed in \cite{Kolesar} and \cite{sloczynski2020not} in  a treatment choice context, accommodating differential sign of a binary treatment response to a binary instrument. To generalize the Local Average Treatment Effect parameter,  \cite{sloczynski2020not} combines the estimates from no-defier and no-complier regions with signs $1$ and $-1$, respectively, while the boundary (i.e.,  the no-defier-and-no-complier region) has zero identification power and, therefore, receives zero weight.   In contrast, the sample selection problem focuses on the always takers -- a different principal strata -- whose selection behavior \textit{ does not change }  in response to treatment.

This paper combines ideas from various branches of economics and statistics, including bounds on causal effects (\cite{Manski89}, \cite{Manski90}, \cite{HorowitzManski}, \cite{HorowitzManski2000}, \cite{FrangakisRubin},  \cite{angrist:2002},  \cite{ZhangRubin}, \cite{angrist:2006},  \cite{ChenFlores}, \cite{feller2016weak}, \cite{APW2013}, \cite{APW2018},  \cite{Honore},  \cite{kamat2021identifying}),  convex analysis and support function   (\cite{CherRigStoker}, \cite{Stoye}, \cite{BM}, \cite{BMM}, \cite{KaidoSantos},  \cite{Kaido}, \cite{StoyeSpread}, \cite{KaidoMolinariStoye}, \cite{Gafarov}, \cite{MolinariStoye}, \cite{MolinariHandbook}),    monotonicity and latent index models (\cite{Vytlacil}, \cite{KWLATE}, \cite{kamat2019identifying}, \cite{sloczynski2020not}, \cite{MTW}, \cite{MTW2}, \cite{Ura}), including the bounds on the same empirical context -- JobCorps job training program -- (\cite{LeeBound}, \cite{FloresLagunes}, \cite{ChenFlores}).

Next, this paper contributes to a large body of work on debiased/orthogonal inference for parameters following regularization or model selection  (\cite{Neyman:1959}, \cite{Neyman:1979}, \cite{HardleStoker1989}, \cite{NeweyStoker},  \cite{andrews:1994}, \cite{Newey1994}, \cite{Robins}, \cite{robinson:88}, \cite{ChenAck}, \cite{ZhangZhang}, \cite{JM}, \cite{chernozhukov2016double}, \cite{LRSP}, \cite{Program}, \cite{sasaki2018estimation}, \cite{sasaki2020unconditional}, \cite{Sasaki}, \cite{chiang2019multiway}, \cite{ning2020doubly}, \cite{chernozhukov2021debiased}, \cite{chernozhukov2021automatic}, \cite{CherSem}, \cite{NSS}, \cite{singh2020debiased}, \cite{Colangelo}, \cite{Lieli}, \cite{ZimLech}).  In many classic cases, such as  \cite{Robins}  or \cite{robinson:88},  orthogonalization expands the set of first-stage parameters to be estimated. In contrast,    the set of first-stage nuisance for the truncated conditional mean functional  does not expand after orthogonalization. Finally, the paper contributes to a growing literature on machine learning for bounds and partially identified models (\cite{kallus2019assessing},  \cite{jeong2020robust}, \cite{Bonvini_2021}, \cite{ZhouSmith}, \cite{SemJoE}).  The causal parameter  is not a special case of a set-identified linear model of \cite{BM,BMM}, and the identification and estimation approaches of \cite{CCMS} and \cite{SemJoE} do not apply. 

An emerging body of research has validated the usefulness of this paper's results by both expanding theoretical framework and/or employing them in applications. For instance, \cite{olma2021nonparametric}  proposes  a nonparametric estimator of truncated conditional expectation functions by plugging an orthogonal moment for the truncated mean into a locally linear regression. While interesting on its own, this parameter also enters the correction term for the unknown propensity score in Section \ref{sec:pscore}. Finally, \cite{Heiler2} proposes an  estimator of heterogeneous treatment effects using least squares regression.

The paper is organized as follows. Section \ref{sec:overview} reviews basic Lee bounds and  Lee's  estimator under the standard unconditional monotonicity assumption. Section \ref{sec:monot} generalizes Lee bounds under conditional monotonicity.  Section \ref{sec:ortho} defines the debiased moment functions for the numerator and the denominator of each boun. Section \ref{sec:estim} overviews the proposed estimator and provides inference results for this parameter. Section \ref{sec:theory} states the asymptotic theory for generalized Lee bounds.  Section \ref{sec:empirical} presents empirical application.   The Supplementary Appendix contains proofs  (Appendix A),   an extension to the multiple outcome case (Appendix B),  and auxiliary empirical details (Appendix C).

\section{Bounds under unconditional monotonicity}
\label{sec:overview}

\subsection{Overview of \cite{LeeBound}'s results}

Consider the  sample selection model in \cite{LeeBound}. Let $D=1$ be an indicator for treatment receipt.  Let $Y(1)$ and $Y(0)$ denote the potential outcomes if an individual is treated or not, respectively.  Likewise, let $S(1)=1$ and $S(0)=1$ be  dummies for whether an individual's outcome is observed  with and without treatment.  The  data vector $W=(D,X,S,S \cdot Y)$  consists of the treatment status $D$, a baseline covariate vector $X$, the selection status $S=D \cdot S(1) + (1-D) \cdot S(0)$ and the outcome $S \cdot Y = S \cdot (D  \cdot Y(1) + (1-D) \cdot Y(0))$ for selected individuals.  \cite{LeeBound} focuses on the average treatment effect (ATE)
 \begin{align}
 \label{eq:truebeta}
 \beta_0 = \E[Y(1) - Y(0)  \mid S(1)=1, S(0)=1] 
 \end{align} 
 for subjects who are selected into the sample regardless of treatment receipt---the \emph{always-takers}. 
 \begin{assumptionp}{1}[Assumptions of \cite{LeeBound}]
 \label{ass:identification:treat}
The following statements hold. 
\begin{compactenum}[(1)]
\item[a] (Complete Independence). The potential outcomes vector $(Y(1),Y(0),S(1),S(0),X)$ is independent of $D$. 
	\item[b] (Monotonicity).  \begin{equation} \label{eq:monot1} S(1) \geq S(0) \quad \text{ a.s. } \end{equation}
\end{compactenum}
 \end{assumptionp}
 The independence assumption holds by random assignment. In addition, it requires all subjects to have the same probability of being treated. 
The monotonicity requires all subjects to have the same direction of selection response. In particular, a subject that is selected into the sample when untreated must remain selected if treated:
$$
S(0) = 1 \quad \Rightarrow \quad S(1) =1.
$$
 As a result, 
$$\E [ Y(0)  \mid S(1)=1, S(0)=1] = \E [ Y(0)  \mid  S(0)=1]. $$
By complete independence,
$$ \E [ Y(0)  \mid  S(0)=1] = \E [ Y \mid S=1, D=0], $$
and $\E [ Y(0)  \mid S(1)=1, S(0)=1] $   is point-identified.

In contrast to the control group, a treated outcome can be either an always-taker's outcome or a complier's outcome. The always-takers' share among the treated outcomes is
\begin{align}
\label{eq:p0}
p_0 = \Pr[ S(1) =1, S(0)=1\mid S(1) =1 ]= \Pr[ S(0)=1\mid S(1) =1 ] =  \dfrac{\Pr [S=1\mid D=0]}{\Pr [S=1\mid D=1]}.
\end{align}
In the best case, the always-takers comprise the top $p_0$-quantile of the treated outcomes.  The largest possible value of $\beta_0$ is basic upper bound
\begin{align}
\label{eq:betabasic}
\bar{\beta}_U = \E [ Y \mid Y \geq Q^1 (1-p_0), D = 1, S=1] - \E[ Y \mid  S=1, D = 0],
\end{align}
where $Q^1 (u)$ is the $u$-quantile of $Y \mid D=1,S=1$  and $p_0$ in \eqref{eq:p0} is the trimming threshold. Likewise, the smallest possible one is
$$
\bar{\beta}_L  = \E [ Y \mid Y \leq  Q^1 (p_0), D = 1, S=1] - \E [ Y \mid  D = 0, S=1].
$$

Lee's identification strategy can be implemented conditional on covariates. Denote the conditional trimming threshold $p(x)$ as
 \begin{align}
  p(x) = \dfrac{\Pr [S=1\mid  D=0, X=x]}{\Pr [S=1\mid D=1, X=x]}=\dfrac{s(0,x)}{s(1,x)} \quad x \in \mathcal{X} \label{eq:condtrim}
 \end{align}
 and  the conditional upper bound  $\bar{\beta}_U (x) $ as
   \begin{align}
   \label{eq:condupperbound}
 \bar{\beta}_U (x) &=\E[ Y \mid D=1, S=1, Y \geq Q^{1}(1-p(x),x), X=x]- \E[ Y \mid S=1, D=0, X=x]
 \end{align} 
 where $Q^{1}(u,x)$ is the conditional $u$-quantile of $Y$ in $S=1,D=1,X=x$ group.   The covariate-based  bound is
   \begin{align}
\beta_U &= \int_{x\in \mathcal{X}} \bar{\beta}_U (x) f_{X}(x \mid S=1, D=0) dx = \int_{x\in \mathcal{X}} \bar{\beta}_U (x) f_{X}(x \mid S(1)=S(0)=1) dx, \label{eq:sharpbetau}
 \end{align}
 which, as Lee has shown, is weakly tighter than the basic one (\cite{LeeBound2}).

The formula for \eqref{eq:sharpbetau} involves a covariate density function. When data has a single continuous covariate, accounting for the estimation error of  the nonparametric density estimate may be challenging both in theory and practice. An alternative approach, proposed in Algorithm \ref{alg:leebound}, is to discretize covariates. In absence of a better name, it is referred to as discrete Lee bounds and summarizes the discretization procedure Lee used to report covariate-based bound (Table 5, \cite{LeeBound}). 

 \begin{algorithm}[H]
\small
\begin{algorithmic}[1]
	\STATE  Partition the covariate space $\mathcal{X}$ into  $J$ discrete cells, indexed by $j=1,2,\dots, J$. 
	\STATE  Estimate cell-specific basic bounds $\{ \widehat{\bar{\beta}}_L(j), \widehat{\bar{\beta}}_U(j)  \}_{j=1}^J$ and densities $\{ \widehat{f}(j| S=1, D=0)\}_{j=1}^J$  in the selected control group.
	\STATE  {Define discrete Lee bounds estimator
	  \begin{align}
	  \widehat{\beta}_L &= \sum_{j=1}^J \widehat{\bar{\beta}}_L(j)  \widehat{f}(j| S=1, D=0), \quad \widehat{\beta}_U = \sum_{j=1}^J \widehat{\bar{\beta}}_U(j)  \widehat{f}(j| S=1, D=0). \label{eq:betadiscrete}
 \end{align}
	}
\end{algorithmic}
\caption{Discrete Lee Bounds}
\label{alg:leebound}
\end{algorithm}

\subsection{Moment-based approach}

This Section describes an alternative --  moment-based -- approach to bounds, which no longer requires estimating the always-takers' covariate density function. 

 \begin{assumptionp}{2(a)}[Conditional Independence]
 \label{ass:indep}
 The potential outcomes vector is independent of the treatment $D$ conditional on $X$:
   $$(Y(1),Y(0),S(1),S(0)) \ci D \mid X.$$ 
 \end{assumptionp}
  Assumption \ref{ass:identification:treat} (a) requires complete independence. Equivalently, the  propensity score 
    \begin{align}
 \label{eq:propscore}
 \mu_1(X) := \Pr (D =1 \mid X), \qquad \mu_0(X) =1 - \mu_1(X)
 \end{align} 
must be constant in $X$.  Assumption \ref{ass:indep} relaxes the complete independence to its conditional analog.

I represent \eqref{eq:sharpbetau} as a ratio of two moments.  Let $W=(D, X, S, S\cdot Y)$ be the data vector.  Define the numerator moment function
\begin{align}
\label{eq:muxi}
m_U(W, \xi)  &=  \dfrac{D}{\mu_1(X)}  \cdot S \cdot Y \cdot 1{\{ Y \geq Q^1(1-p(X),X) \}  } - \dfrac{1-D}{\mu_0(X)} \cdot S \cdot Y
\end{align}
where the true value of the  first-stage  nuisance function $\xi=\xi(x)$ is
\begin{align}
\label{eq:ksi0}
\xi_0(x) =(s(0,x), s(1,x), Q^1(1-p(x),x)), \quad p(x) = s(0,x)/s(1,x). 
\end{align}

\begin{lemma}[Upper bound under unconditional monotonicity]
\label{lem:momentbasic}
Suppose Assumptions \ref{ass:identification:treat}(b)  and \ref{ass:indep} hold. Then, the upper bound $\beta_U$ is a ratio of two moments:
\begin{align} \label{eq:momentu}
\beta_U  = \dfrac{ \E m_U(W, \xi_0) }{  \E s(0,X) }.
 \end{align}
\end{lemma}

Lemma \ref{lem:momentbasic} represents the sharp bound as a ratio of two moments. To get more insight into the result, note that
\begin{align*}
&\E [  \dfrac{D}{\mu_1(X)}  \cdot S \cdot Y \cdot 1{\{ Y \geq Q^1(1-p(X),X) \}  } \mid X] \\
&= \E [ Y \mid Y \geq Q^1(1-p(X),X), D=1, S=1, X] p(X) s(1, X).
\end{align*}
Noting that $p(X) s(1,X)=(s(0,X)/s(1,X)) s(1,X)$ simplifies to  $s(0,X)$ gives 
\begin{align}
\label{eq:condmain}
\E[ m_U(W, \xi_0) \mid X=x]  =  \bar{\beta}_U (x)s(0,x).
\end{align}
Invoking Bayes rule gives 
\begin{align}
\label{eq:condmain2}
\int_{\mathcal{X}}  \bar{\beta}_U (x) f_X(x \mid S(1)=S(0)=1) dx &= \int_{\mathcal{X}}  \bar{\beta}_U (x) \dfrac{s(0,x)f_X(x)} {\E s(0,X)} dx \\
&= \dfrac{\E \bar{\beta}_U (X)s(0,X) }{\E s(0,X)}. \nonumber
\end{align}
Applying Law of Iterated Expectations to \eqref{eq:condmain2} gives the representation \eqref{eq:momentu}.  This representation  makes it possible to employ continuous covariates under various semiparametric assumptions, discussed in Section \ref{sec:estim}.

Remark \ref{rm:kappa} points out an interesting connection between the selection problem, studied here, and the treatment choice. 
\begin{remark}[Lee's trimming function and \cite{Abadie} kappa]
\label{rm:kappa}
The moment function for the numerator $N_U$ is a trimmed version of \cite{Abadie}'s kappa weights, that is, 
$$
m_U(W, \xi) = \left(  \dfrac{D}{\mu_1(X)}  \cdot 1{\{ Y \geq Q^1(1-p(X),X) \}  } - \dfrac{1-D}{\mu_0(X)}  \right) \cdot S \cdot Y,
$$
where $D$ is the exogenous variable (i.e.,   ``instrument'') and $S$ is the endogenous variable (i.e., ``choice'' variable).

\end{remark}

\section{Generalized Lee Bounds} 
\label{sec:monot}

In this Section, I generalize Lee bounds  under conditional monotonicity. Abusing notation, I shall denote the generalized upper bound by $\beta_U$.   The conditional probability of selection is
\begin{align}
\label{eq:sdx}
s(d,x):= \Pr (S=1 \mid D=d, X=x), \quad d \in \{1, 0\}.
\end{align}
The conditional average treatment effect (ATE) on selection is
\begin{align}
\label{eq:cateselection}
\tau(x):=s(1,x)-s(0,x).
\end{align}
The sets of positive and  negative selection response are
\begin{align}
\label{eq:helphurt}
\mathcal{X}_{\text{help}} := \{x: \tau(x)>0 \}, \quad \mathcal{X}_{\text{hurt}} := \{x: \tau(x)<0 \}
\end{align}
and the boundary  is
\begin{align}
\label{eq:boundary}
\mathcal{X}_0:= \{x: \tau(x) = 0 \}.
\end{align} 
Let  $Q^{d}(u,x)$ is the conditional $u$-quantile of $Y$ in $S=1,D=d, X=x$.  Given a quantile level $u \in (0,1)$,  define the upper-trimmed mean for the treated group 
\begin{align}
\label{eq:roundedquantile1}
\beta_{1U} (u,x) :&= \E [ Y \mid D=1, S=1,  Y \geq Q^1(1-u,x), X=x ] 
\end{align}
and the lower-trimmed mean for the control group
\begin{align}
\label{eq:roundedquantile0}
\beta_{0U} (u,x) :&=  \E [ Y \mid D=0, S=1, Y \leq Q^0(u,x),  X=x ].
\end{align}

\begin{assumptionp}{2(b)} [Conditional monotonicity]
\label{ass:identification:treat:cond}
 The  covariate set $\mathcal{X}=\mathcal{X}_{\text{help}} \sqcup \mathcal{X}_{\text{hurt}} \sqcup  \mathcal{X}_0$ can be partitioned into the sets   $\mathcal{X}_{\text{help}} $,   $\mathcal{X}_{\text{hurt}} $ and $\mathcal{X}_0$ so that 
	\begin{align*}
	X &\in \mathcal{X}_{\text{help}} \Rightarrow S(1) \geq S(0) \text{ a.s. }, \\
 X &\in \mathcal{X}_{\text{hurt}} \Rightarrow  S(1) \leq S(0) \text{ a.s. },  \\
	X &\in \mathcal{X}_0 \Rightarrow S(1) = S(0) \text{ a.s. }.
	\end{align*} 
\end{assumptionp} 

Assumption \ref{ass:identification:treat:cond} generalizes the regular unconditional monotonicity assumption to the conditional analog. It  allows the direction of the selection effect to vary across covariate values. Defiers $(S(1) = 0, S(0) = 1)$ are ruled out on the covariate set  $\mathcal{X}_{\text{help}}$,  compliers $(S(1) = 1, S(0) = 0)$ are ruled out on $ \mathcal{X}_{\text{hurt}} $, and both compliers and defiers are ruled out on the set $ \mathcal{X}_0$.  In many practical cases, the boundary set $\mathcal{X}_0 $ has zero mass, but the proposed asymptotic theory accommodates the boundary of arbitrary size.

 Let us show that the always-takers' share is point-identified under  Assumptions \ref{ass:indep} and  \ref{ass:identification:treat:cond}.  For any covariate value $x$ on the positive half-space $\mathcal{X}_{\text{help}}$,  we have 
\begin{align*}
\Pr (S(1) = S(0) = 1\mid X=x)  = \Pr (S=1 \mid X=x, D=0)=s(0,x).
\end{align*}
 Likewise, for any covariate value $x \in \mathcal{X}_{\text{hurt}}$, 
 \begin{align*}
\Pr (S(1) = S(0) = 1\mid X=x) =\Pr (S=1 \mid X=x, D=1)=s(1,x).
\end{align*}
On the boundary $\mathcal{X}_0$, $\tau (x) = 0 \Rightarrow s (1,x) = s (0,x)$. Combining the results
 \begin{align}
 \label{eq:combined}
\Pr (S(1) = S(0) = 1 \mid X=x) =  \min (s(0,x), s(1,x))
 \end{align}
 and  aggregating over the covariate space gives the always-takers' share
 \begin{align}
  \label{eq:atshare}
 \pi_{\text{AT}}   &:=\Pr (S(1) = S(0) = 1)= \E \min (s(0,X), s(1,X)).
 \end{align}
 
 \begin{lemma}[Always-takers' Share]
 \label{lem:atshare}
 Suppose Assumptions  \ref{ass:indep} and \ref{ass:identification:treat:cond} hold. Then, the always-takers' share $ \pi_{\text{AT}}$ is point-identified, and 
  \begin{align}
  \label{eq:atshare2}
 \pi_{\text{AT}}   &:=\Pr (S(1) = S(0) = 1)= \E \min (s(0,X), s(1,X)),
 \end{align}
 where the expectation is taken over the covariate distribution.

 \end{lemma} \qed

 Next, let me generalize the bound itself.   For $x \in \mathcal{X}_{\text{help}}$, the conditional always-takers' share is 
$$  \min (s(0,x), s(1,x)) = s(0,x), \quad x \in \mathcal{X}_{\text{help}}.$$
The conditional upper bound takes the form
  \begin{align}
   \label{eq:condupperbound}
 \beta_U (x)&:=\beta_{1U} (s(0,x)/s(1,x),x) - \beta_{0U}(1,x) = \beta^{\text{basic}}_U(x),
 \end{align} 
 where $\beta^{\text{basic}}_U(x)$ is given in \eqref{eq:betabasic}.  Likewise, for $x \in \mathcal{X}_{\text{hurt}}$,  the roles of the treated and control group are reversed:
\begin{align}
\beta_U(x) := \beta^{\text{hurt}}_U(x) =  \beta_{1U} (1,x) - \beta_{0U}(s(1,x)/s(0,x),x).  \label{eq:betauxhurt}
\end{align}
By Assumption \ref{ass:identification:treat:cond}, both defiers and compliers are ruled out on the  boundary  $\mathcal{X}_0$. Therefore, a selected individual with $S=1$ must be an always-taker, and there is no trimming. The bound reduces to treatment-control difference  
\begin{align}
\label{eq:betax0}
 \beta_U(x)&:= \E [ Y \mid D=1, S=1, X=x] -  \E [ Y \mid D=0, S=1, X=x]. 
\end{align}
Aggregating the conditional sharp bound $\beta_U(x)$ gives 
\begin{align}
\label{eq:sharpbetau2}
\beta_U = \int_{\mathcal{X}} \beta_U(x) f_X( x \mid S(1) = S(0) = 1) dx. 
\end{align}
Invoking Bayes rule   gives
\begin{align*}
\beta_U &= \int_{\mathcal{X}} \beta_U(x) \dfrac{\Pr (S(1) = S(0) = 1\mid X=x)}{\Pr (S(1) = S(0) = 1)} f_X(x)  dx \\
&=\dfrac{ \int_{\mathcal{X}} \beta_U(x)  \min (s(0,x), s(1,x))  f_X(x) dx }{\E\min (s(0,X), s(1,X)) }.
\end{align*}

I conclude this section by representing the generalized bound \eqref{eq:sharpbetau2} as a ratio of two moments. Let $m^{\text{help}}_U (W, \xi)$ be as in \eqref{eq:muxi}.  Define the moment function for the negative half-space $\mathcal{X}_{\text{hurt}}$:
\begin{align}
m^{\text{hurt}}_U (W, \xi)&:=   \dfrac{D}{\mu_1(X)}  \cdot S \cdot Y - \dfrac{1-D}{\mu_0(X)}  \cdot S \cdot Y \cdot 1{\{ Y \leq Q^0(1/p(X),X) \}  }.
\end{align}
If $p(X) = 1$, the moment functions coincide
 $$
 m^{\text{help}}_U (W, \xi) = m^{\text{hurt}}_U (W, \xi) =   \dfrac{D}{\mu_1(X)}  \cdot S \cdot Y - \dfrac{1-D}{\mu_0(X)}  \cdot S \cdot Y
$$
Combining the moment equations gives
\begin{align}
\label{eq:combined}
m_U(W, \xi):= \begin{cases} m^{\text{help}}_U (W, \xi), \qquad  \qquad X \in \mathcal{X}_{\text{help}} \\
m^{\text{hurt}}_U (W, \xi), \qquad  \qquad X \in \mathcal{X}_{\text{hurt}} \\
\left(\dfrac{DS}{\mu_1(X)} - \dfrac{(1-D)S}{\mu_0(X)} \right)Y, \qquad X \in \mathcal{X}_0,
\end{cases}
\end{align}
where the true value $\xi_0(x)$ of the first-stage nuisance parameter is
\begin{align}
\label{eq:ksi0}
\xi_0(x) = \{ s(0,x), s(1,x), Q^1(1-p(x),x) 1\{ x \in \mathcal{X}_{\text{help}} \}  + Q^0(1/p(x),x) 1\{ x \in \mathcal{X}_{\text{hurt}}  \}  \}.
\end{align}
On the boundary $\mathcal{X}_0$, the moment functions $m^{\text{help}}_U (W, \xi)$ and $m^{\text{hurt}}_U (W, \xi)$ coincide. They 
reduce to the treatment-control difference
\begin{align}
m^{\text{help}}_U (W, \xi) = m^{\text{hurt}}_U (W, \xi) = \left(\dfrac{DS}{\mu_1(X)} - \dfrac{(1-D)S}{\mu_0(X)} \right)Y,
 \end{align}
 which involves no trimming.
 
\begin{lemma}[Generalized Lee Bound]
\label{lem:moment}
Suppose Assumptions \ref{ass:identification:treat:cond} and \ref{ass:indep} hold, and $\pi_{\text{AT}}>0$. Then, the upper bound $\beta_U$   in \eqref{eq:sharpbetau2} is a sharp upper bound on the average treatment effect $\beta_0$ in \eqref{eq:truebeta}.  The bound is 
\begin{align} \label{eq:momentu}
\beta_U  &= \dfrac{ \E [\beta_U(X) \min (s(0,X), s(1,X))] }{ \E \min (s(0,X),s(1,X))} = \dfrac{ \E [m_U(W, \xi_0)] }{ \E \min (s(0,X),s(1,X)) }.
 \end{align}
 where $\beta_U(x)$ is given in \eqref{eq:condupperbound}--\eqref{eq:betax0}. 
\end{lemma}

\section{Debiased Moment Functions}
\label{sec:ortho}

Section \ref{sec:ortho} states additional identification results, needed for estimation.  Section \ref{sec:estim1} introduces additional notation for the lower bound.  Section  \ref{sec:nlnugen} describes population moment functions for the numerators $N_U$ and $N_L$ (Section \ref{sec:nlnu}) and the  always-takers' share  (Section \ref{sec:estim2}).

\subsection{Additional Notation for Lower Bound. } 
\label{sec:estim1}

Let me introduce some additional notation for the lower bound. The partition-specific moment functions are
\begin{align}
\label{eq:momentl}
m^{\text{help}}_L (W, \xi) &=  \dfrac{D}{\mu_1(X)}  \cdot S \cdot Y \cdot 1{\{ Y \leq Q^1(p(X),X) \}  } - \dfrac{1-D}{\mu_0(X)} \cdot S \cdot Y \\
m^{\text{hurt}}_L (W, \xi)&=   \dfrac{D}{\mu_1(X)}  \cdot S \cdot Y - \dfrac{1-D}{\mu_0(X)}  \cdot S \cdot Y \cdot 1{\{ Y \geq Q^0(1-1/p(X),X) \}  },
\end{align}
where the true value of the  first-stage nuisance parameter is
\begin{align}
\label{eq:ksi0}
\xi_0(x) = \{ s(0,x), s(1,x), Q^1(p(x),x) 1\{ x \in \mathcal{X}_{\text{help}} \}  + Q^0(1-1/p(x),x) 1\{ x \in \mathcal{X}_{\text{hurt}}  \}  \}.
\end{align}
The combined moment function is 
\begin{align*}
m_L(W, \xi) = 1\{ p(X)\leq 1 \} m^{\text{help}}_L (W, \xi)  + 1\{ p(X)>1 \} m^{\text{hurt}}_L (W, \xi).
\end{align*}
Let $N_L := \E m_L(W, \xi_0)$ and $N_U := \E m_U(W, \xi_0)$. As shown in Lemma \ref{lem:moment}, the upper bound is a ratio.  The same argument applies to the lower bound 
\begin{align}
\label{eq:ratio}
\beta_L = \dfrac{N_L}{\pi_{\text{AT}}}, \qquad \beta_U = \dfrac{N_U}{\pi_{\text{AT}}}.
\end{align}
The bounds (and their numerators) are ordered by construction
\begin{align}
\label{eq:nlnu}
N_L = \E m_L (W, \xi_0) \leq \E m_U  (W, \xi_0) = N_U, \quad \beta_L \leq \beta_U.
\end{align}

\subsection{Second Stage Moment Functions}
\label{sec:nlnugen}

\subsubsection{ The Upper Bound Numerator }
\label{sec:nlnu}

Consider the upper bound numerator $N_U = \E[m_U(W, \xi_0)]$. The moment function $m_U(W, \xi_0)$ is non-orthogonal to the biased estimation of $\xi_0$. To overcome the transmission of this bias, I replace $m^{\text{help}}_U (W, \xi)$ and $m^{\text{hurt}}_U (W, \xi)$ in 
\eqref{eq:combined} by their orthogonal counterparts $g^{\text{help}}_U (W, \xi)$ and $g^{\text{hurt}}_U (W, \xi)$,  defined below. In this Section, the propensity score is assumed known, and the correction term for it is not provided.

\begin{definition}[Orthogonal Moment Function $g^{\text{help}}_U (W, \xi)$ on $\mathcal{X}_{\text{help}}$]
\label{def:help}
Let $X \in \mathcal{X}_{\text{help}}$. Define the bias correction term
\begin{align}
\label{eq:coruhelp}
\text{cor}^{\text{help}}_U (W, \xi) &= Q^1(1-p(X), X)  \bigg( \dfrac{1-D}{\mu_0(X) }\cdot ( S - s(0,X))  \\
&-\dfrac{D}{\mu_1(X) } \cdot p(X)  \cdot ( S - s(1,X)) \nonumber  \\
&+ \dfrac{D S}{\mu_1(X)  }  (  1 \{ Y \leq Q^1(1-p(X), X) \} - 1+ p(X)) \bigg). \nonumber 
\end{align}
and the debiased moment function 
\begin{align}
\label{eq:ghelp}
g^{\text{help}}_U (W, \xi):= m^{\text{help}}_U (W, \xi) + \text{cor}^{\text{help}}_U (W, \xi).
\end{align}
\end{definition}

The bias correction term \eqref{eq:coruhelp} consists of three summands, corresponding to the bias correction of $s(0,x)$, $s(1,x)$, and $Q^1(u,x)$ ( \cite{Newey1994}).   As stated in  \cite{olma2021nonparametric}, adding the correction terms and simplifying gives
\begin{align}
\text{cor}^{\text{help}}_U (W, \xi) &= Q^1(1-p(X), X) \bigg( \dfrac{(1-D) S}{\mu_0(X) } - \dfrac{D S}{\mu_1(X)  } 1 \{ Y \geq Q^1(1-p(X), X) \} \\
&+ s(0,X)  \left( \dfrac{D }{\mu_1(X)  } - \dfrac{1-D }{\mu_0(X)  } \right) \bigg). \nonumber
\end{align}

Remarkably,  the function $Q^1(1-p(X), X)$ is the only nuisance component  of both the original and the debiased moment function. This function maps covariate vector $X$ into  the always-taker's borderline wage in the best case $Q^1(1-p(X), X)$:  the lowest wage earned by an always-taker  in the extreme case  when all always-takers' treated wages are above compliers' treated wages for each covariate value $x \in \mathcal{X}_{\text{help}}$.

\begin{definition}[Orthogonal Moment Function $g^{\text{hurt}}_U(W, \xi)$ on $\mathcal{X}_{\text{hurt}}$]
\label{def:hurt}
Let $X \in \mathcal{X}_{\text{hurt}}$. Define the bias correction term
\begin{align}
\label{eq:coruhurt}
&\text{cor}^{\text{hurt}}_U (W, \xi) = -Q^0(1/p(X), X) \left(- \dfrac{1-D}{\mu_0(X)} (1/p(X)) \cdot (S - s(0,X)) \right. \\
&\quad\quad + \left. \dfrac{D}{\mu_1(X)} \cdot (S - s(1,X)) - \dfrac{(1-D)S}{\mu_0(X)} \left( 1\{Y \leq Q^0(1/p(X), X)\} - 1/p(X) \right) \right) \nonumber
\end{align}
and the debiased moment function 
\begin{align*}
g^{\text{hurt}}_U (W, \xi) := m^{\text{hurt}}_U (W, \xi) + \text{cor}^{\text{hurt}}_U (W, \xi).
\end{align*}
\end{definition}

Definition \ref{def:hurt} describes the bias correction term for the moment function on $\mathcal{X}_{\text{hurt}}$. The respective terms are obtained mirroring those in \eqref{eq:coruhelp} with the roles of the treated and control group reversed. On the boundary $\mathcal{X}_0$, the moment function $m_U(W, \xi)$ involves no trimming, and the correction is not needed. As a result, we have 
\begin{align}
\label{eq:combineddebi}
g_U(W, \xi):= \begin{cases} g^{\text{help}}_U (W, \xi), \quad  X \in \mathcal{X}_{\text{help}} \\
g^{\text{hurt}}_U (W, \xi), \quad  X \in \mathcal{X}_{\text{hurt}} \\
\left(\dfrac{DS}{\mu_1(X)} - \dfrac{(1-D)S}{\mu_0(X)} \right)Y, \qquad X \in \mathcal{X}_0.
\end{cases}
\end{align}

\subsubsection{ The Lower Bound Numerator }

\begin{definition}[Moment Functions for Lower Bound]
\label{def:lowerbound}
Let $X \in \mathcal{X}_{\text{help}}$. Define the bias correction term
\begin{align}
\label{eq:corlhelp}
\text{cor}^{\text{help}}_L (W, \xi) &= Q^1(p(X), X) \bigg( \dfrac{1-D}{\mu_0(X)} \cdot (S - s(0,X)) \\
&-  \dfrac{D}{\mu_1(X)} \cdot p(X) \cdot (S - s(1,X)) - \dfrac{D S}{\mu_1(X)} \left( 1\{Y \leq Q^1(p(X), X)\} - p(X) \right) \bigg) 
\end{align}
Let $X \in \mathcal{X}_{\text{hurt}}$. Define the bias correction term
\begin{align}
\label{eq:corlhurt}
&\text{cor}^{\text{hurt}}_L (W, \xi) = -Q^0(1-1/p(X), X) \left( -\dfrac{1-D}{\mu_0(X)} (1/p(X)) \cdot (S - s(0,X)) \right. \\
&+\quad\quad  \left. \dfrac{D}{\mu_1(X)} \cdot (S - s(1,X)) + \dfrac{(1-D)S}{\mu_0(X)} \left( 1/p(X) - 1\{Y \geq Q^0(1-1/p(X), X)\}  \right) \right) \nonumber
\end{align}
The debiased moment function is
\begin{align}
\label{eq:combineddebilower}
g_L(W, \xi):= \begin{cases} m^{\text{help}}_L (W, \xi) + \text{cor}^{\text{help}}_L (W, \xi) , \quad  X \in \mathcal{X}_{\text{help}} \\
m^{\text{hurt}}_L (W, \xi) + \text{cor}^{\text{hurt}}_L (W, \xi) , \quad  X \in \mathcal{X}_{\text{hurt}} \\
\left(\dfrac{DS}{\mu_1(X)} - \dfrac{(1-D)S}{\mu_0(X)} \right)Y, \qquad X \in \mathcal{X}_0.
\end{cases}
\end{align}

\end{definition}

\subsubsection{The Always-Takers' Share  (Denominator) }
\label{sec:estim2}

In this Section, I state a debiased moment function for the  always-takers' share. Let $d=1$ and $d=0$ denote the treated and the contol state, respectively. The function
\begin{align}
g^d(W, s) := s(d,X) + \frac{1\{D=d\}}{\mu_d(X)} (S - s(d,X)), \quad d=1, 0
\end{align}
is \cite{Robins} debiased moment function for the average potential outcome $\E [ s(d,X )] = \E [ S(d) ]$.  Indeed, for each $d$, we have
$$
E [g^0(W, s) \mid X ] = s(0,X), \quad E [g^1(W, s) \mid X ] = s(1,X).
$$
Combining $g^1(W, s)$ and $g^0(W, s)$ gives 
\begin{align}
\label{eq:atshare3}
g_D(W, s) = g^0(W, s) 1\{ X \in \mathcal{X}_{\text{help}} \} + g^1(W, s) 1\{ X  \in \mathcal{X}_{\text{hurt}} \cup \mathcal{X}_0  \}.
\end{align}
By Law of Iterated Expectations, we have
\begin{align*}
\E g_D(W, s_0) &=\E s(0,X) 1\{\tau(X) > 0 \}  + \E s(1,X) 1\{\tau(X) \leq 0 \}   = \E \min (s(0,X), s(1,X)) = \pi_{\text{AT}},
\end{align*}
and $g_D(W, s)$ is a valid moment function for the always-takers' share.

\section{Overview of Estimation and Inference}
\label{sec:estim}

In this Section, I describe the estimator of the bounds as well as the confidence region for the identified set.   Section \ref{sec:fs}  presents examples of the nonparametric estimators of the first-stage nuisance functions.  Section \ref{sec:estimator} describes the estimator of generalized Lee bounds. Section \ref{sec:demo} explains the use of asymptotic results.

\subsection{Examples of First Stage Estimators }
\label{sec:fs}

In this Section, I provide examples of the first-stage estimators for the selection probability and for the conditional quantile. 

\paragraph{Conditional selection probabilities. } Suppose the selection probability $s(d,x)$ for $d \in \{1, 0\}$ can be approximated by a logistic function
\begin{align}
\label{eq:sel1}
s(d,x) = \Lambda ( B(x)' \gamma^d_{0}) + r_d(x), \quad d \in \{1, 0\},
\end{align}
where $\Lambda(\cdot) = \dfrac{\exp (\cdot)}{1 + \exp (\cdot)}$ is the logistic CDF, $B(x) = (B_1(x), B_2(x), \dots B_{p}(x))'$ is a vector of basis functions (e.g., polynomial series or splines), $\gamma^d_{0} \in \mathrm{R}^{p}$ is the pseudo-true value of the logistic parameter,  and $r_d(x)$ is its approximation error. The logistic likelihood function is 
\begin{align}
\label{eq:ll}
\ell_d(\gamma^d) =\dfrac{1}{N}  \sum_{i=1}^N (D_i = d ) \bigg(  \log ( 1+ \exp (B(X_i) '\gamma^d))  - S_i B(X_i)'\gamma^d \bigg), \quad d \in \{1, 0\}. 
\end{align}
Given an estimate $\widehat{\gamma}^d$ of $\gamma^d$, define  the estimated selection probabilities as
\begin{align}
\label{eq:sel2}
\widehat{s}(d,x) &=  \Lambda ( B(x)' \widehat {\gamma}^d), \quad d \in \{1, 0 \}
\end{align}
and the estimated CATE on selection
\begin{align*}
\widehat{\tau}(x) =  \widehat{s}(1,x) - \widehat{s}(0,x).
\end{align*}

Suppose there exists a vector $\gamma_0 \in \mathrm{R}^{p}$ with only  $s_{\gamma}$ non-zero coordinates such that the approximation error $r_d(x)$ in \eqref{eq:sel1} decays sufficiently fast relative to the sampling error:
\begin{align*}
   \left(\frac{1}{N} \sum_{i=1}^N r_d^2(X_i) \right)^{1/2} \lesssim_P \sqrt{\dfrac{s_{\gamma}^2 \log p }{N}}=:s_N, \quad \forall d=1,0.
\end{align*}
 If this condition holds, the  $\ell_1$-regularized logistic series estimator  of \cite{Program}  applies. It  takes the following form.
 
\begin{examplep}{1}[$\ell_1$-penalized LR, \cite{orthogStructural}]
\label{ex:sel}

Given the penalty parameter $\lambda_S$,   the $\ell_1$-regularized logistic estimator of $\gamma^d$  is
\begin{align}
\label{eq:onepenalty}
\widehat{\gamma}^d_{L}= \arg \max_{\gamma^d \in \mathrm{R}^{p}} \ell_d(\gamma^d) + \lambda_S  \| \gamma^d \|_1.
\end{align}

\end{examplep}

This penalty term $\lambda  \| \gamma \|_1$ prevents overfitting in high dimensions by shrinking the estimate toward zero.  \cite{Program} provides practical choices for the penalty $\lambda$  that provably guard against overfitting.   An imminent cost of applying the penalty $\lambda$ is regularization, or shrinkage, bias, that does not vanish faster than root-$N$ rate. To prevent this bias from affecting the second stage, I construct a Neyman-orthogonal moment equation for each bound.

\paragraph{Conditional outcome quantiles. }  Let $\rho_N = N^{-1/4} \log^{-1} N$ and $U=U_N = [\rho_N, 1-\rho_N]$ be a compact set in $(0,1)$. For each $u \in U$, suppose  the $u$-th conditional quantile can be approximated as
\begin{align}
\label{eq:q}
Q^d(u,x) = B(x)' \xi^{d}_0(u), \quad d \in \{1, 0\},
\end{align}
where the conditional quantile is defined as 
\begin{align*}
\xi^{d}_0 (u) &=   \arg \min_{\xi^d \in \mathrm{R}^{p} }  \E [\rho_u ( Y - B(X)' \xi^d) \mid S=1, D=d, X=x] \quad \forall x.
\end{align*}
where $t \rightarrow \rho_u(t)$ is a check function. The  quantile loss function takes the form
\begin{align*}
\ell_u(\xi^{d}):= \dfrac{1}{N} \sum_{i=1}^N  (D_i = d ) (S_i=1)\rho_u ( Y_i - B(X_i)' \xi^d),
\end{align*}

\begin{examplep}{2}[$\ell_1$-penalized QR, \cite{belloni:11}]
\label{ex:q}
Let $\widehat \sigma^2_j:= N^{-1} \sum_{i=1}^N B^2_j(X_i), \quad j=1,2,\dots, p$. Given the penalty parameter $\lambda_Q$,  the $\ell_1$-penalized quantile regression estimator is 
\begin{align}
\label{eq:onepenaltyquantile}
\widehat{\xi}_L^d(u)=\arg \min_{\xi^d \in \mathrm{R}^{p} }\ell_u(\xi^{d}) + \lambda_{Q}/N \sqrt{ u(1-u) } \sum_{j=1}^{p} \widehat {\sigma}_j | \xi^d  |
\end{align}
and the quantile estimate takes the form
\begin{align*}
\widehat{Q}^d(u,x):= B(x)'\widehat{\xi}^d_L(u), \quad d \in \{1, 0\}.
\end{align*}

\end{examplep}

\setcounter{definition}{0}

\subsection{The Estimator of Generalized Lee Bounds }
\label{sec:estimator}

Section \ref{sec:estimator}  outlines the estimator for the generalized Lee bounds. Definition \ref{def:crossfit} describes cross-fitting. Once the first-stage cross-fitted values are obtained, I estimate parametric components of the bound $N_L, N_U, \pi_{\text{AT}}$, as described in Algorithm \ref{alg:lee}.

\begin{definition}[Cross-Fitted Values]
\mbox{}
 	\label{def:crossfit} 	\begin{compactenum} 
	\item  For a random sample of size $N$, denote a $K$-fold random partition of the sample indices $[N]=\{1,2,...,N\}$ by $(J_k)_{k=1}^K$, where $K$ is  the number of partitions and the sample size of each fold is $n = N/K$. For each $k \in [K] = \{1,2,...,K\}$ define $J_k^c = \{1,2,...,N\} \setminus J_k$.
	\item For each $k \in [K]$, construct 	an estimator $\widehat{\xi}_k = \widehat{\xi}( W_{i \in J_k^c})$   of the nuisance parameter $\xi_0$ using only the data $\{ W_{j}: j \in J_k^c \}$. For any observation $i \in J_k$, define the cross-fitted value $\widehat s_i := (\widehat s_k(1, X_i), \widehat s_k(0,X_i)), \widehat \tau_i := \widehat \tau_k (X_i) = \widehat s_k(1,X_i) - \widehat s_k(0,X_i), \widehat \xi_i := \widehat \xi_k (X_i)$. 
		\end{compactenum}
\end{definition}

\begin{definition}[Debiased Estimator of the Always-Takers' Share]
\label{def:atshare}
Let $\rho_N:= N^{-1/4} \log^{-1} N$. Given the first-stage fitted values $(\widehat s_i)_{i=1}^N$ and $(\widehat \tau_i)_{i=1}^N$, define
\begin{align}
\label{eq:atshare}
\widehat \pi_{\text{AT}}:&= N^{-1} \sum_{i=1}^N g^0(W_i, \widehat s_i) 1{ \{ \widehat \tau_i \geq \rho_N \} } + g^1(W_i, \widehat s_i) 1{ \{ \widehat \tau_i < \rho_N \} } .
\end{align}
\end{definition}

The estimator $\widehat \pi_{\text{AT}}$ in Definition \ref{def:atshare} shifts the classification threshold from zero to a close point with zero mass.  This shift accommodates positive mass at the boundary.  However, if the point mass is assumed to be zero,  the sequence $\rho_N$ should be replaced by zero. In this case, the estimator \eqref{eq:atshare} reduces to the debiased estimator  proposed in \cite{kallus2020assessing} in the context of algorithmic fairness.

\begin{definition}[Debiased Estimator of the  Numerator $N_U$ and $N_L$]
\label{def:nlnu}
Let $\widehat \xi_i = \widehat \xi(X_i)$ be the nuisance parameter cross-fit estimates. Given the sequence $\rho_N:= N^{-1/4} \log^{-1} N$, define the estimated moment
\begin{align}
g_{\star}(W_i, \widehat{\xi}_i) :&= \begin{cases} 
g^{\text{help}}_{\star}(W_i, \widehat{\xi}_i), \qquad \widehat{\tau} (X_i) \geq \rho_N \\
g^{\text{hurt}}_{\star}(W_i, \widehat{\xi}_i), \qquad \widehat{\tau} (X_i) \leq -\rho_N \\
\left(\dfrac{D_i}{\mu_1(X_i)} - \dfrac{1-D_i}{\mu_0(X_i)}\right) S_i Y_i, \qquad |\widehat{\tau} (X_i)| \leq \rho_N,
\end{cases}, \qquad \star \in \{L, U\} \label{eq:nlnu}
\end{align}
where $g^{\text{help}}_{\star}(W, \xi_0)$ and $g^{\text{hurt}}_{\star}(W, \xi_0)$ are the debiased moment functions defined on the covariate partitions $\mathcal{X}_{\text{help}}$ and $\mathcal{X}_{\text{hurt}}$, respectively.
\end{definition}

Definition \ref{def:nlnu} combines the debiased moment functions. The covariate space is divided into three parts, as shown in Equation \eqref{eq:nlnu}. If the fitted value $\widehat{\tau}(X)$ falls outside the range $[-\rho_N, \rho_N]$, the covariate value $X$ is assumed to be classified correctly with high probability. In this case, the moment sample estimate $g_U(W_i, \widehat{\xi}_i)$ is calculated using the debiased moment function in Definition \ref{def:help} or Definition \ref{def:hurt}. Otherwise, if $|\widehat{\tau}(X)|$ is too small, the covariate value $X$ is deemed to be difficult to classify. In this case, the moment sample estimate $g_U(W_i, \widehat{\xi}_i)$ is set to its boundary limit value.

 \begin{algorithm}
 Input: estimated first-stage fitted values $(\widehat s(0,X_i), \widehat s(1,X_i), \widehat \tau(X_i),  \cup_{u \in U} \widehat{Q}(u ,X_i))_{i=1}^N$. Estimate 
\begin{algorithmic}[1]

\STATE The always-takers' share as in Definition  \ref{def:atshare} $$\widehat \pi_{\text{AT}}:=N^{-1} \sum_{i=1}^N g^{\rho_N}_D(W_i, \widehat{\xi}_i).$$ 

\STATE The numerators $N_U$ and $N_L$ as in Definition  \ref{def:nlnu}
$$\widehat N_U:=N^{-1} \sum_{i=1}^N g_U(W_i, \widehat{\xi}_i) , \qquad \widehat N_L:= N^{-1} \sum_{i=1}^N g_L(W_i, \widehat{\xi}_i).$$ 

\STATE The preliminary bounds 
\begin{align}
\label{eq:ratio}
\widehat{\beta}_L:=\dfrac{ \widehat N_L }{\widehat \pi_{\text{AT}}},  \quad \widehat{\beta}_U:= \dfrac{\widehat N_U  }{\widehat \pi_{\text{AT}}}
\end{align}
and the  sorted bounds
\begin{align}
\label{eq:genest}
\widetilde{\beta}_L:= \min (\widehat{\beta}_L, \widehat{\beta}_U), \quad \widetilde{\beta}_U:= \max (\widehat{\beta}_L, \widehat{\beta}_U).
\end{align}

\end{algorithmic}

\caption{Generalized Lee Bounds.}
\label{alg:lee}

\end{algorithm}

\subsection{Asymptotic Distribution of Second-Stage Parameters}
\label{sec:demo}

Consider the vector $(N_L, N_U, \pi_{\text{AT}})$ of second-stage parameters. In the large sample, the asymptotic distribution of $(\widehat N_L, \widehat N_U,  \widehat \pi_{\text{AT}})$ is
 \begin{align*}
    \sqrt{N} \left( \begin{matrix} \widehat{N}_L - N_L \\
     \widehat{N}_U - N_U \\
     \widehat{\pi}_{\text{AT}} - \pi_{\text{AT}} \\
     \end{matrix} \right) \Rightarrow^d N \left (0, \Gamma \right).
\end{align*}
Define the  asymptotic variance matrix 
\begin{align}
\label{eq:gamma}
\Gamma = \text{Var} (g_L(W, \xi_0), g_U(W, \xi_0), g_D(W, \xi_0)).
\end{align}
Suppose $\pi_{\text{AT}}>0$.  Delta method  gives the asymptotic approximation for $(\widehat{\beta}_L, \widehat{\beta}_U)$:
 \begin{align*}
    \sqrt{N} \left( \begin{matrix} \widehat{\beta}_L - \beta_L \\
     \widehat{\beta}_U - \beta_U     \end{matrix} \right) \Rightarrow^d N \left (0, \Omega \right), 
     \end{align*}
where the asymptotic covariance matrix $\Omega $ is
\begin{align}
 \Omega = Q \Gamma Q^{T}, \quad Q = ( \pi_{\text{AT}} )^{-1} \begin{pmatrix} 1 & 0 & -\beta_L \\
0 & 1 & -\beta_U  \\
\end{pmatrix}. \label{eq:omega}
\end{align}   
Let $\widehat \Gamma$ be a consistent estimator of  $\Gamma$, which I assume exists. Define $
\widehat \Omega:= \widehat Q \widehat \Gamma \widehat Q^{T}
$
and let  the diagonal elements of $\widehat \Omega$ by $ \widehat \Omega_{LL}$ and $\widehat \Omega_{UU}$.

\paragraph{Confidence Region for the identified set $[\beta_L, \beta_U]$. }  Given a significance level $\alpha$, a $(1-\alpha)$-Confidence Region for the identified set $[\beta_L, \beta_U]$ takes the form
\begin{align}
\label{eq:cralpha}
CR^{1-\alpha}  := [ \widehat \beta_L - N^{-1/2} \widehat \Omega^{1/2}_{LL} c_{1-\alpha/2}, \quad \widehat \beta_U +  N^{-1/2} \widehat \Omega^{1/2}_{UU} c_{1-\alpha/2}] 
\end{align}
where   $c_{1-\alpha/2}$ is $(1-\alpha/2)$-quantile of $N(0,1)$.   the  endpoints of the confidence region $CR^{1-\alpha}(c_{\alpha/2}, c_{1-\alpha/2})$ may not be ordered by construction. As shown in \cite{chernozhukovmelly},  sorting the endpoints can only improve the coverage\begin{footnote}{This paper focuses on the pointwise coverage, where the true values of $N_U, N_L, \pi_{\text{AT}}$ do not change with sample size. } \end{footnote} property.

\section { Asymptotic Theory for Generalized Lee Bounds }
\label{sec:theory}

In this Section, I describe the assumptions and state the asymptotic results. Section \ref{sec:fs5} describes the regularity conditions on the data generating process. Section \ref{sec:ss5} outlines the first-stage rate requirements. Section \ref{sec:res5} presents the asymptotic results. Section \ref{sec:fsver} verifies Assumption \ref{ass:firststage} in the context of high-dimensional sparse design. Section \ref{sec:remarks} introduces basic generalized bound, a non-sharp alternative to the proposed bound. Section \ref{sec:pscore}  sketches the moment equation for the case of an unknown propensity score.

\subsection{Assumptions}
\label{sec:fs5}

 \begin{assumptionp}{3}[Strict Overlap]
\label{ass:boundedoutcome2}
(SO). There exists an absolute constant  $\kappa \in (0, 1/2)$ so that $s(d,x) \in ( \kappa, 1- \kappa ) \quad$ for all $d \in \{1, 0\}$ and all covariate values $x$. Likewise, the propensity score  $\mu_1(x):= \Pr (D=1 \mid X=x) \in ( \kappa, 1- \kappa )$ for any $x$. 
\end{assumptionp}

Assumption \ref{ass:boundedoutcome2} requires the selection probabilities and the propensity score to be bounded away from zero and one, which is a standard condition in the literature.

  \begin{assumptionp}{4}[Margin Assumption]
\label{ass:boundedoutcome4}
(MA). There exist absolute finite constants $\bar{B}_f$ and $\eta$ so that 
\begin{align}
\label{eq:ma1}
{\Pr} ( 0< | \tau(X)  | \leq t) \leq \bar{B}_f t, \quad 0 \leq t \leq \eta.
\end{align}
\end{assumptionp}

Assumption \ref{ass:boundedoutcome4} assumes that the distribution of the  function $\tau(X)$ is sufficiently smooth. For example, if $\tau(X)$ is continuously distributed with a bounded density,  \eqref{eq:ma1} holds. This assumption is routinely assumed in classification analysis (\cite{MammenTsybakov, Tsybakov}) and empirical welfare maximization (\cite{KitagawaTetenov, MbakopTabord}).  

\begin{assumptionp}{5}[Continuously Distributed Bounded Outcome]
\label{ass:boundedoutcome3}
(BO) Bounded Outcome: There exists a constant $M < \infty$ such that $|Y| \leq M$ almost surely. (REG): For $d \in \{1, 0\}$, there exist constants $C_f$ and $B_f$ such that for the support $\mathcal{Y}^d_x$ of the conditional distribution $Y \mid D=d, X=x$, we have
\begin{enumerate}
  \item The conditional density $f^d(y \mid x) := f_{Y \mid S=1, D=d, X=x}(y \mid x)$ is uniformly bounded from above by $C_f$ for all $y \in \mathcal{Y}^d_x$.
  \item The infimum of $f^d(y \mid x)$ over $x \in \mathcal{X}$ and $y \in \mathcal{Y}_x$ is bounded away from zero by $B_f$.
  \item The derivative of $y \rightarrow f^d(y \mid x)$ is continuous and bounded from above in absolute value by $C_f$ uniformly over $y \in \mathcal{Y}^d_x$.
\end{enumerate}
\end{assumptionp}

Assumption \ref{ass:boundedoutcome3} requires the outcome  to have bounded support and to be continuously distributed without point masses. This condition  is routinely imposed for the consistency of unpenalized (\cite{CherChetQuant}) and $\ell_1$-penalized (\cite{belloni:11}) quantile estimators. Furthermore,  the conditional density must be bounded away from zero on its support.  For example, Assumption \ref{ass:boundedoutcome3} accommodates truncated normal and uniform distributions, but rules out regular normal distribution.

\subsection{First-Stage Rate Requirements }
\label{sec:ss5}

\begin{definition}[Selection Rate]
\label{def:sel}
There exist a sequence of numbers $\phi_N = o(1)$ and a sequence of sets $S^d_N, d \in \{1, 0\}$ such that the first-stage estimates $\widehat{s}(d, x)$ of the true function $s(d, x)$ belong to $S^d_N$ with probability at least $1 - \phi_N$. The sets $S^d_N$ shrink at the following rate:
\begin{align*}
s_N^p:=\sup_{d \in \{1,0\}} \sup_{\bar{s} \in S^d_N} \left(\E_{X} | \bar{s}(d, X) - s(d,X)|^p\right)^{1/p}, \quad 1 \leq p \leq \infty,
\end{align*}
and the functions in $S^d_N$ satisfy $\inf_{x \in \mathcal{X}} \inf_{d \in \{1,0\}} s(d,x) > \kappa/2 > 0$. Let $s_N$ and $s^1_N$ and $s_N^{\infty}$ be the mean square, the $L_1$- and  sup-norm rates, respectively.
\end{definition}

\begin{definition}[Quantile Rate]
\label{def:q}
There exist a sequence of numbers $\phi_N = o(1)$ and a sequence of sets $Q^d_N$ such that the first-stage estimate $\widehat{Q}^d(u,x)$ of $Q^d(u,x)$ shrinks uniformly over $\mathcal{U}_N = [2/\kappa \rho_N, 1- 2/\kappa \rho_N]$ with $\rho_N = N^{-1/4} \log^{-1} N$ at the following rate:
\begin{align*}
q_N^p:= \sup_{d \in \{1, 0\}}\sup_{ \bar{Q}^d \in Q_N^d} \sup_{u \in \mathcal{U}_N} \left(\E | \bar{Q}^d(u, X) - Q^d(u,X) |^p\right)^{1/p}, \quad 1 \leq p \leq \infty,
\end{align*}
where the sets $Q_N^d$ consist of almost surely $M$-bounded functions. Let $q_N$ and $q_N^1$ be the mean square and $L_1$ rates, respectively.
\end{definition}

Assumption \ref{ass:firststage} places bounds on selection and quantile rates in various norms.  For Examples \ref{ex:sel} and \ref{ex:q}, the rates are defined in terms of the model primitives (i.e., the sparsity indices) and are verified below.

\begin{assumptionp}{6}[First-Stage Rates]
\label{ass:firststage}
The sequences $s_N$, $q_N$, $s_N^{\infty}$, $s_N^1$, and $q_N^1$ obey the following bounds:
\begin{enumerate}
\item Mean square rates are sufficiently fast:
\begin{align}
\label{eq:extrarates}
s_N + q_N = o(N^{-1/4}).
\end{align}
\item Worst-case selection rate $s_N^{\infty}$ is sufficiently fast
\begin{align}
s_N^{\infty} = o(N^{-1/4} \log^{-1} N). \label{eq:extrarates2}
\end{align}
\item Estimators are consistent in  $L_1$ norm
\begin{align*}
s_N^1 = o(1), \quad q_N^1 = o(1).
\end{align*}
\end{enumerate}
\end{assumptionp}

Assumption \ref{ass:firststage} states that the functions $s(0,x), s(1,x), Q^1(u,x), Q^0(u,x)$ converge in mean square and sup-rate with sufficiently fast rate.   The first condition  \eqref{eq:extrarates}  controls the higher-order bias; it is a classic  assumption in the semiparametric literature (see, e.g., \cite{Newey1994}, \cite{chernozhukov2016double}).      
\subsection{Main Result}
\label{sec:res5}

\begin{theorem}[Generalized Lee bounds: Asymptotic Theory]
\label{thrm:condmonot}
Suppose Assumptions \ref{ass:identification:treat:cond}, \ref{ass:indep}, \ref{ass:boundedoutcome2}--\ref{ass:firststage} hold. Then, the estimator $(\widehat{N}_L, \widehat{N}_U, \widehat{\pi}_{\text{AT}})$ is consistent and asymptotically normal:
\begin{align*}
\sqrt{N} \left( \begin{matrix} \widehat{N}_L - N_L \\
\widehat{N}_U - N_U \\
\widehat{\pi}_{\text{AT}} - \pi_{\text{AT}} \\
\end{matrix} \right) \Rightarrow N \left(0, \Gamma \right),
\end{align*}
where the asymptotic variance matrix $\Gamma$ is given in \eqref{eq:gamma}. As a result, if $\pi_{\text{AT}} > 0$, the preliminary bounds of Algorithm \ref{alg:lee} are asymptotically Gaussian:
\begin{align*}
\sqrt{N} \left( \begin{matrix} \widehat{\beta}_L - \beta_L \\
\widehat{\beta}_U - \beta_U \\
\end{matrix} \right) \Rightarrow N \left(0, \Omega \right),
\end{align*}
where $\Omega = Q \Gamma Q^T$ as in \eqref{eq:omega}.
\end{theorem}

Theorem \ref{thrm:condmonot} delivers a root-$N$ consistent, asymptotically normal estimator of $(\beta_L, \beta_U)$ assuming the conditional probability of selection and conditional quantile are estimated at a sufficiently fast rate. In particular, this assumption is satisfied when only a few covariates affect selection and the outcome.

\setcounter{remark}{0}

\begin{remark}[Strong\begin{footnote}{The 2020 version of the manuscript was based on this assumption. The author thanks to the  discussants and referees who pointed out its weaknesses. } \end{footnote} separation from the boundary]
\label{rm:discrete}
Consider Assumption \ref{ass:identification:treat:cond} holding with $\Pr(\mathcal{X}_0) = 0$. Given a fixed $\epsilon > 0$, a separation condition
\begin{align}
\label{eq:strong}
\inf_{x \in \mathcal{X}} |\tau(x)| = \inf_{x \in \mathcal{X}} |s(1,x) - s(0,x)| > \epsilon
\end{align}
may be plausible in settings with discrete covariates. If $s_N^{\infty} = o(1)$, the subjects are correctly classified into $\mathcal{X}_{\text{help}}$ and $\mathcal{X}_{\text{hurt}}$ with probability approaching one. Then, the statement of Theorem \ref{thrm:condmonot} holds under Assumptions \ref{ass:boundedoutcome2}, \ref{ass:boundedoutcome3} (REG), and Assumption \ref{ass:firststage} (1)-(2), while (MA) and (BO) are no longer required. Furthermore, the relevant set of estimated quantiles $U$ reduces to $U := [2\epsilon/\kappa, 1-2\epsilon/\kappa]$ and no longer approaches $(0,1)$ as the sample size grows. As a result, unbounded outcome distributions such as Gaussian satisfy Assumption \ref{ass:boundedoutcome3} (REG).
\end{remark}

\subsection{Verification of Assumption \ref{ass:firststage}}
\label{sec:fsver}

In this Section, I verify Assumption \ref{ass:firststage} in the context of high-dimensional sparse models.

\begin{examplep}{1'}[Example \ref{ex:sel}, cont.]
\label{rm:sel}
Consider the model \eqref{eq:sel1} with $p=\dim(B(X)) \gg N$. Suppose there exists a vector $\gamma^d_{0} \in \mathrm{R}^{p}$ with only $s_{\gamma}$ non-zero coordinates such that the approximation error $r_d(x)$ in \eqref{eq:sel1} decays sufficiently fast relative to the sampling error:
\begin{align*}
\sup_{d \in {1, 0 }} \left(\frac{1}{N} \sum_{i=1}^N r_d^2(X_i) \right)^{1/2} \lesssim_P \sqrt{\dfrac{s_{\gamma}^2 \log p }{N}}.
\end{align*}
Then, the $\ell_1$-regularized LR of Example \ref{ex:sel} with the data-driven choice of penalty as in \cite{Program} attains the mean square rate $s_N: = O\left(\sqrt{s_{\gamma} \log p /N} \right)$ and $s^1_N :=s^{\infty}_N: = O \left(\sqrt{s_{\gamma}^2 \log p /N} \right)$. Thus,
\begin{align*}
s_{\gamma}^2 \log^2 N \log p = o (N^{1/2})
\end{align*}
is sufficient for $s^{\infty}_N = o(N^{-1/4} \log^{-1} N)$.
\end{examplep}

A major challenge of this paper is to verify the mean square quantile rate on the set of quantile levels $U_N = [2 \rho_N/\kappa, 1-2 \rho_N/\kappa]$ which involves extreme quantiles. Here, Assumption \ref{ass:boundedoutcome3} focuses on  bounded outcomes whose density is bounded away from zero on the support. For example,  $Y \sim U[0, M] \mid X, S=1, D=1$, we have
\begin{align*}
\bar{B}_f:= \lim_{t \rightarrow 0+} f_{Y \mid D=1, S=1, X} (t) = \lim_{t \rightarrow M-} f_{Y \mid D=1, S=1, X} (t)  = \dfrac{1}{M} >0.
\end{align*}
As a result, extreme quantiles of level $\rho_N$ and $1-\rho_N$ 
  can be consistently estimated at a mean square rate $\sqrt{ s\log p/ (N \rho_N)}$, where $N \rho_N$ acts as an effective sample size. Choosing the trimming threshold $\rho_N = N^{-1/4} \log^{-1} N$ makes the mean square quantile rate $q_N = o(N^{-1/4})$ plausible in Example  \ref{rm:q}.

\begin{examplep}{2'}[Example \ref{ex:q}, cont.]
\label{rm:q}
Consider the model \eqref{eq:q}. Suppose Assumption \ref{ass:boundedoutcome3} holds.  Then, the $\ell_1$-regularized quantile regression of \cite{belloni:11} with data-driven choice of penalty attains the mean square rate and $L_1$ rates:
\begin{align}
\label{eq:meanrate}
q_N= O\left( \sqrt{ \dfrac{ s_Q \log p }{N \rho_N} }\right), \quad q^1_N=O \left( \sqrt{ \dfrac{ s^2_Q \log p }{N \rho_N} } \right),
\end{align}
where $\rho_N$ is the trimming threshold in Definition \ref{def:nlnu}, and $U_N = [2 \rho_N/\kappa, 1-2 \rho_N/\kappa]$. Here, the quantity $N \rho_N$ is the effective sample size used to estimate the conditional $(2 \rho_N/\kappa)$-quantile. The proposed choice $\rho_N:= N^{-1/4} \log^{-1} N$ ensures that
\begin{align*}
\log N s^2_Q \log p = o(N^{1/4})
\end{align*}
holds, which suffices for $q_N = o(N^{-1/4})$ and $q^1_N = o(1)$.
\end{examplep}

\subsection{Basic Generalized Bound}
\label{sec:remarks}

In this Section, I present an alternative generalization of Lee bounds under conditional monotonicity, which does not require trimming (and, therefore, quantiles) to be conditional on covariates. Suppose Assumptions \ref{ass:identification:treat:cond} and \ref{ass:identification:treat}(b) hold.  Let $\bar{\beta}^{\text{help}}_U$ and $\bar{\beta}^{\text{hurt}}_U$ be the basic Lee bounds of Section \ref{sec:overview}, defined on $\mathcal{X}_{\text{help}}$ and $\mathcal{X}_{\text{hurt}}$, respectively. Focusing on the boundary $ \mathcal{X}_0$, define the treatment-control difference as 
$$\bar{\beta}^0 := \E[ Y \mid D=1, S=1, X \in \mathcal{X}_0] - \E[ Y \mid D=0, S=1, X \in \mathcal{X}_0].$$
Likewise, let
\begin{align}
\label{eq:shelpshurt}
S_{\text{help}}:= \int_{\mathcal{X}: \tau(x)>0} s(0,x) f_X(x) dx, \quad S_{\text{hurt}}:= \int_{\mathcal{X}: \tau(x)<0} s(1,x) f_X(x) dx
\end{align}
and 
\begin{align}
\label{eq:0}
S_0:= \int_{\mathcal{X}: \tau(x)=0} s(1,x) f_X(x) dx = \int_{\mathcal{X}: \tau(x)=0} s(0,x) f_X(x) dx=\bar{\beta}^0.
\end{align}
By construction, $S_{\text{help}} +S_{\text{hurt}}+ S_0 = \pi_{\text{AT}}$. Aggregating  over the covariate space gives  basic generalized bound:
\begin{align}
\label{eq:nocovariate}
\bar{\beta}_U = \dfrac{ \bar{\beta}^{\text{help}}_U S_{\text{help}} + \bar{\beta}^{\text{hurt}}_U S_{\text{hurt}}+ \bar{\beta}^0 }{ \pi_{\text{AT}}}.
\end{align}
If Assumption \ref{ass:identification:treat} holds, $\bar{\beta}_U  = \bar{\beta}^{\text{help}}_U$ reduces to  basic Lee bound in \eqref{eq:betabasic}. This bound  is a direct generalization of basic Lee bound to the case of conditional monotonicity.

\begin{lemma}[Basic Generalized Bound]
\label{lem:basicbound}
Suppose Assumptions \ref{ass:identification:treat:cond} and \ref{ass:identification:treat}(b) hold. Then, $\bar{\beta}_U $ is a valid bound on $\beta_0$ obeying 
$$
\beta_0 \leq \beta_U \leq \bar{\beta}_U.
$$
\end{lemma}

\subsection{Unknown propensity score}
\label{sec:pscore}

In this Section, I consider the case when the propensity score is unknown and needs to be estimated.   Let  $\beta^{\text{help}}_U(x)$ be the conditional Lee bound defined in \eqref{eq:condupperbound}, and let $\beta^{1\text{help}}_U(x)$ and $\beta^{0\text{help}}_U(x)$  be its first and second summand, respectively. Likewise, let $\beta^{\text{hurt}}_U(x) $ be the conditional Lee bound defined in \eqref{eq:betauxhurt}, and let $\beta^{1\text{hurt}}_U(x) $  and $ \beta^{0\text{hurt}}_U(x)$ be its first and second summand. Below, I describe the debiased moment function for the  upper bound $\beta_U$ in \eqref{eq:sharpbetau}.

For $x \in \mathcal{X}_{\text{help}}$,  define the Riesz representer function 
\begin{align}
\Lambda_U(x):=\Lambda^{\text{help}}_U(x):= -\left(\dfrac{\beta^{1\text{help}}_U(x)}{\mu_1(x)} + \dfrac{\beta^{0\text{help}}_U(x)}{\mu_0(x)} \right) s(0,x).
\end{align}
For $x \in \mathcal{X}_{\text{hurt}}$, define 
\begin{align}
\Lambda_U(x):=\Lambda^{\text{hurt}}_U(x):= -\left(\dfrac{\beta^{1\text{hurt}}_U(x)}{\mu_1(x)} + \dfrac{\beta^{0\text{hurt}}_U(x)}{\mu_0(x)} \right) s(1,x).
\end{align}
On the boundary, there is no trimming, and the two functions coincide
$$
\Lambda_U(x) = \Lambda^{\text{help}}_U(x) =\Lambda^{\text{hurt}}_U(x), \quad x \in \mathcal{X}_0.
$$
The bias correction term for the propensity score is  
\begin{align}
\label{eq:corpscore}
\text{cor}_{U\mu} (W, \Lambda_U, \mu_1):= \Lambda_U(X)( D- \mu_1(X)). 
\end{align}
The  debiased moment function takes the form
\begin{align}
\label{eq:debpscore}
g_U(W, \xi, \mu_1, \Lambda_U):= m_U(W, \xi) + \text{cor}_U(W, \xi) + \text{cor}_{U\mu} (W, \mu_1, \Lambda_U).
\end{align}
In particular, the propensity score correction term depends on the conditional trimmed mean function $\Lambda_U(x)$.    In a low-dimensional smooth setting,  the function 
$\Lambda_U(x)$ can be estimated by  the local linear regression estimator proposed in \cite{olma2021nonparametric}.  In a high-dimensional sparse setting, one could use the automatic debiasing approach of \cite{chernozhukov2021automatic}. 

 \section{JobCorps revisited  }
\label{sec:empirical}

\subsection{Overview of JobCorps data}

\label{sec:overviewjobcorps}

\paragraph{Data description. } \cite{LeeBound} studies the effect of winning a lottery to attend JobCorps, a federal vocational and training program, on applicants' wages. In the mid-1990s, JobCorps used lottery-based admission to assess its effectiveness. The control group of $5, 977$ applicants was essentially embargoed from the program for three years, while the remaining applicants (the treated group) could enroll in JobCorps as usual.  The sample consists of $9,145$ JobCorps applicants and has data on lottery outcome, hours worked and wages for 208 consecutive weeks after random assignment.  In addition, the data contain educational attainment, employment, recruiting experiences, household composition, income, drug use, arrest records, and applicants' background information. These data were collected as part of a  baseline interview, conducted by Mathematica Policy Research (MPR) shortly after randomization (\cite{Schochet}). Lee has condensed this information to 28 covariates, including demographic characteristics, parental education, and income, wages, and hours of work at baseline (see Table \ref{tab:lee} in Appendix or Table 2, \cite{LeeBound}). This section 
considers a richer specification, which includes frequency and type of drug use, arrest experiences, reasons for joining JobCorps, and occupation at baseline. I shall refer to these covariate choices  as Lee's covariates (28) and All covariates (>1,000), respectively.

\begin{figure}
\centering
\caption{ Treatment-control differences in employment rate by week. }\par\medskip
   \includegraphics[scale = 0.45]{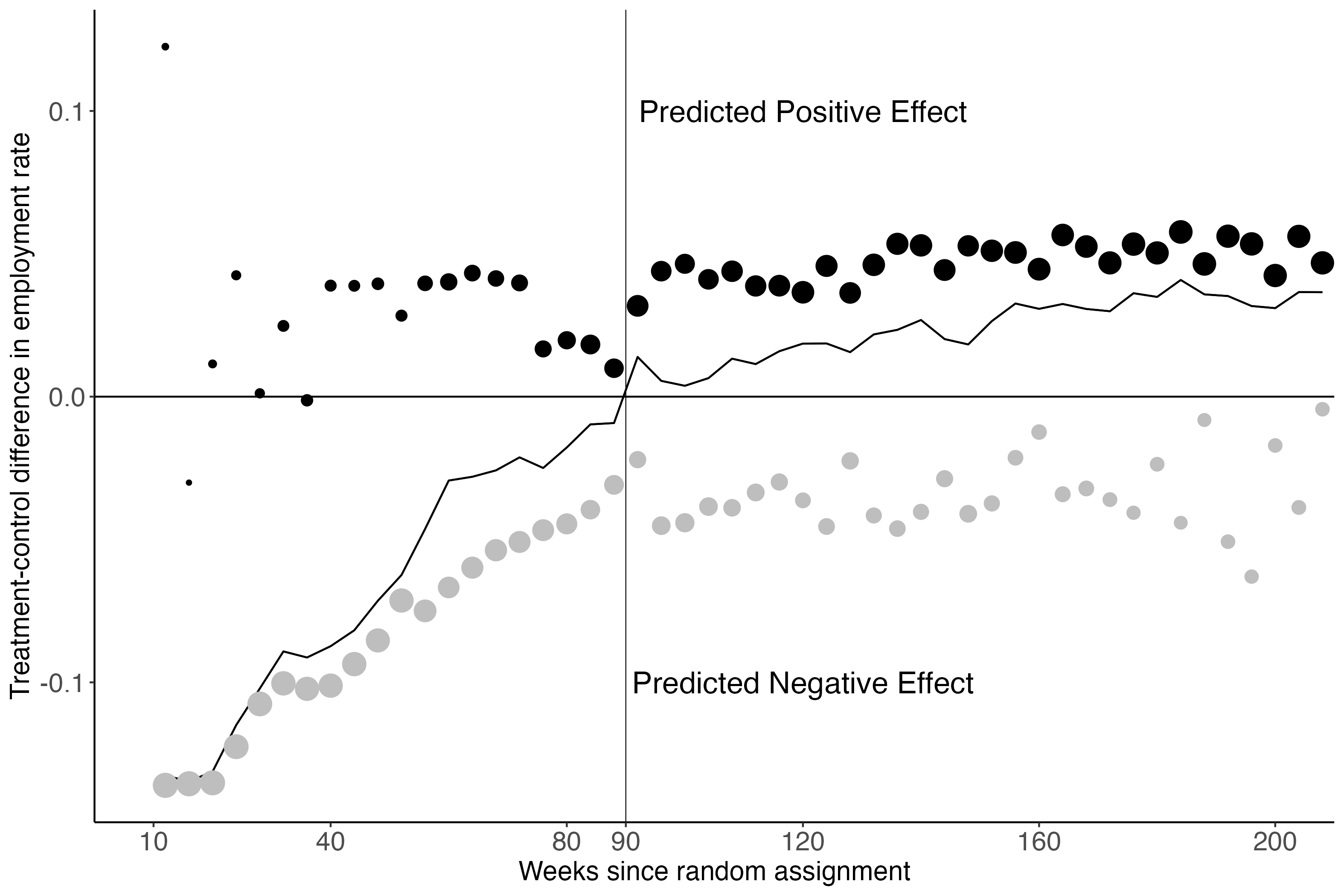}
   \caption*{  Notes. The horizontal axis shows the number of weeks since random assignment. The vertical axis shows the treatment-control difference in employment rate.  The black dot represents applicants whose conditional employment effect $\tau(x)$ is positive, and  the gray dot is its complement. The black line represents all $9, 145$ applicants.      (For each week, $\tau(x)$ is defined as in equation \eqref{eq:cateselection} and estimated as in equation \eqref{eq:sel}).   The size of each dot is proportional to the fraction of applicants.  The sample size $N=9, 145$.     Computations use design weights. }
    	\label{fig:Figure1}
     \end{figure}

\subsection{Testing unconditional monotonicity. }

\label{sec:testing}

Baseline covariates can detect violations of  unconditional monotonicity.  If this assumption holds, the conditional average treatment effect on employment 
    $\tau(x)$ in \eqref{eq:cateselection} must be either non-positive or non-negative for all covariate values. Consequently, it cannot be the case that 
\begin{align} \label{eq:prob} \text{Prob} ( \tau(X) >0) >0  \quad \text{ and }  \quad \text{Prob} ( \tau (X) <0) >0 \end{align} 
for any $X$ taken to be the subset of all covariates. 

\begin{figure}
\centering
\caption{  Fraction of JobCorps applicants with positive conditional employment effect  by week.} \par\medskip
         \includegraphics[scale = 0.45]{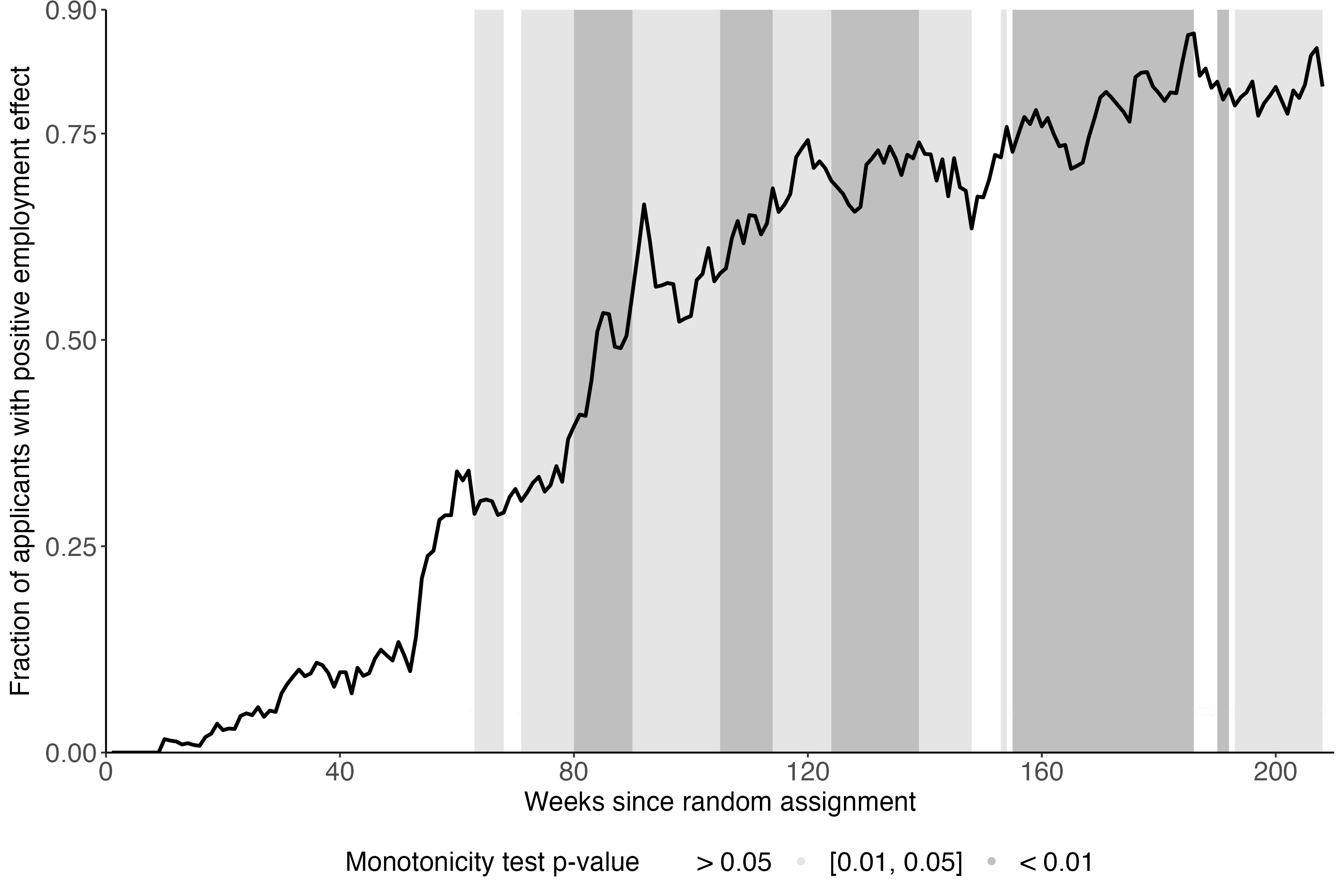}
	 \caption*{ Notes. The horizontal axis shows the number of weeks since random assignment. The vertical axis shows  the fraction of  applicants whose conditional employment effect $\tau(x)$ is positive.  Following week 60, a week is shaded  if the   test statistic $T$ exceeds the  critical value  at  the $p=0.01$ (dark gray) or $p \in [0.05, 0.01)$ (light gray) significance level. The covariate vector $X$ consists of 28 Lee's covariates. For each week, $\tau(x)$ is defined in equation \eqref{eq:cateselection} and estimated as in equation \eqref{eq:sel}, the null hypothesis is as in  equation \eqref{eq:hypot1},  the	 test statistic $T$ is as in equation \eqref{eq:teststat}, and  the test cells and critical values are as defined in Table \ref{tab:covgroups}. The sample size $N=9, 145$.     Computations use design weights. 
	 	 }
		\label{fig:Figure2}

 \end{figure}

The first exercise is to estimate $s(1,x)$ and $s(0,x)$ by a week-specific cross-sectional logistic regression. Let
\begin{align}
\label{eq:sel}
s(D,X) = \Lambda (X ' \alpha_0 +  D \cdot X' \gamma_0), 
\end{align}
where $\Lambda(\cdot) = \dfrac{\exp (\cdot)}{1 + \exp (\cdot)}$ is the logistic CDF, $X$ is a vector of baseline covariates that includes a constant and 28 covariates Lee selected, $D \cdot X$ is a vector of covariates interacted with treatment,  and $\alpha$ and $\gamma$ are fixed vectors. Figures \ref{fig:Figure1} and  \ref{fig:Figure2} show the results: the share of subjects with positive selection effect (Figure \ref{fig:Figure2}, solid black line) and the average employment effect for subjects with $\tau(X)<0$ and $\tau(X)>0$ (Figure \ref{fig:Figure1}).

The second exercise is to test monotonicity without relying on a particular logistic specification\begin{footnote}{On p. 1085, Lee says ``when $\Pr (S=1 \mid D=1) - \Pr (S=1 \mid D=0)=\E[ \tau(X) ] =0$, there is a limited test of whether monotonicity holds''. Lee considers a logistic regression of treatment $D=1$ (as outcome) on $X$ in the selected sample $S=1$. In contrast to Lee, this paper tests monotonicity using covariates, which does not require assuming $\E[ \tau(X) ] =0$. Furthermore, the covariate-based test \eqref{eq:hypot1} may have higher power since it uses the full sample (and not only observations with $S=1$).  } \end{footnote}. For each week, I select a small number of discrete covariates and partition the sample into discrete cells $C_j, \quad j \in \{1,2,\dots, J\}$, determined by covariate values. For example, one binary covariate corresponds to $J=2$ two cells. By monotonicity, the vector of cell-specific treatment-control differences in employment rates, $\mathbf{\mu}= (\E[\tau(X) | X \in C_j])_{j=1}^J$, must be non-negative:
\begin{align}
\label{eq:hypot1}
H_0: \quad  (-1) \cdot \mathbf{\mu} \leq 0. 
\end{align}
The test statistic for the hypothesis in equation \eqref{eq:hypot1} is 
\begin{align}
\label{eq:teststat}
T= \max_{1 \leq j \leq J} \dfrac{ (-1) \cdot \widehat{\mu}_j}{\widehat{\sigma}_j}, 
\end{align}
and the critical value is the self-normalized critical value of \cite{CCK}.

Figure  \ref{fig:Figure2}  plots the fraction of subjects with  a positive JobCorps effect on employment in each week (that is, the fraction of applicants in black dots in Figure \ref{fig:Figure1}).  In the first weeks after  random assignment, there is no evidence of a positive JobCorps effect on employment for any group. By the end of the second year (week 104),  JobCorps increases employment for nearly $75\%$ of the individuals, and this fraction rises to $0.9$ by the end of the study period (week 208).  This pattern is consistent with the JobCorps program description. While being enrolled in JobCorps, participants cannot hold a job, which is known as the lock-in effect (\cite{FloresLagunes}). After finishing the program,  JobCorps graduates may have gained employment skills that help them outperform the control group. However, the share of subjects with positive employment effect never reaches $100\%$, even   four years after RA.

Figure \ref{fig:Figure2} shows the results of testing the inequality in \eqref{eq:hypot1} for each week. The direction of the employment effect varies  with socio-economic factors. For example, the applicants who received AFDC benefits during the 8 months before RA     or who belonged to median income and yearly earnings groups experience a significantly positive ($p \leq 0.05$) employment effect at weeks $60$--$89$, although the average effect is significantly negative. As another example,  the applicants who  answered ``1: Very important'' to the question  ``How important was getting away from community on the scale from $1$ (very important) to $3$ (not important)?'' ($\text{R\_Home}=1$)  and who smoke marijuana or hashish a few times each months experience a significantly negative ($p\leq 0.05$) employment effect at week $117$--$152$ despite the average effect being positive.      Finally,  at week $153$--$186$,  the average JobCorps effect is significantly negative for subjects whose most recent arrest occurred less than 12 months ago ($\text{MARRCAT=1}$), despite the average effect being positive.  

Figure   \ref{fig:Figure1}  plots the average effects on employment rates across weeks.  The effect is shown to be negative for weeks $1-89$ and positive thereafter.  Remarkably, week 90 is also the  only week whose average wage effect was found significant out of four weeks  considered (\cite{LeeBound}). For this reason, the rest of the section focuses on week 90 as the most interesting one.

\begin{table}
   \captionof{table}{Bounds on the JobCorps  effect  on week 90 log wages under unconditional monotonicity}
 \label{tab:JCuncond}
        \centering
       \begin{tabularx}{\linewidth}{p{3cm} *{3}{Y}}
      \toprule

    \\
            \multicolumn{1}{p{3cm}}{ \centering   Target parameter }  & $[\bar{\beta}_L, \bar{\beta}_U]$ in \eqref{eq:betabasic} & $[\beta_L, \beta_U]$ in  \eqref{eq:sharpbetau}              \\
\\

 \multicolumn{1}{p{3cm}}{ \centering   Covariates }   & None & 1 (split into 5 groups) \\
    & (1) & (2)    \\
    \midrule
      \multicolumn{1}{p{3cm}}{ \centering   Bounds  } & [0.048, 0.049] & [0.046, 0.055]    \\       
        \multicolumn{1}{p{3cm}}{ \centering  $95 \%$ Confidence Region  }          &    (0.012, 0.087) & (0.014, 0.087) \\

  \bottomrule

\end{tabularx}
\caption*{
Notes. Table shows estimated  bounds  in square brackets and  the $95 \%$ confidence region for the identified set  in parentheses. 
Column (1) reports basic Lee bounds as in \eqref{eq:betabasic}. Column (2) reports discrete Lee bound as in \eqref{eq:sharpbetau}.  The five discrete groups are formed according to whether the predicted week 90 wage potential is within intervals defined by $\$6.75$, $\$7$, $\$7.50$, and $\$ 8.50$. This exercise replicates the calculation in Table 5,  \cite{LeeBound}, with week 90 data in place of week 208 data.  Computations use design weights.The sample size $N=9, 145$.  The asymptotic variance for the $95 \%$ Confidence Region is based on $B=1,000$ bootstrap repetitions. }
\end{table}

\begin{landscape}
   \captionof{table}{Bounds on the JobCorps  effect  on week 90 log wages under conditional monotonicity}
 \label{tab:JC}
        \centering 
      
      \begin{tabularx}{\linewidth}{p{3cm} *{6}{Y}}
\toprule
\multicolumn{1}{p{3cm}}{\centering  Covariates }  & \multicolumn{2}{c}{28 (Lee)} & \multicolumn{2}{c}{>1000 (All)} & \multicolumn{2}{c}{43 (Lasso)} \\
        \cmidrule(lr){2-3} \cmidrule(lr){4-5} \cmidrule(lr){6-7}  
 & (1) & (2) & (3) & (4) & (5) & (6) \\
\midrule
\multicolumn{1}{p{3cm}}{\centering Always-takers' share} 
& \multicolumn{2}{c}{0.44} &  \multicolumn{2}{c}{0.44} &  \multicolumn{2}{c}{0.44}  \\

\midrule

\multicolumn{1}{p{3cm}}{\centering Bounds} 
& [-0.060, 0.124] & [-0.072, 0.004] & [-0.021, 0.165] & [-0.088, 0.039] & [-0.180, 0.231] & [-0.055, 0.015] \\ 

\multicolumn{1}{p{3cm}}{\centering $95\%$ CR} 
&   & (-0.106, 0.036) &  & (-0.116, 0.071) &  & (-0.096, 0.046) \\ 

\multicolumn{1}{p{3cm}}{\centering Width Reduction} & &  41.3\%  & & 68.3\% & & 17.0\% \\
%

\bottomrule
\end{tabularx}
\caption*{
Notes. Table shows estimated  bounds in square brackets and  the $95 \%$ confidence region for the identified set  in parentheses. Columns (1), (3) and (5) report basic generalized bound in \eqref{eq:nocovariate} under Assumption \ref{ass:identification:treat:cond} with different choice of covariate sets. Columns (2), (4) and (6) report generalized Lee bound.  First Stage: The employment equation is estimated using  logistic regression (LR) with Lee's covariates (Columns (1)-(2)), post-Lasso-logistic regression (post-Lasso LR) with all covariates (Columns (3)-(4)), and LR with $43$ Lasso-selected covariates (Columns (5)-(6)). The wage quantile regression is estimated using quantile regression (QR) with Lee's covariates (Column (2)), $\ell_1$-QR  with all covariates (Column (4)),  and QR with $43$ covariates Lasso selected in (Column (6)).  The automated penalty choice for $\ell_1$-QR is in equation (2.6) of \cite{belloni:11}. Second stage. The always-takers' share is estimated as in Definition \ref{def:atshare}.  The bounds are estimated in Algorithm \ref{alg:lee}.   Computations use design weights. The sample size $N=9, 145$.  The asymptotic variance for the $95 \%$ Confidence Region is based on $B=1,00$ bootstrap repetitions. Width reduction is defined as a ratio of sharp bounds' width in an even-numbered column $2j$ to its basic analog in  $2j-1$ for $j \in \{1,2,3\}$. }
\end{landscape}

 
\paragraph{Results under unconditional monotonicity.}
Table \ref{tab:JCuncond} replicates Lee's results under unconditional monotonicity. The estimated effect on week-90 employment is $0.001$. Among treated individuals, approximately 99.9\% of wages are attributed to always-takers. As a result, the basic Lee bounds collapse to a near-point estimate, which coincides with the observed treatment-control difference in log wages. Assuming JobCorps does not reduce employment, the wage effect lies between $4.8\%$ and $4.9\%$ on average. The discrete bounds in Column (2) are wider than the basic ones. The sharpness property of Lee bounds holds in population but may fail in sample, as cell-specific employment effects are positive in some cells and negative in others. Under unconditional monotonicity, such sign reversals can only arise from sampling noise; negative effects are truncated at zero.

\paragraph{Results under conditional monotonicity.}
Table \ref{tab:JC} presents results under conditional monotonicity using different covariate sets. Columns (1)–(2) use only the original covariates from Lee’s analysis, relying on standard nonparametric assumptions. Columns (3)–(4) incorporate the full set of available covariates, imposing sparsity assumptions on employment (both columns) and wage (Column (4) only) equations. Columns (5)–(6) restrict attention to covariates selected by Lasso from the full set in Columns (3)–(4). The confidence region in Column (6) does not account for uncertainty due to covariate selection. The resulting bounds are not directly comparable since the assumptions underlying the different specifications are not nested.
  

Our empirical findings are as follows. First, the estimated share of always-takers is 44\%. This estimate remains remarkably stable across different covariate specifications. Second, the upper bound on the wage effect ranges from 0.04\% in Column (2) to 3.9\% in Column (4). Notably, it does not overlap with the basic lower bound reported in Table \ref{tab:JCuncond}, presenting further evidence against unconditional monotonicity. Finally,  the generalized upper bound is substantially tighter than its basic counterpart. In terms of width, using covariates to tighten the bound reduces the width by approximately 40\% in Columns (3)–(4) and nearly 80\% in Columns (5)–(6). Assuming structure on the employment equation—such as sparsity or smoothness—is essential for recovering the direction of the selection effect. Further assuming smoothness or sparsity in the wage equation helps rule out implausibly large wage effects. Because the wage effect is close to zero, its sign remains unknown.

\begin{figure}
 \centering
 \caption{Estimated bounds on the average always-takers' control wage  by week. }\par\medskip
   \includegraphics[scale = 0.5]{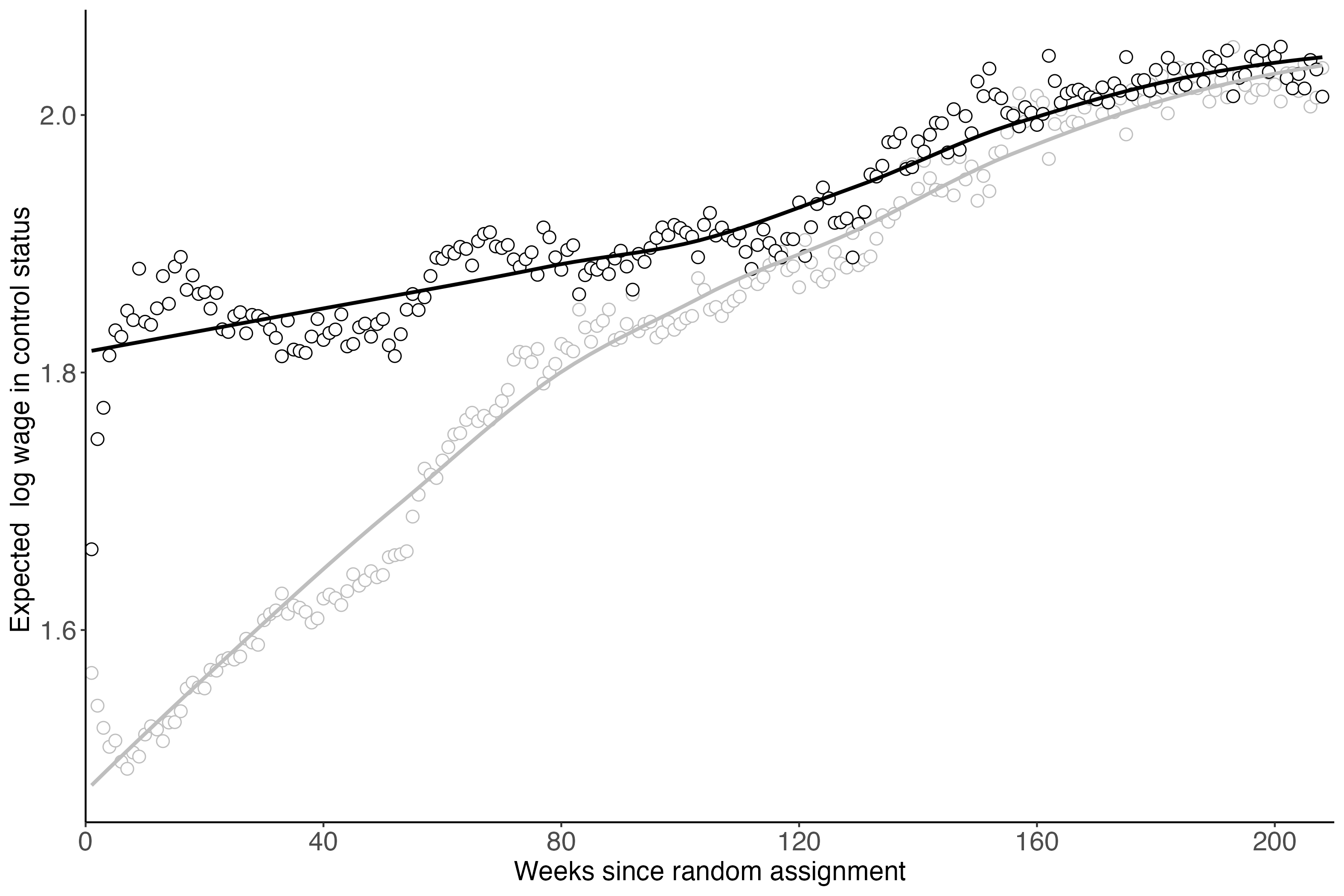}
       	\centering
    \caption*{ Notes. The horizontal axis shows the number of weeks since random assignment. The black (gray) circles show the upper (lower) bound on the average control  log wage  for the always-takers.  The covariate vector $X$ consists of 28 Lee's covariates. The estimator is based on the Algorithm \ref{alg:lee}, where the first-stage always-takers' share and the borderline wage are estimated using logistic and quantile regression (see Column (4), Table \ref{tab:JC}).   The sample size $N=9, 145$.    Computations use design weights.      }
    \label{fig:Figure3}
 \end{figure}

 Figure \ref{fig:Figure3} reports the upper and lower bounds on the average log wage for the always-takers in the control state. The lower (upper) bound grows from $1.63$ ($1.92$) in week $5$ to $1.96$ ($1.96$) in week $208$. The bounds' width decreases from $0.3$ in week 14 to $0.01$ in week 208. The gap between the lower and the upper bound shrinks over time as the share of applicants with a positive employment effect, where the average control log wage is point-identified, increases.  The upward trend in the control wages suggests that evaluating JobCorps would have been very difficult without a randomized experiment, as one would need to explicitly model mean reversion in the baseline potential wage.
 

\paragraph{Conclusion. } Lee bounds are a popular empirical strategy for addressing post-randomization selection bias, assuming treatment's effect on selection has the same sign (unconditional monotonicity). This paper generalizes Lee bounds under conditional monotonicity, which allows the direction of selection effect to differ only with observed covariates. This generalization has proven especially useful for JobCorps job training program, where unconditional monotonicity is unlikely to hold. Relaxing conditional monotonicity is left for the future work.

\appendix

\section {Appendix A. Proofs}

\label{sec:proofs}

\renewcommand{\theequation}{A.\arabic{equation}}
\renewcommand{\thesection}{A.\arabic{section}}
\renewcommand{\thesubsection}{A.\arabic{subsection}}
\renewcommand{\thedefinition}{A.\arabic{definition}}
\renewcommand{\thesubsubsection}{A.\arabic{subsubsection}}
\renewcommand{\thetheorem}{A.\arabic{theorem}}
\renewcommand{\thelemma}{A.\arabic{lemma}}
\renewcommand{\theassumption}{A.\arabic{assumption}}
\renewcommand{\thetable}{A.\arabic{table}}
\renewcommand{\thefigure}{A.\arabic{figure}}
\renewcommand{\theremark}{A.\arabic{remark}}
\setcounter{section}{0}
\setcounter{table}{0}
\setcounter{theorem}{0}
\setcounter{remark}{0}
\medskip

\subsection{Notation, Definitions, and Technical Lemmas}

\paragraph{Notation and Definitions. } Consider the following functions describing the outcome distribution in the selected treated and selected control groups. The conditional CDF is
\begin{align}
\label{eq:cdf}
F_d (t \mid x):= \Pr ( Y \leq t \mid S=1, D=d, X=x), \quad d \in \{1, 0\}, \quad t \in \mathrm{R},
\end{align}
the conditional density is
\begin{align*}
   f_d(t|x) = f_{Y \mid S=1,D=d, X=x}(t \mid x), \quad d \in \{1,0\}, \quad t \in \mathrm{R},
\end{align*}
the conditional quantile function is
\begin{align}
    \label{eq:condquantil0}
    Q^d(u,x): \quad \Pr (Y \leq Q^d(u,x)| S=1, D=d, X=x) = u,  \quad u \in [0,1], \quad d \in \{1, 0\}.
\end{align}

%
%
%
%
%
%
%
%

%
%
%

\subsection{Proofs for Section \ref{sec:monot}}

\begin{lemma}[Corollary 4.1, \cite{HorowitzManski}]
\label{lem:cormanski}
Let $Y$ be a continuous random variable and a mixture of two random variables, with CDFs $M^{*} (y)$ and $N^{*}(y)$,  and a known mixing proportion $p^{*} \in (0,1)$, so we have
$F^{*}(y) = p^{*} M^{*} (y) + (1-p^{*}) N^{*} (y)$. Let $G^{*}(y)$ be the CDF of  $Y$ after truncating the $p^{*}$ lower tail of $Y$. Then,
\begin{align*}
 \int_{-\infty}^{\infty} y d N^{*} (y) \leq \int_{-\infty}^{\infty} y d G^{*} (y).
\end{align*}
Furthermore, $\int_{-\infty}^{\infty} y d G^{*} (y)$ is the sharp upper bound on $\int_{-\infty}^{\infty} y d N^{*} (y)$.

\end{lemma}

\begin{proof}[Proof of Lemma \ref{lem:atshare}]
For any $x \in \mathcal{X}_{\text{help}}$, 
\begin{align*}
\Pr (S(1) = S(0) = 1\mid X=x)  = \Pr (S=1 \mid X=x, D=0)=s(0,x).
\end{align*}
 Likewise, for any $x \in \mathcal{X}_{\text{hurt}}$, 
 \begin{align*}
\Pr (S(1) = S(0) = 1\mid X=x) =\Pr (S=1 \mid X=x, D=1)=s(1,x).
\end{align*}
 For any  $x \in \mathcal{X}_0$, $\tau(x) = 0 \Rightarrow s(1,x) = s(0,x)$.  Bayes rule for conditional density gives
\begin{align}
\label{eq:density}
f_{X}(x \mid S(1) = S(0) = 1) = {\pi_{\text{AT}}}^{-1} \begin{cases} s(0,x) f_X(x) & x \in \mathcal{X}_{\text{help}} \\
s(1,x) f_X(x)& x \in \mathcal{X}_{\text{hurt}},\\
s(1,x) f_X(x) & x \in \mathcal{X}_{0},\\
\end{cases}
\end{align}
where the denominator is 
\begin{align}
{\pi_{\text{AT}}} &= \int_{\mathcal{X}_{\text{help}} }   s(0,x) f_X(x) d x +  \int_{\mathcal{X}_{\text{hurt}} }   s(1,x) f_X(x) d x +  \int_{\mathcal{X}_{0} }   s(1,x) f_X(x) d x  \label{eq:mins0s1} \\
&=\int_{\mathcal{X}}  \min (s(0,x), s(1,x)) f_X(x) d x= \E \min (s(0,X), s(1,X)). \nonumber
\end{align}
\end{proof}

\begin{proof}[Proof of Lemma \ref{lem:moment}]

\textbf{Step 1. }  The proof of sharpness is similar to \cite{LeeBound} (more precisely, the working paper version \cite{LeeBound2}).  Consider the following data generating process (d.g.p.). For each $x \in \mathcal{X}_{\text{help}}$, the always-takers' treated outcomes comprise the top $p(x) = s(0,x)/s(1,x)$ quantile of the distribution of $Y \mid D=1, S=1, X=x$ in the treated selected group. For this group, $\beta_0(x) = \beta_U(x)$ holds by  Lemma \ref{lem:cormanski}.  Likewise, for each  $x \in \mathcal{X}_{\text{hurt}}$, the always-takers' untreated outcomes comprise the bottom $1/p(x)$-quantile of the distribution of $Y \mid D=0, S=1, X=x$ in the untreated selected group. For this group, $\beta_0(x) = \beta_U(x)$.  Finally, for $x \in \mathcal{X}_0$, $\beta_L(x) = \beta_U(x)$, which implies $\beta_L(x) = \beta_0(x) = \beta_U(x)$. Therefore, the largest attainable value of $\E[ Y(1) - Y(0) \mid S(1) = S(0) = 1]$ is 
$$
\int_{\mathcal{X}}   \beta_U(x) f_{X}(x \mid S(1) = S(0) = 1) dx.
$$
Since the conditional density $f_{X}(x \mid S(1) = S(0) = 1)$ is identified in \eqref{eq:density}, the statement holds. For any $\beta' < \beta_U$, $\beta'$ cannot be a valid bound because there exists a d.g.p. for which $\beta_0 = \beta_U>\beta'$.  

%
%
\textbf{Step 2. } By the law of conditional probability,  $$\Pr (Y \geq Q^1(1-p(x),x) \mid D=1, S=1, X=x) = 1- (1-p(x)) = p(x).$$ For any $x \in \mathcal{X}_{\text{help}}$, 
\begin{align*}
&\E [ m^{\text{help}}_U(W,\xi_0) \mid X=x] \nonumber \\
&= \E[ Y 1 { \{ Y  \geq Q^1(1-p(X),X) \}}  \mid D=1, S=1, X=x] s(1, x) \\
&- \E[ Y \mid D=0, S=1, X=x] s(0,x). \nonumber \\
&= \E[ Y \mid Y \geq Q^1(1-p(x),x), D=1, S=1, X=x] p(x)  s(1,x)  \\
&- \E[ Y \mid D=0, S=1, X=x] s(0,x) \\
&=  \beta^{\text{help}}_U(x) s(0,x) =  \beta^{\text{help}}_U(x)  \min (s(0,x), s(1,x)). \nonumber
\end{align*}
For $x \in \mathcal{X}_{\text{hurt}}$, a similar argument gives
\begin{align*}
&\E [ m^{\text{hurt}}_U(W,\xi_0) \mid X=x] \\
&= \E[ Y \mid D=1, S=1, X=x] s(1,x) \\
& - \E[ Y 1 { \{ Y  \leq Q^0(1/p(X),X) \}}  \mid D=0, S=1, X=x] s(1,x)\\
&=\beta^{\text{hurt}}_U(x) s(1,x) \\
&=   \beta^{\text{hurt}}_U(x)  \min (s(0,x), s(1,x)). 
\end{align*}
Finally, for $x\in \mathcal{X}_0$, we have $s(0,x)=s(1,x)$, and

$$\E[ m_U(W, \xi_0) \mid X=x]=\E[ m_L(W, \xi_0) \mid X=x] = \beta_U(x) \min (s(0,x), s(1,x)).$$

\end{proof} 

\subsection{Proofs for Section \ref{sec:estim}}

In this Section, we shall denote the true value of nonparametric functions using a different notation. Let $s_0(d,x)$ be the true value of $s(d,x)$ and $\tau_0(x)$ be the true value of $\tau(x)$.  Likewise, let $Q^1_0(\cdot,x), Q^0_0(\cdot,x)$ denote the true value of $Q^1(\cdot, x)$ and $Q^0(\cdot, x)$. Finally, we shall write the true value of $$\xi_0(\cdot, x):= \{ s_0(1,x), s_0(0,x), Q^1_0(\cdot,x), Q^0_0(\cdot,x)\}.$$

\begin{definition}[Nuisance Realization Set]
Let $\{S^d_N\}_{N \geq 1}$ and  $\{Q^d_N\}_{N \geq 1}$ be the sequences of sets in Definitions \ref{def:sel} and \ref{def:q}. Define the
set $\Xi_N$ as the cartesian product of $S^1_N, S^0_N, Q^1_N, Q^0_N$ consisting of $\{\xi(\cdot, x), x \in \mathcal{X} \}$ that shrinks towards $\{ s_0(1,x), s_0(0,x), Q^1_0(\cdot,x), Q^0_0(\cdot,x)\}$. 

\end{definition}

\subsubsection{Asymptotic Theory for Always-Takers' Share $\pi_{\text{AT}}$ }

 A covariate value $x$ is misclassified if and only if it belongs to the misclassification event
\begin{align}
\label{eq:dtau1}
\mathcal{D}_{\tau}: =\{ \tau(x) < 0 < \tau_0(x) \} \cup \{  \tau_0(x) < 0 < \tau(x) \}.
\end{align}
On this event, the true absolute value of $|\tau_0(x)|$ cannot be large:
\begin{align}
\label{eq:dtau}
\mathcal{D}_{\tau} \Rightarrow \bigg\{ 0 < | \tau_0(x) | < | \tau(x)  - \tau_0(x)| \bigg\}=:\mathcal{D}^2_{\tau}.
\end{align}
Define an additional misclassification event
\begin{align}
\mathcal{D}^3_{\tau} &= \{x \in \mathcal{X}: 0 < |\tau_0(x)| \leq 2\max(s^{\infty}_N, \rho_N)\}. 
\end{align}
Observe that
$$
\mathcal{D}_{\tau} \subseteq \mathcal{D}^2_{\tau} \subseteq \mathcal{D}^3_{\tau}.
$$

\begin{lemma}[Misclassification Events]
\label{lem:oracledenom}
Suppose Assumption \ref{ass:firststage} holds, and the sample size $N$ is large enough. Given $\rho_N>0$, define misclassification events
$$
\mathcal{D}^{\text{help}}_{\tau} : = \{ x \in \mathcal{X}: \tau(x) \geq \rho_N, \tau_0(x) \leq 0  \}.   
$$
and
$$
\mathcal{D}^{\text{hurt}}_{\tau} : = \{ x \in \mathcal{X}: \tau(x) < \rho_N, \tau_0(x)  >0  \}.   
$$
Then, $\mathcal{D}^{\text{help}}_{\tau} \cup \mathcal{D}^{\text{hurt}}_{\tau}   \subseteq   \mathcal{D}^3_{\tau}$. Furthermore, if $\rho_N=0= \Pr(\mathcal{X}_0)$, for the event $\mathcal{D}_{\tau} \subseteq  \mathcal{D}^3_{\tau}$. 
\end{lemma}

\begin{proof}[Proof of Lemma \ref{lem:oracledenom}]
For any $\tau(x) = s(1,x) - s(0,x) \in S^1_N - S^0_N$, for $N$ large enough, 
\begin{align}
\sup_{x \in \mathcal{X}} |\tau(x) - \tau_0(x)| &\leq \sup_{x \in \mathcal{X}} (|s(1,x) - s_0(1,x)| + |s(0,x) - s_0(0,x)|) \leq 2s^{\infty}_N  \label{eq:pbound}
\end{align}
The rest of the proof proceeds in two steps. Step 1 focuses on the case $\rho_N > 0$, while Step 2 focuses on the case $\rho_N = 0$.

\textbf{Step 1.}  Suppose $x$ is misclassified and  $x \in \mathcal{D}^{\text{help}}_{\tau}$. We have $2s^{\infty}_N< \rho_N$ for $N$ large enough. Therefore, it must be 
$$ \tau_0(x) \leq 0  < 2s^{\infty}_N < \rho_N \leq \tau(x),$$ which contradicts \eqref{eq:pbound}, and $\mathcal{D}^{\text{help}}_{\tau} = \emptyset$. Therefore, if $x$ is misclassified, it must be that $0 < \tau_0(x)$.  If $\tau_0(x) > 2\rho_N$, then $\tau(x) \geq  \rho_N$, and $x \not \in \mathcal{D}^{\text{hurt}}_{\tau} $. Therefore, $ \mathcal{D}^{\text{help}}_{\tau} \cup  \mathcal{D}^{\text{hurt}}_{\tau}  \subseteq  \mathcal{D}^3_{\tau}$.

\textbf{Step 2.} Consider the case $\rho_N = 0$ and $\Pr(\mathcal{X}_0) = 0$. A covariate value is misclassified if and only if $x \in \mathcal{D}_{\tau}$ in \eqref{eq:dtau1}. 
\end{proof}

Define the error terms as 
$$
S_1 (s):= g^0(W, s)- g^0(W, s_0), \quad S_2 (s):= g^1(W, s)- g^1(W, s_0)
$$
and
\begin{align*}
S_3 (\tau)&=g^1(W, s_0) 1\{ \tau (X) < \rho_N \} - g^0(W, s_0) 1\{ \tau_0(X) > 0\} \\
&+ g^0(W, s_0) 1\{ \tau(X) \geq \rho_N \}  - g^1(W, s_0) 1\{ \tau_0 (X) <0  \}.
\end{align*}

\begin{lemma}[Bound on Bias]
\label{lem:oracledenom1}
Under Assumptions \ref{ass:boundedoutcome3} and \ref{ass:firststage}, the bias is second-order
\begin{align}
\sup_{s(d,x) \in S^d_N   } | \E [g^{\rho_N}_D(W, s) - g^0_D(W, s_0)] | &\leq  \bar{B}_f  (2\max (s^{\infty}_N, \rho_N )  )^{2}.  \label{eq:bias} 
\end{align}
\end{lemma}

\begin{proof}[Proof of Lemma \ref{lem:oracledenom1}]
Decompose the estimation error of moment function
\begin{align}
g^{\rho_N}_D(W, s) - g^{0}_D(W, s_0) &=   \left(g^0(W, s)- g^0(W, s_0) \right)  1\{ \tau(X) \geq \rho_N \} \nonumber \\
&+ \left(g^1(W, s)- g^1(W, s_0) \right)  1\{ \tau(X) < \rho_N \} + S_3(\tau) \label{eq:decompose} \\
&= S_1 (s) 1\{ \tau(X) \geq \rho_N \}   + S_2 (s)  1\{ \tau(X) < \rho_N \} + S_3 (\tau)  \nonumber
\end{align} 

\textbf{Step 1. } Expectation of $S_1 (s)$ and $S_2 (s)$. Noting that for $d \in \{1, 0\}$
$$
g^d(W, s)- g^d(W, s_0)  = \left( 1 -\dfrac{1\{ D=d\}}{\mu_d(X)} \right) (s(d,X) - s_0(d,X))=S_{d+1}(s), 
$$
which implies $\E [S_1(s) 1\{ \tau(X) \geq \rho_N \}    + S_2(s) 1\{ \tau(X) < \rho_N \}  ]  =0.$
\textbf{Step 2. } 
\begin{align*}
 &\E [ g^1(W, s_0) - g^0(W, s_0) \mid X ] \\
 &= (s_0(1,X) - s_0(0,X)) + \E \bigg[ \dfrac{1\{ D=1\} S}{\mu_1(X)} - \dfrac{1\{ D=0\} S}{\mu_0(X)} \mid X \bigg]\\
 & - s_0(1,X) + s_0(0,X) \\
& = \tau_0(X) + \tau_0(X) -\tau_0(X)  = \tau_0(X),
\end{align*}
which implies a bound  $| \tau_0(X) |$ on the magnitude of expected mistake. If covariate value $x$ is misclassified, $ x \in \mathcal{D}^3_{\tau} $ by Lemma \ref{lem:oracledenom}. Therefore, 
\begin{align*}
\E [S_3(\tau) \mid X ] \leq | \tau_0(X) | 1 \{ X \in \mathcal{D}^3_{\tau} \}.
\end{align*}
Taking unconditional expectation gives
\begin{align}
|  \E S_3(\tau) | \leq \E1 \{ X \in \mathcal{D}^3_{\tau} \}  | \tau_0(X) | \leq  2 \Pr (X \in \mathcal{D}^3_{\tau}) \max ( s^{\infty}_N, \rho_N) \label{eq:bias1}.
 \end{align}
By Assumption \ref{ass:boundedoutcome4}, for sufficiently large $N$, invoking margin assumption \eqref{eq:ma1}
\begin{align}
\label{eq:maapp}
 \Pr (X \in \mathcal{D}^3_{\tau}) \leq  2\max ( s^{\infty}_N, \rho_N)\bar{B}_f.
 \end{align}
 Combining \eqref{eq:bias1} and \eqref{eq:maapp} gives the bound on bias \eqref{eq:bias}
 $$
 \sup_{\tau \in S^1_N - S^0_N } | \E  S_3(\tau) |  \leq \bar{B}_f (2\max ( s^{\infty}_N, \rho_N))^2.
 $$

\end{proof}

\begin{lemma}[Bound on Variance]
\label{lem:oracledenom2}
Under Assumptions  \ref{ass:boundedoutcome2}--\ref{ass:firststage}, the second moment is bounded
\begin{align}
\sup_{s(d,x) \in S^d_N } (\E (g^{\rho_N}_D(W, s) - g^0_D(W, s_0))^2)^{1/2} &=  O(  \max (s^{\infty}_N, \rho_N )^{1/2} + s_N  ). \label{eq:var}
\end{align}
\end{lemma}

\begin{proof}[Proof of Lemma \ref{lem:oracledenom2}]
Applying the  inequality $(a+b+c)^2 \leq 3 (a^2+b^2+c^2)$ to \eqref{eq:decompose} gives
\begin{align}
(g^{\rho_N}_D(W, s) - g^0_D(W, s_0))^2 = (S_1(s) + S_2(s) + S_3 (\tau) )^2  \leq 3 (S_1^2(s) + S_2^2(s) + S^2_3(\tau)).
\end{align}
Noting that 
$$
\E  \bigg[\left( 1 -\dfrac{1\{ D=d\}}{\mu_d(X)} \right)^2 \mid X \bigg]  \leq (1-\kappa)\kappa/\kappa^2 = (1-\kappa)/\kappa.
$$
The first and second term are bounded as
\begin{align*}
\E [S_{d+1}^2(s) ] &\leq \E  \left( 1 -\dfrac{1\{ D=d\}}{\mu_d(X)} \right)^2 (s(d,X) - s_0(d,X))^2 \\
&\leq \kappa^{-1} \E (s(d,X) - s_0(d,X))^2, \quad d=0,1.
\end{align*}
The variance of misclassification is bounded as
\begin{align*}
\sup_{\tau \in S^1_N - S^0_N } \E S^2_3(\tau)  = O( \Pr (X \in \mathcal{D}^3_{\tau})) = O(\max ( s^{\infty}_N, \rho_N)).
\end{align*}
\end{proof}

\subsubsection{Asymptotic Theory for Numerators $N_U$ and $N_L$}

In what follows, I shall write $\xi$ and $\xi_0$ to denote the estimated and the true value, respectively. 

\begin{lemma}[Primary moment error at the boundary]
\label{lem:cont}

Under Assumptions \ref{ass:boundedoutcome2}--\ref{ass:firststage}, 
\begin{align}
\sup_{ \xi \in \Xi_N}  | \E 1\{  |\tau(X)|  \leq \rho_N \}  (m_U (W, \xi) - m_U  (W, \xi_0)) |  &= O ((2\rho_N)^{2}  ) \label{eq:boundl2}  \\
\sup_{ \xi \in \Xi_N}  \E 1\{  |\tau(X)|  \leq \rho_N \}  (m_U  (W, \xi) - m_U  (W, \xi_0))^2   &= O (  (2\rho_N)^{2}  ) \label{eq:boundl3} 
\end{align}
\end{lemma}

\begin{proof}[Proof of Lemma \ref{lem:cont}]
Step 1 bounds the conditional expectations given $X$. Step 2 bounds unconditional moments.

\textbf{Step 1. } Consider the $X: | \tau(X) | \leq \rho_N$.  The moment function is
$$
m_U (W, \xi) = \dfrac{D S Y}{\mu_1(X) } -  \dfrac{(1-D) S Y}{\mu_0(X) }.
$$
The estimation error $m_U (W, \xi) - m_U (W, \xi_0)$ takes the form
\begin{align}
m_U (W, \xi) - m_U (W, \xi_0) = \begin{cases}
\dfrac{D S Y}{\mu_1(X)}  1\{ Y \leq Q_0^1(1-p_0(X),X)\}, \quad X \in \mathcal{X}_{\text{help}} \\
-\dfrac{(1-D) SY}{\mu_0(X)}  1\{ Y \geq Q_0^0(1/p_0(X),X)\}, \quad X \in \mathcal{X}_{\text{hurt}}  \\
0, \quad p_0(X)=1. \\ 
\end{cases}
\end{align}
The  expected  estimation error 
$
R(X, \xi_0):=\E[ m_U (W, \xi) - m_U (W, \xi_0) \mid X]
$
reduces to
\begin{align*}
R(X, \xi_0)=\begin{cases} \E [ Y 1\{ Y \leq Q_0^1(1-p_0(X),X)\} \mid D=1,S=1,X] s_0(1,X), \quad X \in \mathcal{X}_{\text{help}}   \\
-\E [ Y 1\{ Y \geq Q_0^0(1/p_0(X),X)\} \mid D=0,S=1,X] s_0(0,X), \quad X \in \mathcal{X}_{\text{hurt}} \\
0, \quad p_0(X) =1.
\end{cases}
\end{align*}
Consider the covariate value $X \in \mathcal{X}_{\text{help}}$. By Assumption \ref{ass:boundedoutcome4}, the expected mistake is bounded
\begin{align*}
| R (X,\xi_0) | &\leq M  F_1(Q_0^1(1-p_0(X),X) \mid X)  s_0(1,X)  \\
&= M | 1- p_0(X) | s_0(1,X) = M | \tau_0(X) |,
\end{align*}
where $F_1(\cdot)$ is the outcome CDF in \eqref{eq:cdf}. Reversing the roles of treated and control group gives the bound for  $X \in \mathcal{X}_{\text{hurt}}$
\begin{align*}
| R (X,\xi_0) | &\leq M | 1-F^0(Q_0^0(1/p_0(X),X) \mid X) | s_0(0,X) \\
&= M | 1- s_0(1,X)/s_0(0,X) | s_0(0,X) = M | \tau_0(X) |.
\end{align*}
Since $R(X, \xi_0) =0$ on the boundary $\mathcal{X}_0$, $R(X, \xi_0) \leq M | \tau_0(X) |$ for any $X: | \tau(X) | \leq \rho_N$. Note that
$$
\E [(m_U (W, \xi) - m_U (W, \xi_0))^2 \mid X] \leq M^2/ \kappa | \tau_0(X) |.
$$

\textbf{Step 2. } As shown in the proof  Lemma \ref{lem:oracledenom},  \eqref{eq:pbound} $|\tau(X)|  \leq \rho_N $ implies $ |\tau_0(X)|  \leq 2\rho_N$. Therefore, 
\begin{align*}
 & | \E 1\{  |\tau(X)|  \leq \rho_N \}  (m_U (W, \xi) - m_U (W, \xi_0)) | \\
&\leq | \E 1\{  0< |\tau_0(X)|  \leq 2\rho_N \} \E[ (m_U (W, \xi) - m_U (W, \xi_0)) \mid X ] | \\
&\leq  M \E 1\{  0< |\tau_0(X)|  \leq 2\rho_N \}   | \tau_0(X) | \\
&\leq 2 \rho_N M  \Pr (0 < | \tau_0(X)| \leq 2\rho_N ),
\end{align*}
By Assumption  \ref{ass:boundedoutcome4} \eqref{eq:ma1}, the term above is bounded by $M \bar{B}_f (2\rho_N)^{2}$.  Likewise, the second moment is bounded 
\begin{align*}
&\sup_{ \xi \in \Xi_N}  \E 1\{  |\tau(X)|  \leq \rho_N \}  (m_U (W, \xi) - m_U (W, \xi_0))^2   \\
&= (2\rho_N) M^2/\kappa \Pr ( 0 < | \tau_0(X) | \leq 2\rho_N) \leq M^2/\kappa \bar{B}_f  (2\rho_N)^{2}.
\end{align*}

\end{proof}

\begin{lemma}[Orthogonal correction error at the boundary]
\label{lem:orthocor}
Under Assumptions \ref{ass:boundedoutcome2}--\ref{ass:firststage},
\begin{align}
\sup_{ \xi \in \Xi_N}  | \E 1\{  |\tau(X)|  \leq \rho_N \} (  \text{cor}_U (W, \xi) -  \text{cor}_U (W, \xi_0)) |  &= 0 \label{eq:boundl4}  \\
\sup_{ \xi \in \Xi_N}    \E 1\{  |\tau(X)|  \leq \rho_N \}   (  \text{cor}_U (W, \xi) -  \text{cor}_U (W, \xi_0))^2   &= O ( \rho_N ) \label{eq:boundl5} 
\end{align}

\end{lemma}

\begin{proof}[Proof of Lemma \ref{lem:orthocor}]
For any $X:  |\tau(X)|  \leq \rho_N$, $ \text{cor}_U(W, \xi):=0$. I shall write
$$
 \text{cor}_U(W, \xi_0) =  \text{cor}^{\text{help}}_U(W, \xi_0) 1\{ \tau_0(X)>0\} + \text{cor}^{\text{hurt}}_U(W, \xi_0) 1\{ \tau_0(X)<0\} + 0 \cdot 1\{ \tau_0(X)=0\}.
$$
For $X: \tau_0(X) = 0$,
 $$ \quad \text{cor}_U(W, \xi) -  \text{cor}_U(W, \xi_0) = 0 \quad \text{a.s.}$$ 
 Therefore, it suffices to focus on $X:  |\tau(X)|  \leq \rho_N$ and $X: |\tau_0(X)|>0$. For $C(M, \kappa)$ large enough, 
\begin{align*}
\E [  (  \text{cor}_U(W, \xi) -  \text{cor}_U(W, \xi_0))^2  \mid X ] = \E[  \text{cor}^2_U(W, \xi_0) \mid X ] \leq C(M, \kappa) 1 \{ | \tau_0(X) | > 0 \}.
\end{align*}
 Finally, since correction term is a sum of zero mean residuals, 
\begin{align*}
\E [  \text{cor}_U(W, \xi) -  \text{cor}_U(W, \xi_0) \mid X] = \E [ 0 -  \text{cor}_U(W, \xi_0) \mid X] = 0.
\end{align*}
Therefore, 
\begin{align*}
&\sup_{ \xi \in \Xi_N}   \E 1\{  |\tau(X)|  \leq \rho_N \}   (  \text{cor}_U(W, \xi) -  \text{cor}_U(W, \xi_0))^2 \\
&\leq C(M, \kappa)  \Pr ( 0 < | \tau_0(X) | \leq  2\rho_N) = O ( 2\rho_N ).
\end{align*}

\end{proof}

Next, we shall proceed with covariate values $X$ such that $| \tau(X) | \geq \rho_N$.  Let $ r\in (0,1)$ and
\begin{align*}
 s_r(0,x) &= s_0(0,x) + r (s(0,x) -s_0(0,x)) \\
 s_r(1,x) &= s_0(1,x) + r (s(1,x) -s_0(1,x)) \\
b^r_1(x) &= Q_r(1-p(x),x) = Q_0(1-p(x), x) + r(  Q(1-p(x), x) - Q_0(1-p(x), x) )
\end{align*}
For $k=1,2,3$, define
\begin{align*}
\psi_k(r):&=\dfrac{D S} {\mu_1(X) } Y 1 \{ Y \geq b^r_{k}(X), X) \},
\end{align*} 
and
\begin{align*}
b^1_2(x) &= Q_0(1-s(0,x)/s(1,x),x) \\
b^1_3(x) &= Q_0(1-s_0(0,x)/s(1,x),x)
\end{align*}
is the true and estimated quantile values. By construction, $b^0_1(x) = b^1_2(x)$ and $b^0_2(x) = b^1_3(x)$. Noting that
$$
m^{\text{help}}_U(W, \xi) = \psi_1(1) - \dfrac{1-D}{\mu_0(X)} SY, \quad m^{\text{help}}_U(W, \xi_0) = \psi_3(0) - \dfrac{1-D}{\mu_0(X)} SY,
$$
and $$\psi_1(0) = \psi_2(1), \quad \psi_2(0) = \psi_3(1), $$ we have
\begin{align*}
m^{\text{help}}_U(W, \xi) - m^{\text{help}}_U(W, \xi_0) =\psi_1(1) - \psi_3(0) = \sum_{k=1}^3 [\psi_k(1) - \psi_k(0)].
\end{align*}
The estimated correction is 
\begin{align*}
\text{cor}_U(W, \xi) &= b^1_1(X) ( \dfrac{(1-D)}{\mu_0(X)} \left(S - s(0,X) \right)  - p(X) \dfrac{D}{\mu_1(X)} \left(S - s(1,X) \right) \\
&+ \dfrac{D S}{\mu_1(X)}  \left ( 1 \{ Y \leq b^1_1(X) \}  - (1 - p(X)) \right)) \\
&= \psi_4(1) + (\psi_5(1)+\psi_7(1)) + (\psi_8(1)+\psi_6(1)) + \psi_9(1),
\end{align*}
where the correction terms are decomposed into first-order terms
\begin{align}
\psi_4(1):&= \dfrac{D S}{\mu_1(X)} b^1_1(X) \left ( 1 \{ Y \leq b^1_1(X) \}  - (1 - p(X)) \right)\\
\psi_5(1):&=  b^1_2(X) \dfrac{(1-D)}{\mu_0(X)} \left(S - s(0,X) \right) \\
\psi_6(1):&= -b^1_3 (X)  p(X) \dfrac{D}{\mu_1(X)} \left(S - s(1,X) \right),
\end{align}
and the higher-order error terms
\begin{align}
\psi_7(1):&=  ( b^1_1(X)- b^1_2(X)) \dfrac{(1-D)}{\mu_0(X)} \left(S - s(0,X) \right) \\
\psi_8(1):&= -( b^1_1(X)- b^1_2(X)) p(X) \dfrac{D}{\mu_1(X)} \left(S - s(1,X) \right) \\
\psi_9(1):&= -(b^1_2(X)- b^1_3(X)))  p(X) \dfrac{D}{\mu_1(X)} \left(S - s(1,X) \right).
\end{align}
The true analogs of $\psi_4(1), \psi_5(1),\psi_6(1)$ are
\begin{align}
\psi_4(0):&= b^0_3(X) \dfrac{D S}{\mu_1(X)}  \left ( 1 \{ Y \leq b^0_3(X)   \}  - (1 - p_0(X)) \right)\\
\psi_5(0):&=  b^0_3(X) \dfrac{(1-D)}{\mu_0(X)} \left(S - s_0(0,X) \right) \\
\psi_6(0):&=  -b^0_3(X) p_0(X) \dfrac{D}{\mu_1(X)}\left(S - s_0(1,X) \right).
\end{align}
By construction,
\begin{align*}
\text{cor}^{\text{help}}_U(W, \xi) - \text{cor}^{\text{help}}_U(W, \xi_0) &=  \sum_{k=4}^6 [\psi_k(1) - \psi_k(0)] + \sum_{k=7}^9\psi_k(1)
\end{align*}
and
$$
g^{\text{help}}_U(W, \xi) - g^{\text{help}}_U(W, \xi_0)= \sum_{k=1}^6 [\psi_k(1) - \psi_k(0)] + \sum_{k=7}^9\psi_k(1).
$$

\begin{lemma}[Verification of small bias condition]
\label{lem:class}
Under Assumptions \ref{ass:boundedoutcome2}--\ref{ass:firststage}, 
\begin{align}
\sup_{ \xi \in \Xi_N}  | \E    (g_U(W,\xi) - g_U(W,\xi_0)) 1\{ | \tau(X) | > \rho_N \} |  = O (s_N^2 + q_N^2 ) \label{eq:class21} \\
\sup_{ \xi \in \Xi_N}  \E (g_U(W,\xi) - g_U(W,\xi_0))^2   1\{ | \tau(X) | > \rho_N \} = O (s^1_N + q^1_N +s_N^2 + q_N^2 )   \label{eq:class22} 
\end{align}

\end{lemma}

\begin{proof}[Proof of Lemma \ref{lem:class}]

\textbf{ Step 1.  Intermediate value theorem. }   Define
\begin{align*}
\Gamma_1(r,x):&= y f_1(y\mid x) |_{y = b^r_1(x) }. \\
\Gamma_2(r,x):&= y^2 f_1(y\mid x)|_{y = b^r_1(x)  }. \\
 \zeta(r,x) &:= (f_1(y\mid x)  + y \partial_{y}  f_1(y\mid x))  |_{y = b^r_1(x) }.
\end{align*}
Recall that $ \psi_k(1)$ and $ \psi_k(0)$ are trimmed means for $k=1,2,3$. Taking expectations gives
\begin{align*}
\E [ \psi_k(1) - \psi_k(0)\mid X=x ] &=   -s_0(1,x) \int_{ b^0_k (x) }^{ b^1_k (x)  }y f_1(y\mid x) dy \\
\E [ (\psi_k(1) - \psi_k(0))^2\mid X=x ] &=   s_0(1,x) \bigg| \int_{ b^1_k (x) }^{ b^0_k (x)  }y^2 f_1(y\mid x) dy \bigg|
\end{align*}
By intermediate value theorem, there exists $r_1(x) \in (0,1)$ such that
$$
\E [ \psi_1(1) - \psi_1(0)\mid X=x ] =  -s_0(1,x)\Gamma_1(r_1(x), x) (b^1_1(x) - b^0_1(x)).
$$
Likewise,  for some   $r_2=r_2(x)$ and $r_3=r_3(x)$, we have
\begin{align*}
\E [ \psi_2(1) - \psi_2(0)\mid X=x ] &=  (-1)^2 s_0(1,x) \cdot  b_2^{r_2(x)} (x) \cdot  (s(0,x) -s_0(0,x))/s(1,x) \\
\E [ \psi_3(1) - \psi_3(0)\mid X=x ] &=  (-1)^3 s_0(1,x) b_3^{r_3(x)} (x)  \cdot (s(1,x)-s_0(1,x) )/s(1,x) \cdot p_0(x)
\end{align*}

For the second powers, for some $r'_1=r'_1(x)$, $r'_2=r'_2(x)$ and $r'_3=r'_3(x)$, 
\begin{align}
\E [ (\psi_1(1) - \psi_1(0))^2\mid X=x ] &=  s_0(1,x)\Gamma_2(r'_1(x), x) |b^1_1(x) - b^0_1(x)|. \label{eq:power2} \\
\E [ (\psi_2(1) - \psi_2(0))^2 \mid X=x ] &=  s_0(1,x)/s(1,x)  \cdot  (b_2^{r_2'(x)} (x))^2  \cdot  |s(0,x) -s_0(0,x)| \nonumber  \\
\E [ (\psi_3(1) - \psi_3(0))^2\mid X=x ] &= (b_3^{r_3'(x)} (x))^2  \cdot  s_0(1,x)/s(1,x) \cdot p_0(x)\cdot |s(1,x)-s_0(1,x) | \nonumber 
\end{align}
For some $r_4 = r_4(x)$ and $r_5 = r_5(x)$, the following equalities hold
\begin{align}
F_1 (b^1_1(x) \mid x) -F_1 (b^0_1(x) \mid x)&= f_1 ( b^{r_4(x)}_1 (x) \mid x) (b^1_1(x) - b^0_1(x) ) \label{eq:power6} \\
b^1_2 (x) - b^0_2(x) &= f_1^{-1} ( b^{r_5(x)}_2  \mid x ) ( s(0,x) -s_0(0,x)) s^{-1}(1,x) \label{eq:power7} \\
b^1_3 (x) - b^0_3(x) &=  f_1^{-1} ( b^{r_6(x)}_3  \mid x )  ( s(1,x) -s_0(1,x)) p_0(x) s^{-1}(1,x)  \label{eq:power8}
\end{align}

\textbf{ Step 2.a.  Bounding the bias $\psi_1(1) - \psi_1(0) + \psi_4(1)$. }
\begin{align*}
&\E[ \psi_4(1) \mid X=x] =s_0(1,x) b^1_1(x)  ( F_1 (b^1_1(x) \mid x) -F_1 (b^0_1(x) \mid x)) \\
&= s_0(1,x) b^1_1(x)  f_1( b^{r_4(x)}_1(x)  \mid x)  ( b^1_1(x)  - b^0_1(x)) \\
&=  s_0(1,x)( b^1_1(x)  - b^{r_4(x)}_1(x))  f_1( b^{r_4(x)}_1(x)  \mid x)  ( b^1_1(x)  - b^0_1(x)) \\
&+ s_0(1,x)\Gamma_1(r_4(x), x)  ( b^1_1(x)  - b^0_1(x)).
\end{align*}
Therefore, 
\begin{align*}
| \E [ \psi_1(1) - \psi_1(0) + \psi_4(1) \mid X=x ] |&\leq | (\Gamma_1(r_4(x),x) - \Gamma_1(r_1(x),x)) ( b_1^1(x)  - b_1^0(x) ) | \\
&+|f_1( b^{r_4(x)}_1(x)  \mid x) (r_4(x) -r_1(x)) ( b^1_1(x)  - b^0_1(x))^2 |\\
&\leq (C_f + M C_f ) ( b^1_1(x)  - b^0_1(x))^2 + C_f ( b^1_1(x)  - b^0_1(x))^2.
\end{align*}

\textbf{ Step 2.b.  Bounding the bias $\psi_2(1) - \psi_2(0) + \psi_5(1)$.  } 
\begin{align*}
\E [ \psi_5(1) \mid X=x ] &=  b^1_2(x) ( s_0(0,x) - s(0,x)).
\end{align*}
Summing $\E [ \psi_2(1) - \psi_2(0)\mid X=x ] $ and $\E [ \psi_5(1) \mid X=x ] $ gives
\begin{align*}
&| \E [ \psi_2(1) - \psi_2(0) \mid X=x ] + \E [ \psi_5(1) \mid X=x ]  | \\
&= | -b^1_2(x) ( 1- s_0(1,x)/s(1,x)) (s (0,x) -s_0(0,x) ) \\
&+ (  b^1_2(x) -b^{r_2(x)}_2(x) ) (s (0,x) -s_0(0,x) ) s_0(1,x)/s(1,x)  | \\
&= | -b^1_2(x) (s(1,x) -s_0(1,x))  (s (0,x) -s_0(0,x) ) s^{-1}(1,x) \\
&+ (    b^1_2(x) -b^{r_2(x)}_2(x) )(s (0,x) -s_0(0,x) )s_0(1,x)/s(1,x) |.
\end{align*}
Invoking $| b^0_1(x) | \leq M$ and  $s^{-1}(1,x) \leq 2/\kappa$ and 
\begin{align*}
&| b^1_2(x) -b^{r_2(x)}_2(x) ) | \leq  C_f  |(s(0,x)-s_0(0,x))| 
\end{align*}
\begin{align*}
&| \E [ \psi_2(1) - \psi_2(0) \mid X=x ] + \E [ \psi_5(1) \mid X=x ]  | \\
&\leq 2(C_f + M)/\kappa ((s(0,x)-s_0(0,x))^2 + |(s(0,x)-s_0(0,x)) (s(1,x) -s_0(1,x))| ).
\end{align*}

\textbf{ Step 2.c.  Bounding the bias $\psi_3(1) - \psi_3(0) + \psi_6(1)$.  } 
\begin{align*}
\E [ \psi_6(1) \mid X=x ] &=  -b^1_3 (x)  \cdot p(x) \cdot (  s(1,x)-s_0(1,x) ).
\end{align*}
Summing $\E [ \psi_3(1) - \psi_3(0)\mid X=x ] $ and $\E [ \psi_6(1) \mid X=x ] $ gives
\begin{align*}
&| \E [ \psi_3(1) - \psi_3(0) \mid X=x ] + \E [ \psi_6(1) \mid X=x ]  | \\
&\leq | b^{r_3(x)}_3(x) - b^{1}_3(x)  |  \cdot p_0(x) \cdot (s(1,x)-s_0(1,x) ) s_0(1,x)/s(1,x) | \\
&+ |b^{1}_3(x)  (s(0,x) -s_0(0,x)) \cdot (s (1,x) -s_1(0,x) ) s^{-1} (1,x) |
\end{align*}

\textbf{ Step 2.d. Conclusion}  For $k=7,8$, there exist some constant $ C(C_f, \kappa) $ such that
\begin{align*}
& | \E \psi_k(1) |  \leq C(C_f, \kappa) (\sup_{u \in U} (\E (Q(u, X) - Q_0(u,X))^2)^{1/2}  + C_f  \sup_{d \in \{1, 0\}}   {\E (s(d, X) - s_0(d,X) )^2}^{1/2})  \\
 &| \E \psi_9(1) |  \leq   C(C_f, \kappa) \sup_{d \in \{1, 0\}}   \E (s(d, X) - s_0(d,X) )^2,
 \end{align*}
 which implies
 $$
\sup_{\xi \in \Xi_N} \sum_{k=1}^9 \E [ (\psi_k (1) - \psi_k(0)) ] = O(s^2_N +q^2_N).
$$
 
\textbf{ Step 3. Bounding second powers  } Invoking \eqref{eq:power2} gives

Observe that
\begin{align*}
\psi_4(1) - \psi_4(0) &=  \dfrac{D S}{\mu_1(X)} (b^1_1(X) - b^0_3(X)) \left(  1 \{ Y \leq b^1_1(X) \}  - (1 - p(X)) \right) \\
&+ \dfrac{D S}{\mu_1(X)}  b^0_3(X) \left(  1 \{ Y \leq b^1_1(X) \}  -  1 \{ Y \leq b^0_3(X) \}  \right) \\
&+  \dfrac{D S}{\mu_1(X)} b^0_3(X) \left( p(X) -p_0(X) \right) =: I_{41} + I_{42} + I_{43}.
\end{align*}
and
\begin{align*}
\psi_5(1) - \psi_5(0) &= (b^1_2(X) - b^0_3(X)) \dfrac{(1-D)}{\mu_0(X)}  (S -s(0,X)) \\
&+ b^0_3(X) \dfrac{(1-D)}{\mu_0(X)}  (s_0(0,X)-s(0,X)) =: I_{51} + I_{52}.
\end{align*}
and 
\begin{align*}
\psi_6(1) - \psi_6(0) &= -(b^1_3(X) - b^0_3(X)) p(X) \cdot \dfrac{D}{\mu_1(X)} \left(S - s(1,X) \right) \\
&-b_0^3(X) (p(X) - p_0(X)) \cdot   \dfrac{D}{\mu_1(X)} \left(S - s(1,X) \right) \\
&- b_0^3(X) p_0(X) \cdot \dfrac{D}{\mu_1(X)} \left (s_0(1,X) -s(1,X) \right) =: I_{61} + I_{62} + I_{63}.
\end{align*}
Observe that 
\begin{align*}
 \E [ (1 \{ Y \leq b^1_1(X) \}  -  1 \{ Y \leq b^0_3(X) \})^2 \mid X=x] = | F_1 (b^1_1(x) ) - F_1 (b^0_3(x)) | \leq C_f | b^1_1(x)  - b^0_3(x)|.
\end{align*}
Invoking \eqref{eq:power7}-\eqref{eq:power8} gives
\begin{align}
| b^1_1(x) - b^0_1(x) | &\leq \sup_{u \in U} \sup_{d \in \{1,0\}} | Q^d(u,x) -Q^d_0(u,x) | \label{eq:power9} \\
 | b^1_2(x)  - b^0_3(x) | &\leq \sum_{k=2}^3  |b^1_k(x) - b^0_k(x)|  \leq  4C_f/\kappa \sup_{d \in \{1, 0\}} | s(d,x) -s_0(d,x) | \\
 | p(x) - p_0(x) | &\leq 2/ \kappa (| s(0,x) -s_0(0,x) | +  | s(1,x) -s_0(1,x) |).\label{eq:power10}
\end{align}
Thus,
\begin{align*}
\sup_{\xi \in \Xi_N}  \E [I^2_{42} ]  \leq  \sup_{Q^d \in Q^d_N} \sup_{u \in U}  \E |Q^d(u,X) -Q^d_0(u,X)| \leq q^1_N.
\end{align*}
Let $k \in \{1,2,3\}$. For all terms $I_{4k}, I_{5k}, I_{6k}$ not $I_{42}$, the term is a product of a difference of functions of $X$, including $s(d,x) -s_0(d,x), \quad d \in \{1, 0\}$ and \eqref{eq:power9}-\eqref{eq:power10}  and a random variable that is bounded a.s. Therefore, 
\begin{align*}
\sup_{\xi \in \Xi_N} \E I^2_{lk}  \leq O (q^2_N +s^2_N), \quad l \in \{4,5,6\}, k \{1,2,3\}, \quad I_{lk} \neq I_{42}.
\end{align*}
\textbf{ Step 4. Summary} Steps 1-3 imply 
\begin{align*}
\sup_{ \xi \in \Xi_N} |  \E  (g_U (W, \xi) - g_U (W, \xi_0)) 1\{ | \tau(X) | > \rho_N \} | = O (s^2_N + q^2_N). \\
\sup_{ \xi \in \Xi_N}  \E (g_U (W, \xi) - g_U (W, \xi_0))^2 1\{ | \tau(X) | \leq \rho_N \}   = O (q^1_N + s^1_N + s^2_N + q^2_N). 
\end{align*}

\end{proof}

\subsubsection{Proofs from Section \ref{sec:theory}}

\begin{lemma}[Lemma A.3, \cite{CherSem}]
\label{lem:markov}
Let $R(W, \xi)$ be a known function of the data vector $W$ and the nuisance parameter $\xi_0$. Let  $\{ \Xi_N: N \geq 1 \}$ be a sequence of sets that contain the first-stage estimate $\widehat{\xi}$ w.p. approaching one. The sets shrink at the following rates
\begin{align*}
\sup_{\xi \in \Xi_N} | \E [R(W, \xi) - R(W, \xi_0)] | = O(B_N)= o(N^{-1/2}) \\
\sup_{\xi \in \Xi_N}  (\E (R(W, \xi) - R(W, \xi_0))^2)^{1/2}  = O(V_N) =o(1).
\end{align*}
Then, $\sqrt{N} \E_N [R(W_i; \widehat{\xi}_i) -  R(W_i, \xi_0)]= o_P(1)$.
\end{lemma}

\begin{proof}[Proof of Theorem \ref{thrm:condmonot}]
\textbf{Step 1.}  Invoking Lemma \ref{lem:markov} with $B_N = \max^2 (s^{\infty}_N, \rho_N) $ and $V^2_N =  \max (s^{\infty}_N, \rho_N) + s_N$ gives
 \begin{align*}
\sqrt{N} (\E_N g^{\rho_N}_D(W_i; \widehat{s}_i) - g_D(W_i, s_0) ) = o_P (1).
\end{align*}

\textbf{ Step 2. } Combining Lemma \ref{lem:cont}, \ref{lem:orthocor} and  Lemma \ref{lem:class}, the first moments are bounded as
\begin{align*}
| \E  (g_U (W, \xi) - g_U (W, \xi_0)) | &\leq  | \E  (g_U (W, \xi) - g_U (W, \xi_0)) 1 \{ | \tau(X)| > \rho_N  \} | \\
&+  | \E  (g_U (W, \xi) - g_U (W, \xi_0)) 1 \{ | \tau(X)| \leq  \rho_N  \} | \\
 \E  (g_U (W, \xi) - g_U (W, \xi_0))^2  &\leq   \E  (g_U (W, \xi) - g_U (W, \xi_0))^2 1 \{ | \tau(X)| > \rho_N  \}  \\
 &+ \E  (g_U (W, \xi) - g_U (W, \xi_0))^2 1 \{ | \tau(X)| \leq \rho_N  \}.
\end{align*}
Invoking Lemma \ref{lem:markov} with $B_N =   s_N^2 + q_N^2 + \max (s^{\infty}_N, \rho_N)^2 $ and $V^2_N = s_N^1 + q_N^1 + s^2_N +q^2_N + \max (s^{\infty}_N, \rho_N)^2 $ gives  $$\sqrt{N} (\E_N g_U(W_i; \widehat{\xi}_i) - \E_N g_U(W_i, \xi_0) ) = o_P (1).$$
\textbf{ Step 3. } Consider a continuous function  $$\psi( x, y, z) := ( x/z,  \quad y/z)'$$ on $[ N_L/2, 2 N_L] \times [ N_U/2, 2 N_U] \times [\pi_{\text{AT}}/2,  3/2 \pi_{\text{AT}}]$.  Invoking Delta method with $$\begin{pmatrix} x \\  y  \\ z  \end{pmatrix} = \begin{pmatrix}  \E g_L(W, \xi_0) \\  \E g_U(W, \xi_0)  \\ \E g_D(W, s_0)  \end{pmatrix}$$ gives the statement of Theorem.

\end{proof}

\begin{proof}[Proof of Lemma \ref{lem:basicbound}]
I shall decompose the average treatment effect $\beta_0$ in \eqref{eq:truebeta} as 
\begin{align*}
&\E[ Y(1) - Y(0) \mid S(1) = S(0) =1 ] \\
&= \E[ Y(1) - Y(0) \mid S(1) = S(0) =1, X \in \mathcal{X}_{\text{help}} ]  \Pr (X \in \mathcal{X}_{\text{help}} \mid S(1) = S(0) =1 )\\
&+ \E[ Y(1) - Y(0) \mid S(1) = S(0) =1, X \in \mathcal{X}_{\text{hurt}} ]  \Pr (X \in \mathcal{X}_{\text{hurt}}  \mid S(1) = S(0) =1) \\
&+ \E[ Y(1) - Y(0) \mid S(1) = S(0) =1, X \in \mathcal{X}_0 ]  \Pr ( X \in \mathcal{X}_0 \mid S(1) = S(0) =1) \\
&=: S_{\text{help}}  \Pr (X \in \mathcal{X}_{\text{help}} \mid S(1) = S(0) =1 ) + S_{\text{hurt}} \Pr (X \in \mathcal{X}_{\text{hurt}}  \mid S(1) = S(0) =1)  + S_0  \Pr ( X \in \mathcal{X}_0 \mid S(1) = S(0) =1).
\end{align*}
By Assumption \ref{ass:identification:treat:cond}, $X \in \mathcal{X}_{\text{help}} \Rightarrow S(1) \geq S(0)$, and the covariate space is a mixture of always-takers and compliers. By Lemma \ref{lem:cormanski},
$$
S_{\text{help}}    \leq \bar{\beta}^{\text{help}}_U, \quad S_{\text{hurt}} \leq \bar{\beta}^{\text{hurt}}_U.
$$
Finally, if $X \in \mathcal{X}_0$, an observed individual must be an always-taker, and $$ S_0 = \E[ Y(1) - Y(0) \mid S(1) = S(0) =1, X \in \mathcal{X}_0 ] =\bar{\beta}_0.$$ As discussed in the proof of Lemma \ref{lem:atshare}, the conditional always-takers' share is point-identified as $\min (s(0,x), s(1,x))$.  Invoking Bayes rule gives
\begin{align*}
 \Pr (X \in \mathcal{X}_{\text{help}} \mid S(1) = S(0) =1 ) &= \dfrac{  \Pr (S(1) = S(0) =1,  X \in \mathcal{X}_{\text{help}} ) }{ \Pr (S(1) = S(0) =1)} = \dfrac{S_{\text{help}}}{\pi_{\text{AT}}} \\
  \Pr (X \in \mathcal{X}_{\text{hurt}} \mid S(1) = S(0) =1 ) &= \dfrac{  \Pr (S(1) = S(0) =1,  X \in \mathcal{X}_{\text{hurt}} ) }{ \Pr (S(1) = S(0) =1)} = \dfrac{S_{\text{hurt}}}{\pi_{\text{AT}}}.
\end{align*}
Aggregating  over the covariate space gives  basic generalized bound:
\begin{align*}
\bar{\beta}_U = \dfrac{ \bar{\beta}^{\text{help}}_U S_{\text{help}} + \bar{\beta}^{\text{hurt}}_U S_{\text{hurt}}+ \bar{\beta}^0 }{S_{\text{help}} + S_{\text{hurt}}+ S_0}.
\end{align*}

\end{proof}

\section*{Appendix B: Multiple Outcomes}
\label{sec:extensions}
\renewcommand{\theequation}{B.\arabic{equation}}
\renewcommand{\thesection}{B.\arabic{section}}
\renewcommand{\thesubsection}{B.\arabic{subsection}}
\renewcommand{\thelemma}{B.\arabic{lemma}}
\renewcommand{\thetheorem}{B.\arabic{theorem}}
\renewcommand{\thetable}{B.\arabic{table}}
\renewcommand{\thefigure}{B.\arabic{figure}}
\setcounter{equation}{0}
\setcounter{section}{0}
\setcounter{table}{0}
\medskip

In this section, I extend the trimming bounds to accommodate multiple outcomes. Section \ref{sec:singleselecscalar} considers the case of a scalar selection outcome and multiple outcomes.
 Section \ref{sec:singleselecmulti} considers the case of multiple selection outcomes. Section \ref{sec:multiexamples} describes the examples. Section \ref{sec:multitheory} sketches the estimator of the support function and proposes the asymptotic theory. For the sake of exposition, I assume unconditional monotonicity and complete independence (Assumption \ref{ass:identification:treat}) instead of their conditional analogs.

\subsection{Single selection outcome}
\label{sec:singleselecscalar}
Consider the sample selection problem with multiple outcomes.  The data  $W = (X, D, S, S \cdot \mathbf{Y} )$ consist of the covariates $X$, the treatment $D$, the  scalar selection indicator $S$ and the \textbf{multidimensional} outcome $$\mathbf{Y} = (\mathbf{Y}_1, \mathbf{Y}_2, \dots, \mathbf{Y}_{d_{\beta}})' \in \mathrm{R}^{d_{\beta}}$$ observed if and only if $S=1$.  The parameter of interest is the  ATE for the always-takers
\begin{align}
\label{eq:beta00}
\beta_0 = \E[ \mathbf{Y} (1) - \mathbf{Y} (0) \mid S(1) = S(0) = 1].
\end{align}
Since $\beta_0$ is not point-identified, the target parameter is the  identified set $\mathcal{B}$ for $\beta_0$ as well as its projections onto various directions of economic interest.

A standard approach to describing the identified set is to use its support function. Define the $d_{\beta}$-dimensional unit sphere
 \begin{align}
\label{eq:unitsphere}
\mathcal{S}^{d_{\beta}-1} := \{q \in \mathrm{R}^{d_{\beta}}, \quad \| q\| = 1\}.
\end{align}
For any direction $q \in \mathcal{S}^{d_{\beta}-1}$, define the support function as the upper bound on $q' \beta_0$ 
 \begin{align}
\label{eq:suppfun}
    \sigma(q):= \sup_{b \in \mathcal{B}} q' b.
\end{align}

The support function is derived in two steps.  First, I conjecture the formula for $\sigma(q)$.  Second, I verify that $\sigma(q)$ is convex, positive homogenous of degree one, and lower hemicontinuous. Therefore, the set
\begin{align}
\label{eq:idset}
\mathcal{B} = \cap_{q \in \mathrm{R}^{d_{\beta}}: \| q \| =1} \{ b \in \mathrm{R}^{d_{\beta}}: q'b \leq \sigma (q) \}.
\end{align}
is a compact and convex set and $\sigma(q)$ is its support function.

Let me first describe $\sigma(q)$ in the model without covariates. Define the outcome projection $Y_q$ and the data vector $W_q$
$$
Y_q:= q' \mathbf{Y},  \quad W_q:=  (D, S, S\cdot Y_q)
$$
Consider a function
 \begin{align}
\label{eq:suppfun2}
\sigma(q) = \E[ Y_q \mid Y_q \geq Q^1_{Y_q} (1-p_0), D=1, S=1] - \E[ Y_q \mid  D=0, S=1].
\end{align}
By the properties of trimmed mean, $\sigma(q)$ is a convex, positive-homogeneous of degree one, and lower-hemicontinuous functions of $q$. In addition, if  $\{ Y_q, q \in \mathcal{S}^{d_{\beta}-1} \}$ is continuously distributed, $\sigma(q)$ is differentiable. Its gradient is 
 \begin{align}
 \label{eq:gammaq}
\gamma(q) := \partial_q \sigma(q) = \E[ Y \mid Y_q \geq Q^1_{Y_q} (1-p_0), D=1, S=1] - \E[ Y \mid  D=0, S=1].
\end{align}

Next, I describe the support function $\sigma(q)$ for the sharp identified set $\mathcal{B}$ in the model $(X, D, S, \mathbf{Y})$ with covariates. Plugging $W_q$ in place of $W$ into \eqref{eq:momentu} gives the function
\begin{align}
\label{eq:sigmaq}
&\sigma(q) = \dfrac{\E m_U(W_q, \xi_0(q)) }{\E s(0,X)},
\end{align}
where the nuisance parameter $\xi_0(q)$ is
$$
\xi_0(q,x) = \{ s_0(0,x), s_0(1,x), Q^1_{Y_q}(u,x), \quad q \in \mathcal{S}^{d_{\beta}-1}, u \in U \}.
$$
Theorem \ref{prop:idset} shows that $\sigma(q)$ is a convex, positive-homogeneous of degree one, and lower-hemicontinuous functions of $q$. In addition, if  $\{ Y_q, q \in \mathcal{S}^{d_{\beta}-1} \}$ is continuously distributed, $\sigma(q)$ is differentiable and its gradient is equal to
 \begin{align}
 \label{eq:gammaq2}
\gamma(q) := \partial_q \sigma(q) =  \dfrac{\partial_q  \E m_U(W_q, \xi_0(q)) }{\E s(0,X)}.
\end{align}

Theorem \ref{prop:idset} characterizes the sharp identified set $\mathcal{B}$  for the causal parameter $\beta_0$ in \eqref{eq:beta00}. 

\begin{theorem}[Lee's Identified Set]
\label{prop:idset}
Suppose there exists a finite $M$ such that $\| \mathbf{Y} \| \leq M \text{ a.s. }$ and (2) $\{ Y_q, q \in \mathcal{S}^{d_{\beta}-1} \}$ has a conditional density  bounded by $\bar{B}_f$ uniformly over $\mathcal{S}^{d_{\beta}-1}$ and $\mathcal{X}$, namely
\begin{align}
\label{eq:boundedq}
\inf_{q \in \mathcal{S}^{d_{\beta}-1}} \inf_{t \in \text{Conv} (Q_q(1-p), p \in U, q \in \mathcal{S}^{d_{\beta}-1})} \inf_{x \in \mathcal{X}} |f_q (t \mid x) | \geq \bar{B}_f>0.
\end{align}
 Then, 
$\sigma(q)$ in \eqref{eq:suppfun2}  is a convex, positive-homogeneous of degree one, and differentiable  function of $q$. Therefore, the set $\mathcal{B}$ in \eqref{eq:idset} is a compact and strictly convex set, and $\sigma(q)$ is its support function. The boundary of $\mathcal{B}$ consists of the support vectors:
$$
\partial \mathcal{B} = \{ \gamma(q), q \in \mathcal{S}^{d_{\beta}-1} \}.
$$
\end{theorem}

Theorem \ref{prop:idset} shows that the sharp identified set for $\beta_0$ is compact and convex, and  thus can be summarized by its projections on various directions of economic interest. For any point $q$ on the unit sphere, the largest admissible value $\sigma(q)$ of $q'\beta_0$ consistent with the observed data, is  commonly referred to as the \textit{support function}. The sharp bounds for the projection $q' \beta_0$ are given by
$$
[-\sigma(-q), \sigma(q)].
$$
The support function is determined by the moment function \eqref{eq:sigmaq}. Like in the single-dimensional case,  including a wider covariate set will weakly tighten the bounds. 

 The support function is frequently used in econometrics (see, e.g. \cite{BM}, \cite{BMM}, \cite{KaidoSantos}).  A usual approach, applied for the models with bracketed outcome $Y \in [Y_L, Y_U]$,  is to verify convexity and compactness using random set theory and then to derive the closed-form solution for $\sigma(q)$. However, the parameter $\beta_0$ in \eqref{eq:beta00} is not a special case of a set-identified linear model of \cite{BM,BMM}. As a result, we apply a different strategy -- we first establish that $\sigma(q)$ is a support function of some convex and compact set and then  define the identified set as an intersection of supporting hyperplanes.

\subsection{Multiple selection outcomes}
\label{sec:singleselecmulti}

Consider the sample selection problem with multiple outcomes and multiple selection outcomes. In the section below,  the data vector $W = (D,X,  \mathbf{S}, \mathbf{S} \cdot \mathbf{Y} )$ consists of treatment $D$, covariates $X$,  the multi-dimensional outcome $\mathbf{Y} \in \mathrm{R}^{d_{\beta}}$ and the vector of selection indicators 
$\mathbf{S} \in \mathrm{R}^{d_{\beta}}$. Define the scalar selection outcome
\begin{align}
\label{eq:reduced}
S(d) := \begin{cases}1, \quad \mathbf{S}(d) = \mathbf{1} \\
0, \quad \text{otherwise}
\end{cases}, \quad d \in \{1,0\}.
\end{align}
Thus, the problem with multiple selection outcomes  is reduced to the problem with single selection outcome $S:= 1\{ \mathbf{S} =1 \}$. The target parameter $\beta_0$ in \eqref{eq:beta00} reduces to the average treatment effect on subjects who are selected into the sample for each scalar outcome $j: 1 \leq j \leq d$. For example, if $d=2$, the target population is 
$$
S_1(1) = S_1(0) = S_2(1) = S_2(0) = 1.
$$
In what follows, I refer to the \eqref{eq:monot1} as reduced scalar monotonicity.

A sufficient condition for monotonicity \eqref{eq:monot1} of the reduced outcome $S$ is to have vector monotonicity:
\begin{align}
\label{eq:monot2}
\mathbf{S}(1) \geq \mathbf{S}(0),
\end{align}
frequently employed in causal inference literature (e.g., actual monotonicity in \cite{MTW,MTW2}). Indeed, if \eqref{eq:monot2} holds for $\mathbf{S}$, \eqref{eq:monot1} holds for $S$. For another example, the following inequality
\begin{align*}
S_1(1) \leq S_1(0) \leq  S_2(0) \leq S_2(1), 
\end{align*}
also implies reduced scalar monotonicity with $S$ in \eqref{eq:reduced} and   $S(0) \geq S(1)$.

\subsection{Examples}
\label{sec:multiexamples}

\begin{example}[Wage Growth]
\label{ex:wagegrowth}
 Let $\mathbf{S}=(S_{t_1}, S_{t_2})$ be a vector of employment outcomes for the period $t \in \{ t_1,t_2 \}$, $\mathbf{Y}=(Y_{t_1}, Y_{t_2})$ be a vector of log wages.  Let 
 $$
 \beta_0  = (\beta_{t_1}, \beta_{t_2})= \E[  \mathbf{Y}(1) - \mathbf{Y}(0) \mid S(1) = S(0) = 1] 
 $$
be the effect on log wage in time periods $t_1$ and $t_2$.   The target parameter is  the effect on \textbf{average wage growth}  from $t_1$ to $t_2$, that is, $ \beta_{t_2}-\beta_{t_1}$.   
For $\sigma(q)$ in \eqref{eq:sigmaq}, the sharp  bounds on $\beta_{t_2} - \beta_{t_1}$ are
\begin{align} \label{eq:growth} [-\sqrt{2} \sigma (-q_0), \quad \sqrt{2} \sigma(q_0)], \quad q_0= (1/\sqrt{2}, -1/\sqrt{2}). \end{align} 
\end{example}

Example \ref{ex:wagegrowth} describes the bounds on the effect on the average wage growth.  A simplistic approach to construct an upper bound on $\beta_{t_2} -  \beta_{t_1}$ is to subtract the lower bound on $ \beta_{t_1}$ from the upper bound on $\beta_{t_2}$. Since wages in weeks $t_1$ and $t_2$ are likely to be correlated, this upper bound may not correspond to any data generating process consistent with  the observed data. In contrast, the sharp  bounds on $ \beta_{t_2}-\beta_{t_1}$  are obtained by projection of $\mathcal{B}$ onto the $-45$ degree line, as formalized in \eqref{eq:growth}.

 \begin{example}[Aggregated Treatment Effect]
\label{ex:ste}
Let $\mathbf{Y}$ be a vector of related outcomes from a shared domain and $\mathbf{S} = (S_1, S_2, \dots, S_j, S_{d_{\beta}} )$ be a vector such that $S_j =1$ if $Y_j$ is observed. Let 
 $$
 \beta_0  = (\beta_1, \dots, \beta_{d_{\beta}})= \E[  \mathbf{Y}(1) - \mathbf{Y}(0) \mid S(1) = S(0) = 1] 
 $$
 be the average causal effect on each component of the outcome.   A common approach to summarize findings is to consider the
\textit{aggregated treatment effect} 
\begin{align}
 \label{eq:ste}
 \text{ATE} = \dfrac{1}{d_{\beta} } \sum_{j=1}^{d_{\beta}} \dfrac{\beta_j}{\zeta_j},
\end{align}
where $\zeta_j$ is the standard deviation of  the outcome $j$ in the control group $Y_j \mid D=0, S=1$. The sharp lower and upper bounds on $\text{ATE}$ are given by
\begin{align} \label{eq:ste} [ -  C_{\zeta} \sigma(-q), \quad  C_{\zeta}  \sigma(q)], \end{align}
where  $\widetilde{q}= 1/\zeta$ and  $q=\widetilde{q}/\| \widetilde{q} \|$ and  $C_{\zeta}=\| \widetilde{q} \|/d_{\beta}$. 
\end{example} 

Example \ref{ex:ste} describes the bounds on the effect on a linear combination of related outcomes. In contrast to Example \ref{ex:wagegrowth}, the target direction $q$ is  \textit{an unknown population parameter} that needs to be estimated. Therefore, the estimator $\widehat \sigma(q)$ must be approximated \textit{in some neighborhood of $q$} rather than just at a specific $q$ itself. This parameter  calls for the uniform Gaussian approximation of $\widehat \sigma(q)$ established in Theorem \ref{thrm:condmonot2}  in addition to the pointwise one established in Theorem \ref{thrm:condmonot}.

\subsection{Overview of the Estimator and the Results}
\label{sec:multitheory}

The proposed estimator of $\widehat \sigma(q)$ consists of two stages. In the first stage, I estimate $s(d,x)$ and $Q^d_{Y_q}(u,x)$ such as in Examples \ref{ex:sel} and \ref{ex:q}, respectively, and construct the first-stage fitted values. In the second stage, I estimate $\widehat \sigma(q)$  as in Definition \ref{alg:leesupp}.  

\begin{definition}[Support Function]
\label{alg:leesupp}
Given the estimated first-stage values $(\widehat s(0,X_i), \widehat s(1,X_i), \widehat \tau(X_i)=(\widehat s(1,X_i)-\widehat s(0,X_i)), \widehat{Q}^d_{Y_q}(u,X_i))_{i=1}^N$,  the support function estimator is
\begin{align}
\label{eq:genestsupp}
\widehat \sigma(q) := \dfrac{N^{-1} \sum_{i=1}^N g_U(W_{q,i}, \widehat{\xi}_i (q))}{\widehat \pi_{\text{AT}}}.
\end{align}p

\end{definition}

\begin{assumption}[Regularity conditions for multiple outcomes]
\label{ass:firststageq}
Suppose the conditions of Theorem  \ref{prop:idset} holds. In addition, suppose the $\ell_1$-regularization estimated quantile $\widehat Q_{Y_q} (p(X),X) = Z(X)' \widehat \zeta(q)$ belongs to the realization  set
$$
Q_N^d:= \bigg\{  W \rightarrow Z(X)' \zeta(q) , \quad \zeta(q) \in \mathrm{R}^{p},  \sup_{q} \| \zeta(q) \|_0 \leq C_X s_N,   \sup_{q} \| \zeta(q) \| \leq C_X, \quad q \in \mathcal{S}^{d_{\beta}-1}  \bigg\}, \quad d \in \{1, 0\}.
$$ 
w.p. $1-o(1)$. Furthermore, the rates $q_N$ and $q^1_N$, redefined as 
\begin{align*}
 \sup_{d \in \{1, 0\}}\sup_{Q \in Q_N^d} \sup_{u \in U} \sup_{q \in \mathcal{S}^{d_{\beta}-1}} (\E  (Q_{Y_q} ^d(u, X) -Q_{Y_q,0} ^d(u,X) )^2)^{1/2} =:q_N \\
  \sup_{d \in \{1, 0\}}\sup_{Q \in Q_N^d} \sup_{u \in U}  \sup_{q \in \mathcal{S}^{d_{\beta}-1}} \E | Q_{Y_q} ^d (u, X) - Q_{Y_q,0} ^d (u, X) | =:q^1_N.
\end{align*}
obey a condition with $V^2_N:=s^2_N + q^2_N + s^1_N + q^1_N$ and $a_N = p + N$ 
\begin{align}
\label{eq:cond}
V_N\log^{1/2} (a_N/V_N)+ N^{-1/2 } \log^{1/2} (a_N/V_N) = o(1).
\end{align}
The conditional CDF is Lipschitz in $q$, namely, for some $L_2$-integrable $\bar{F}(x)$,
\begin{align}
\label{eq:bound2}
 \sup_{t \in \mathcal{R}} \sup_{x \in \mathcal{X}} | F_{q_1}(t \mid x) - F_{q_2}(t \mid x)| \leq \bar{F}(x) \| q_1 - q_2 \|, \quad q_1, q_2 \in \mathcal{S}^{d_{\beta}-1}
\end{align}

\end{assumption}

Assumption \ref{ass:firststageq} states the conditions for uniform inference on the support function. First, it redefines the convergence rates $q_N$ and $q^1_N$ and places an additional restriction \eqref{eq:cond}, which comes from maximal inequality to bound estimation error uniformly over $q \in \mathcal{S}^{d_{\beta}-1}$. Finally, the quantile  realization set is restricted to linear sparse combinations of covariates.

Theorem \ref{thrm:condmonot2} shows that the Support Function Estimator is asymptotically equivalent to a tight Gaussian process with a nondegenerate covariance function.

\begin{theorem}
\label{thrm:condmonot2}
Suppose Assumptions   \ref{ass:boundedoutcome2}--\ref{ass:boundedoutcome3} hold for  $\{Y_q, q \in \mathcal{S}^{d_{\beta}-1} \}$ uniformly over $\mathcal{S}^{d_{\beta}-1}$. Suppose Assumption \ref{ass:firststageq} hold and $s^2_N + q^2_N = o (N^{-1/2})$.  Then,   the support function process $S_N(q):= \sqrt{N}  (\widehat{\sigma}(q) - \sigma(q))$
is asymptotically linear uniformly on $\mathcal{S}^{d_{\beta}-1}$,
$$
S_N(q) = \G_N [ h(W, q)] + o_P(1) \text{ uniformly on } \mathcal{S}^{d_{\beta}-1}, 
$$
where 
\begin{align*}
h(W, q) :&= \pi^{-1}_{\text{AT}} (g_U(W_q, \xi_0(q)) - \sigma(q) - \pi_{\text{AT}}^{-1} \sigma(q)( g_D(W, \tau_0) - \pi_{\text{AT}})).
\end{align*}
Furthermore,   the process $S_N(q)$ admits the following approximation
  $$
S_N(q) =_d \G[h(q) ] + o_P(1) \quad \text{ in } \ell^{\infty} (\mathcal{S}^{d_{\beta}-1}),$$
where the process $ \G[h(q)] $ is a tight $P$-Brownian bridge in $ \ell^{\infty} (\mathcal{S}^{d_{\beta}-1})$ with a non-degenerate covariance function
\begin{align*}
    \Omega(q_1,q_2 ) = \E [h(W,q_1)h(W,q_2 )] - \E[h(W,q_1)] \E[h(W,q_2 )], \quad q_1, q_2  \in \mathcal{S}^{d_{\beta}-1}.
\end{align*}

\end{theorem}

\begin{lemma}[Lipschitz CDF implies Lipschitz quantile]
\label{quantile}
Let $\{F_{q}(\cdot), q \in \mathcal{S}^{d_{\beta}-1} \}$ be the CDF of $Y_q$  such that
\begin{align}
\label{eq:bound}
 \sup_{t \in \mathcal{R}} | F_{q_1}(t) - F_{q_2}(t)| \leq \bar{F} \| q_1 - q_2 \|
\end{align}
and the PDF such that
$$
\inf_{q \in \mathcal{S}^{d_{\beta}-1}} \inf_{t \in \text{Conv} (Q_q(1-p), p \in U, q \in \mathcal{S}^{d_{\beta}-1})} |f_q (t) | \geq \bar{B}_f>0.
$$
Then, 
\begin{align}
\label{eq:Lipschitz}
 \sup_{q_1,q_2 \in \mathcal{S}^{d_{\beta}-1}} \sup_{p \in U}  | Q_{q_2}(1-p) - Q_{q_1}(1-p) | &\leq \sup_{t \in \mathcal{R}} | F_{q_1} (t) - F_{q_2} (t)  | \sup_{t \in \mathcal{R}} \sup_{q \in \mathcal{S}^{d_{\beta}-1}} f_{q}^{-1}(t) \nonumber \\
    &\leq   \| q_1 - q_2 \| \bar{F}/\bar{B}_f.
\end{align}
\end{lemma}

\begin{proof}[Proof of Lemma \ref{quantile}]
I use an implicit function theorem-type argument. By definition,
$$
F_{q_1} (Q_{q_1}(1-p)) = 1-p, \quad F_{q_2} (Q_{q_2}(1-p)) = 1-p,
$$
which implies
\begin{align*}
F_{q_1} (Q_{q_1}(1-p)) - F_{q_2} (Q_{q_2}(1-p)) &= F_{q_1} (Q_{q_1}(1-p))  - F_{q_2} (Q_{q_1}(1-p))  \\
&+  F_{q_2} (Q_{q_1}(1-p)) - F_{q_2} (Q_{q_2}(1-p)) = 0.
\end{align*}
Mean value theorem gives for some $t^{*} \in (Q_{q_1}(1-p), Q_{q_2}(1-p))$ 
\begin{align*}
  | F_{q_2} (Q_{q_1}(1-p)) - F_{q_2} (Q_{q_2}(1-p)) | = f_{q_2} (t^{*}) |Q_{q_1}(1-p)) - Q_{q_2}(1-p)|,
\end{align*}
and
\begin{align*}
|Q_{q_1}(1-p) - Q_{q_2}(1-p)| \leq \sup_{t} \sup_{t^{*}} \dfrac{  | F_{q_2} (t) - F_{q_2} (t) |}{  f_{q_2} (t^{*}) } \leq \bar{F} \| q_1 - q_2 \|/ \bar{B}_f.
\end{align*}
Invoking \eqref{eq:bound} gives \eqref{eq:Lipschitz}.
\end{proof}

\begin{assumption}[Concentration]
\label{ass:concentration:chap1}
Let $\{ R(W, \xi(q)), q \in \mathcal{S}^{d_{\beta}-1}\}$ be a known function of the data vector $W_q$ and the nuisance parameter $\xi_0(q)$.  The following conditions hold for the function class 
\begin{align}
\label{eq:fxi}
    \mathcal{F}_{\xi} &= \{ R(W, \xi(q)), q \in \mathcal{S}^{d_{\beta}-1}\}
\end{align}
(1). There exists  a  measurable envelope function  $F_{\xi} = F_{\xi}(W)$ that almost surely bounds all elements in the class $ \sup_{q\in \mathcal{S}^{d_{\beta}-1}} | R(W, \xi(q)) | \leq F_{\xi}(W) \quad \text{a.s.}$. There exists $c>2$ such that $\|F_{\xi}\|_{L_{P,c}} := \left(\int_{w \in \mathcal{W}}  |F_{\xi}(w)|^c \right)^{1/c} < \infty$. There exist constants $a,v$ that may depend on $N$ such that the uniform covering entropy of the function class $\mathcal{F}_{\xi}$ is  bounded  
\begin{align}
\label{eq:UCE}
 \log \sup_{Q} N(\epsilon \| F_{\xi} \|_{Q,2}, \mathcal{F}_{\xi} , \| \cdot \|_{Q,2}) \leq  v \log (a/\epsilon), \quad \text{ for all } 0 < \epsilon \leq 1.
\end{align}
and \eqref{eq:cond} holds.
(2) For the true parameter value $\xi_0$, the function class $   \mathcal{F}_{\xi_0}$ is $P$-Donsker, and \eqref{eq:UCE} holds for some constant $\bar{v}$ and $\bar{a}$  that do not change with $N$. 
\end{assumption}
 
\begin{lemma}[Negligible First-Stage Error]
\label{lem:markovq}
Let $\{ R(W_q, \xi(q)), q \in \mathcal{S}^{d_{\beta}-1}\}$ be a known function of the data vector $W_q$ and the nuisance parameter $\xi(q)$. Suppose the following conditions hold. (1) There exist sequences $B_N = o(N^{-1/2})$ and $V_N = o(1)$ such that
\begin{align*}
\sup_{\xi \in \Xi_N} \sup_{q \in \mathcal{S}^{d_{\beta}-1}} | \E [R(W_q, \xi(q)) - R(W_q, \xi_0(q))] | = O(B_N)= o(N^{-1/2}) \\
\sup_{\xi \in \Xi_N} \sup_{q \in \mathcal{S}^{d_{\beta}-1}}  (\E (R(W_q, \xi (q)) - R(W_q, \xi_0(q)))^2)^{1/2}  = O(V_N) =o(1).
\end{align*}
(2) Assumption \ref{ass:concentration:chap1} holds. Then, $\sup_{q \in \mathcal{S}^{d_{\beta}-1}} |\sqrt{N} \E_N [R(W_{qi}; \widehat{\xi}_i (q)) -  R(W_{qi}, \xi_0(q))]|= o_P(1)$.
\end{lemma}

Lemma \ref{lem:markovq} follows from the Steps  2--5 of the proof of Theorem 3.1 in \cite{SemJoE} as well as the maximal inequality (Lemma 6.2 in \cite{chernozhukov2016double}).

  \begin{lemma}[Entropy Bounds]
\label{lem:maxineq:suppfun2}
(1) Let $Z(X)$ be a vector of basis functions.  The function class
\begin{align}
\mathcal{M} &= \bigg\{ W \rightarrow Z(X)' \zeta, \quad \zeta \in \mathrm{R}^{p},  \| \zeta \|_0 \leq C_X s_N,  \| \zeta \| \leq C_X \bigg\} \nonumber
\end{align}
obeys  $a_N = p +N$,
 \begin{align}
\label{eq:entropyboundapp2}
 \log \sup_{Q} N(\epsilon \| M \|_{Q,2}, \mathcal{M} , \| \cdot \|_{Q,2}) \lesssim 1+  s_N \log (a_N/\epsilon), \quad \text{ for all } 0 < \epsilon \leq 1.
 \end{align}
 (2) Suppose $\mathbf{Y} \in \mathrm{R}^{d_{\beta}}$ is a.s. bounded: $\| \mathbf{Y} \| \leq M \text{ a.s. }$ for some finite $M$. (3) There exists an integrable function $\bar{F}(x)$, so that
\begin{align*}
   \sup_{t \in \mathrm{R}^{d_{\beta}}}  |F_{q_1} (t \mid x) - F_{q_2} (t \mid x) | \leq \bar{F}(x) \| q_1 - q_2 \|, \quad \forall x \in \mathcal{X}
\end{align*}
and $\inf_{q \in \mathcal{S}^{d_{\beta}-1}} \inf_{t \in \text{Conv} (Q_q(1-p), p \in U, q \in \mathcal{S}^{d_{\beta}-1})} |f_q (t \mid x) | \geq \bar{B}_f>0$ for all $x \in \mathcal{X}$. Then, \eqref{eq:UCE} holds with $\bar{v} = d$ and $\bar{a}$ that do not depend on $N$, and the function class $\mathcal{M}$ 
\begin{align*}
    \mathcal{M}:= \{ X \rightarrow Q_{Y_q}(p(X), X), \quad q \in \mathcal{S}^{d_{\beta}-1} \}
\end{align*}
is a VC class.

\end{lemma}

\begin{proof}[Proof of Lemma \ref{lem:maxineq:suppfun2}]
(1) is stated without proof. (2) Invoking \eqref{eq:Lipschitz} gives 
\begin{align}
\label{eq:boundqq}
    | Q_{q_1} (p(x), x) - Q_{q_2}(p(x), x) | &\leq   \bar{F}(x)/ \bar{B}_f  \| q_1 - q_2 \| 
\end{align}
By Example 19.7 from \cite{vdv}, the bracketing numbers of the function class $ \mathcal{M}$ obey
\begin{align*}
    N_{[]}( \epsilon \|  \bar{F}(x)/ \bar{B}_f \|_{P,r},  \mathcal{M}, L_r(P)) \lesssim \left(\dfrac{2}{\epsilon}\right)^d, \text{ every } 0< \epsilon < 2.
\end{align*}
Finally, since $\mathbf{Y} \in \mathrm{R}^{d_{\beta}}$ is an a.s. bounded vector,  each element of the class $ \mathcal{M}$ is bounded by $\| \mathbf{Y}  \| \leq M \text{ a.s. }$, and $M$ can be taken as the envelope of $ \mathcal{M}$. Therefore, $ \mathcal{M}$ is $P$-Donsker and obeys \eqref{eq:UCE} with $v=d$ and $a=2$. 

\end{proof}

\begin{lemma}[Verification of Assumption \ref{ass:concentration:chap1}]
\label{lem:uce}
The class
$\mathcal{F}_{\xi_0} = \{ g_U(W_q, \xi_0(q)), q \in \mathcal{S}^{d_{\beta}-1} \}$ obeys \eqref{eq:entropyboundapp2} with $v$ and $a$ that do not change with $N$. Furthermore, the class
$\mathcal{F}_{\xi} = \{ g_U(W_q, \xi(q)), q \in \mathcal{S}^{d_{\beta}-1} \}$ obeys \eqref{eq:entropyboundapp2} with some $v_N$ and $a_N$. 

\end{lemma}

\begin{proof}[Proof of Lemma \ref{lem:uce}]

Consider the class of true quantile functions
$$
\mathcal{L}_{\xi_0}:=\bigg\{ L(q,X),  \quad q \in \mathcal{S}^{d_{\beta}-1} \bigg\}=:\{ X \rightarrow Q_{Y_q}(p(X), X), \quad q \in \mathcal{S}^{d_{\beta}-1} \}
$$
and the class of estimates 
$$
\mathcal{L}_{\xi}:= \bigg\{  W \rightarrow Z(X)' \zeta, \quad \zeta \in \mathrm{R}^{p},  \| \zeta \|_0 \leq C_X s_N,  \| \zeta \| \leq C_X  \bigg\}
$$

\textbf{ Step 1. } The function class $\mathcal{H}'_{\xi} = \bigg\{ W \rightarrow Y_q -  L(q,X),  \quad q \in \mathcal{S}^{d_{\beta}-1} \bigg\}$ is the sum of 2 classes obeying \eqref{eq:UCE}. Therefore, by \cite{andrews1994}, $\mathcal{H}'$ is a VC class itself. Therefore, the class of indicators 
\begin{align*}
    \mathcal{H}:= \bigg\{ W \rightarrow 1{\{ Y_q -  L(q,X) \leq 0 \}}, \quad q \in \mathcal{S}^{d_{\beta}-1} \bigg\}.
\end{align*}
also obeys \eqref{eq:UCE} with possibly different constants.

\textbf{ Step 3 }.  The function class 
$$\mathcal{H}_1 = \bigg \{ W \rightarrow \dfrac{D \cdot S \cdot 1{\{ Y_q \leq L(q,X) \}}}{\mu_1(X)}  \bigg \}$$
is obtained by multiplying each element of $ \mathcal{H}$  by an a.s. bounded random variable  $D\cdot S/\mu_1(X)$.  The function class $$\mathcal{H}_2 =  \bigg \{ W \rightarrow L(q,X) \left(\dfrac{(1-D)}{\mu_0(X)} - s(0,X) \right)  \bigg \}$$ is obtained from  $ \mathcal{M}$  by multiplying each element of $ \mathcal{M}$  by an a.s. bounded random variable $\left(\dfrac{(1-D)}{\mu_0(X)} - s(0,X) \right)$.  The same argument applies to the function class
$$\mathcal{H}_3 = \bigg\{ W \rightarrow L(q,X)p(X) \bigg(\dfrac{D}{\mu_1(X)} - s(1,X ) \bigg) \bigg \}.$$ The function class 
$$\mathcal{H}_4 =  \bigg \{ W \rightarrow L(q,X)s(1,X) \left( \dfrac{D \cdot S  1{\{ Y_q \leq L(q,X) \}}}{\mu_1(X) s(1,X)} - p(X) \right)  \bigg \} $$ is obtained 
as a product of function classes $\mathcal{M}$ and $\mathcal{H}$, multiplied by a random variable $s(1,X)$. Finally, the function class $\mathcal{F}_{\xi}$ in \eqref{eq:fxi} is obtained by adding the elements of $\mathcal{H}_k, \quad k=1,2,3,4$.
Since  entropies obey the rules of addition and multiplication by a random variable (\cite{andrews1994}), the argument follows.

\end{proof}

\subsection{Proofs for Appendix B}

\begin{proof}[Proof of Theorem \ref{prop:idset}]

Suppose $\sigma(q): \mathcal{S}^{d_{\beta}-1} \rightarrow \mathrm{R}$  is (1) convex, (2) positive homogenous of degree one and (3) lower-semicontinuous function of $q$. By Corollary 13.2.1 from \cite{Rock},  the properties (1)-(3) imply that $\mathcal{B}$ in \eqref{eq:idset} is a convex and compact set and $\sigma(q)$ is its support function.  Steps 1-3 verify these properties for the trimming functional $q \rightarrow \sigma(q)$ in \eqref{eq:suppfun}. Steps 4 shows that $ \sigma(q)$ in \eqref{eq:suppfun} is differentiable if $Y \mid D=1,S=1$ is continuously distributed. Relying on Steps 1--4, Steps 5-6 verify the properties (1)--(3) and differentiability for $ \sigma(q)$ in \eqref{eq:suppfun2}, which establishes Theorem \ref{prop:idset}.

\textbf{ Step 1. } By construction, $q' \beta_0 = q' \E[ \mathbf{Y} (1) -  \mathbf{Y} (0) \mid S(1)=S(0)=1] =   \E [ Y_q (1) - Y_q(0) \mid S(1) = S(0) = 1]$, which coincides with the one-dimensional ATE in the model 
$(D,S,Y_q)$. Invoking Lee bound for one-dimensional case gives
\begin{align*}
q_1' \beta_0 \leq \sigma(q_1) \quad \text{ and } \quad q_2' \beta_0 \leq \sigma(q_2).
\end{align*}
Let $\lambda \in [0,1]$. Multiplying the inequalities by $\lambda$ and $1-\lambda$ gives $$(\lambda q_1 + (1-\lambda)q_2)' \beta_0 \leq \lambda  \sigma(q_1)  + (1-\lambda) \sigma(q_2).$$ Next, take $q:=\lambda q_1 + (1-\lambda) q_2$. By sharpness,  $\sigma( \lambda q_1 + (1-\lambda) q_2)$ is the smallest bound on $(\lambda q_1 + (1-\lambda)q_2)' \beta_0$ in the model without covariates.  Therefore,
\begin{align*}
\sigma( \lambda q_1 + (1-\lambda) q_2) \leq \lambda \sigma(q_1) + (1-\lambda) \sigma(q_2),
\end{align*}
which implies that $\sigma(q)$ is a convex function of $q$.

\textbf{ Step 2. } Let $\lambda >0$.  Observe that the  event $\{ \lambda q' \mathbf{Y} \geq Q_{\lambda q' \mathbf{Y}}( u, X) \} $ holds if and only if $\{Y_q \geq Q_{Y_q}( u, X) \} $. Since $Y_q = q' \mathbf{Y}_q$ is a linear function of $q$, $\sigma(q)$ defined in \eqref{eq:sigmaq} is positive homogenous of degree $1$. 

\textbf{ Step 3. } Consider a sequence of vectors $q_k \rightarrow q, k \rightarrow \infty$. Suppose $\sigma(q_k) \leq C$. Then,
$ q_k' \beta_0 \leq \sigma(q_k)  \leq C$, which implies that $q' \beta_0 \leq C$ must hold. Therefore, $C$ is a  bound on $q' \beta_0$. By sharpness, $\sigma(q)$ is the smallest bound on $q'\beta_0$, which implies $\sigma(q) \leq C$.  

\textbf{ Step 4. }  I  show that  $\sigma(q)$ is differentiable in $q$, which implies that $\mathcal{B}$ in \eqref{eq:idset} is strictly convex.   Recall that $\gamma(q)$ is defined as
\begin{align*}
\gamma (q):&= \E [Y \mid Y_q \geq Q^1_{Y_q} (1-p_0), D=1, S=1] - \E[ Y \mid D=0, S=1]\\
&=p_0^{-1}  \E  [Y 1\{ Y_q \geq Q^1_{Y_q} (1-p_0) \} \mid D=1, S=1]- \E[ Y \mid D=0, S=1].
\end{align*}
Let $q_1, q_2 \in \mathcal{S}^{d_{\beta}-1}$. Define
\begin{align*}
G(q_1, q_2):&= \sigma(q_2) - \sigma(q_1) - \gamma (q_1)' (q_2 - q_1) \\
&=p_0^{-1}  \E  [Y_{q_2} (1\{ Y_{q_2}   - Q^1_{Y_{q_2}} (1-p_0) \geq 0  \} - 1\{ Y_{q_1}   - Q^1_{Y_{q_1}} (1-p_0) \geq 0 \}) \mid D=1, S=1 ]  \\
&=p_0^{-1} \E  [ (Y_{q_2} -  Q^1_{Y_{q_2}} (1-p_0))   (1\{ Y_{q_2}   - Q^1_{Y_{q_2}} (1-p_0)  \geq 0  \} - 1\{ Y_{q_1}   - Q^1_{Y_{q_1}} (1-p_0)  \geq 0 \}) \mid D=1, S=1 ]  \\
&+ p_0^{-1}  Q^1_{Y_{q_2}} (1-p_0) \E [ (1\{ Y_{q_2}   - Q^1_{Y_{q_2}} (1-p_0) \geq 0  \} - 1\{ Y_{q_1}   - Q^1_{Y_{q_1}} (1-p_0) \geq 0 \}) \mid D=1, S=1 ] \\
&= G_1(q_1, q_2) + G_2(q_1, q_2).
\end{align*}
The second term is zero by construction
$$
G_2(q_1, q_2) = p_0^{-1} Q^1_{Y_{q_1}} (1-p_0) (p_0 - p_0) =0.
$$
To bound the first term, I invoke the following bound
\begin{align}
\label{eq:xyz}
|  1{\{ x \geq  z \}} -1{\{ y \geq z \}} | \leq 1 \{ | z -y | < | x - y | \}
\end{align}
with $y:= Y_{q_2}   - Q^1_{Y_{q_2}} (1-p_0) $ and $x:= Y_{q_1}   - Q^1_{Y_{q_1}} (1-p_0)$ and $z:=0$. As a result, 
\begin{align*}
|G(q_1, q_2)| &\leq p_0^{-1} \E | Y_{q_2} -  Q^1_{Y_{q_2}} (1-p_0) | 1 \{  | Y_{q_2} -  Q^1_{Y_{q_2}} (1-p_0)   | \\
&< | Y_{q_2} - Y_{q_1} -  ( Q^1_{Y_{q_2}} (1-p_0)  - Q^1_{Y_{q_1}} (1-p_0)) |  \} \mid D=1, S=1 ].
\end{align*}
Invoking \eqref{eq:Lipschitz} from Lemma \ref{quantile} gives for  $\bar{M}:= M+ \bar{F}/\bar{B}_f$
\begin{align*}
 | Y_{q_2} - Y_{q_1} -  ( Q^1_{Y_{q_2}} (1-p_0)  - Q^1_{Y_{q_1}} (1-p_0)) | &\leq  \| q_2 - q_1 \| M + | ( Q^1_{Y_{q_2}} (1-p_0)  - Q^1_{Y_{q_1}} (1-p_0))| \\
 &\leq  \| q_2 - q_1 \| \bar{M}.
\end{align*}
which is symmetric in $(q_1, q_2)$. An upper bound for $G(q_2, q_1)$ gives 
\begin{align*}
|G(q_2, q_1)|  &\leq p_0^{-1}     \E[ | Y_{q_1} -  Q^1_{Y_{q_1}}  (1-p_0) | 1 \{  0 \leq | Y_{q_1} -  Q^1_{Y_{q_1}} (1-p_0) | <  \bar{M} \| q_2 - q_1 \|  \} \mid D=1, S=1 ]  \\
&\leq p_0^{-1} \bar{M} \| q_2 - q_1 \| \Pr (0 \leq | Y_{q_1} -  Q^1_{Y_{q_1}} (1-p_0)) | <  \bar{M} \| q_2 - q_1 \|  ) \\
&\leq p_0^{-1} \bar{M} \| q_2 - q_1 \|  (F_{q_1} ( Q^1_{Y_{q_1}} (1-p_0) +  \bar{M} \| q_2 - q_1 \| ) - F_{q_1} ( Q^1_{Y_{q_1}}  (1-p_0)  - \bar{M} \| q_2 - q_1 \| )) \\
&= O ( \| q_2 - q_1 \|^2)= o (\| q_1 - q_2 \|), \quad q_1 - q_2 \rightarrow 0.
\end{align*}
A similar argument applies for $G(q_1, q_2)$.

\textbf{ Step 5. }  The trimmed mean functions are
\begin{align*}
\beta (q,x) = \E[ Y_q \mid Y_q \geq Q^1_{Y_q} (1-p(x), x), D=1, S=1, X=x] -\E[ Y_q \mid  D=0, S=1, X=x]
\end{align*}
and
$$
\sigma(q) = \dfrac{ \E_{X} \beta (q,X)  s(0,X)}{\E s(0,X)}.
$$
By construction, for each $x \in \mathcal{X}$, $\beta (q,x)$ is a convex and positive homogenous of degree one function of $q$. Therefore, a weighted average of these functions (with non-negative weighting function) must retain the properties. 

\textbf{ Step 6. } I invoke the argument of Step 5 conditional on covariates. Define the conditional gradient as 
$$
\gamma (q,x)=\E[ Y \mid Y_q \geq Q^1_{Y_q} (1-p_0), D=1, S=1,X=x] - \E[ Y \mid  D=0, S=1,X=x].
$$
and the derivative as 
\begin{align*}
G(q_2, q_1,x):&= \beta(q_2,x) - \beta(q_1,x) - \gamma (q_1,x)' (q_2 - q_1) =: G_1(q_1, q_2,x) + G_2(q_1, q_2,x),
\end{align*}
where $G_2(q_1, q_2,x)=0$ for all $q_1$ and $q_2$.  Invoking \eqref{eq:boundedq} gives
\begin{align*}
&\sup_{q_1, q_2 \in \mathcal{S}^{d_{\beta}-1} , x \in \mathcal{X}} |G(q_1, q_2,x)| = O ( \| q_2 - q_1 \|^2) = o (\| q_1 - q_2 \|), \quad q_1 - q_2 \rightarrow 0.
\end{align*}
Therefore, $q \rightarrow \beta(q,x)$ is differentiable in $q$ almost surely in $\mathcal{X}$ with the gradient $\gamma(q,x)$. Note that $\sigma(q)$ in \eqref{eq:suppfun2} reduces to
$$\sigma(q)=\dfrac{  \int_{\mathcal{X}} \beta(q,x) s(0,x) f_X(x) dx}{ \int_{\mathcal{X}}s(0,x) f_X(x) dx } = \dfrac{\E m_U (W_q, \xi_0(q))}{\E s(0,X)}$$
The following dominance condition holds
\begin{align*}
\sup_{q \in \mathcal{S}^{d_{\beta}-1}} \| \gamma(q,x) \|&=\sup_{q \in \mathcal{S}^{d_{\beta}-1}} \|  \nabla_q \beta(q,x)  \|  \leq 2 \bar{M}, \quad \forall x \in \mathcal{X}.
\end{align*}
By dominated conference theorem,  the function 
$$
\partial_q   \dfrac{  \int_{\mathcal{X}} \beta(q,x) s(0,x) f_X(x) dx}{ \int_{\mathcal{X}}s(0,x) f_X(x) dx }  =   \dfrac{  \int_{\mathcal{X}} \partial_q  \beta(q,x) s(0,x) f_X(x) dx}{ \int_{\mathcal{X}}s(0,x) f_X(x) dx } =   \dfrac{  \int_{\mathcal{X}} \gamma(q,x) s(0,x) f_X(x) dx}{ \int_{\mathcal{X}}s(0,x) f_X(x) dx }.
$$
 is differentiable (and, therefore, lower-hemicontinuous). Invoking Step 5 implies that $\sigma(q)$ in \eqref{eq:suppfun2} obeys the properties (1)--(3) outlined at the beginning of the proof.

\end{proof}

\begin{proof}[Proof of Theorem \ref{thrm:condmonot2}]

Invoking Lemma \ref{lem:class} with $B_N:= s^2_N + q^2_N$ and $V_N^2 = s^1_N + q^1_N + s_N + q_N$ verifies  condition (1) in Lemma \ref{lem:markovq}. Assumption \ref{ass:concentration:chap1} 
 is verified in Lemma \ref{lem:uce}. Invoking Lemma \ref{lem:markovq} concludes the proof.
\end{proof}

\section*{Appendix C: Empirical details}

\renewcommand{\theequation}{C.\arabic{equation}}
\renewcommand{\thesection}{C.\arabic{section}}
\renewcommand{\thesubsection}{C.\arabic{subsection}}
\renewcommand{\thetable}{C.\arabic{table}}
\renewcommand{\thefigure}{C.\arabic{figure}}
\setcounter{equation}{0}
\setcounter{section}{0}
\setcounter{table}{0}
\label{sec:mc}
\medskip

\label{sec:data}

\setcounter{subsection}{0}
\setcounter{table}{0}

\subsection{JobCorps Data description.} 
\label{sec:datadescription}
In this section, I describe baseline covariates for the JobCorps empirical application. The data is taken from \cite{Schochet}, who provides covariate descriptions in Appendix L. All covariates  describe experiences before random assignment (RA). Most of the covariates represent answers to multiple choice questions; for these covariates I list the question and the  list of possible answers. An answer is highlighted in boldface if is selected by post-lasso-logistic regression for one of employment equation specifications, described below. Table \ref{tab:lee} lists the covariates selected by \cite{LeeBound}.  A full list of numeric covariates, not provided here, includes $p=781$ numeric covariates.

\textbf{Covariates selected by \cite{LeeBound}}.  \cite{LeeBound} selected 28 baseline covariates to estimate parametric specification of the sample selection model. They are given in Table \ref{tab:lee}. 

\begin{table}[H]
\small
    \centering
    \caption{ Baseline covariates selected by \cite{LeeBound}.}
    \begin{tabular}{|c|c|}
    \toprule
    Name & Description  \\
    \toprule
      $\text{FEMALE}$ & female  \\
      $\text{AGE}$ & age \\
     $\text{BLACK}$, $\text{HISP}$, $\text{OTHERRAC}$ & race categories \\
       $\text{MARRIED}$,  $\text{TOGETHER}$,  $\text{SEPARATED}$ & family status categories \\
     $\text{HASCHILD}$ & has child \\
       $\text{NCHILD}$ & number of children \\
        $\text{EVARRST}$ & ever arrested \\
        $\text{HGC}$ & highest grade completed \\
        $\text{HGC\_MOTH}$, $\text{HGC\_FATH}$ & mother's and father's HGC \\
         $\text{HH\_INC}1-\text{HH\_INC}5$ & five household income groups with cutoffs $3, 000, 6, 000, 9, 000, 18, 000$ \\
         $\text{PERS\_INC}1-\text{PERS\_INC}4$  & four personal income groups with cutoffs $3, 000, 6, 000, 9, 000$ \\
         $\text{WKEARNR}$ & weekly earnings at most recent job \\
         $\text{HRSWK\_JR}$ & ususal weekly work hours at most recent job\\
         $\text{MOSINJOB}$ & the number of months employed in past year \\
          $\text{CURRJOB}$ & employed at the moment of interview \\
         $\text{EARN\_YR}$ & total yearly earnings \\
         $\text{YR\_WORK}$ & any work in the year before RA \\
    \bottomrule
     \bottomrule
    \end{tabular}
    \label{tab:lee}
\end{table}
\textbf{Reasons for joining JobCorps ($\text{R\_X}$)}. Applicants were asked a question ``How important was reason X on the scale from $1$ (very important) to $3$ (not important),  or $4$ (N/A), for joining JobCorps?''. Each reason X was asked about in an independent question.
\begin{table}[H]
    \centering
    \small
       \caption{Reasons for joining JobCorps}
    \begin{tabular}{|c|c|c|c|}
    \toprule
    Name & description & Name & description \\
    \toprule
      $\text{R\_HOME}$ & \textbf{ getting away from home }  & $\text{R\_COMM}$ & \textbf{getting away from community}  \\
        $\text{R\_GETGED}$  & getting a GED  & $\text{R\_CRGOAL}$ & desire to achieve a career goal \\
       $\text{R\_TRAIN}$ & getting job training & $\text{R\_NOWORK}$  & not being able to find work   \\
    \bottomrule
    \end{tabular}
 
    \label{tab:r}
\end{table}
For example, a covariate $\text{R\_HOME1}$ is a binary indicator for the reason $\text{R\_HOME}$ being ranked as a very important reason for joining JobCorps.

\textbf{Sources of advice  about the decision to enroll in JobCorps ($\text{IMP\_X}$)}. Applicants were asked a question ``How important was advice of X on the scale from $1$ (important) to $0$ (not important) ?''. Each source of advice was asked about in an independent question. 
\begin{table}[H]
    \centering
        \small
     \caption{Sources of advice  about the decision to enroll in JobCorps.}
    \begin{tabular}{|c|c|c|c|}
    \toprule
    Name & description & Name & description \\
    \toprule
       $\text{IMP\_PAR}$ & parent or legal guardian  & $\text{IMP\_FRD}$ & friend \\
        $\text{IMP\_TCH}$  & teacher  & $\text{IMP\_CW}$ & case worker  \\
       $\text{IMP\_PRO}$ & \textbf{probation officer} & $\text{IMP\_CHL}$ & church leader \\
    \bottomrule
    \end{tabular}
    \label{tab:source}
\end{table}

\textbf{Main types of worry about joining JobCorps ($\text{TYPEWORR}$)}.
Applicants were asked  to select one main type of worry about joining JobCorps. 
\begin{table}[H]
    \centering
     \caption{Types of worry about joining JobCorps}
    \begin{tabular}{|c|c|c|c|}
    \toprule
    $\#$ & description & $\#$ & description \\
    \toprule
        1 & not knowing anybody or not fitting in  & 2 & violence $/$ safety \\
        3  & homesickness  & 4 & not knowing what it will be like \\
        5 & \textbf{dealing with other people} & 6 & living arrangements \\
        7 & strict rules and highly regimented life & 8 & racism \\
        9 & not doing well in classes & 10 & none \\
    \bottomrule
    \end{tabular}
   
    \label{tab:worry}
\end{table}

\textbf{Drug use summary  ($\text{DRUG\_SUMP}$)}. Applicants were asked to select one of $5$ possible answers best describing their drug use in the past year before RA.
\begin{table}[H]
    \centering
           \small
    \caption{Summary of drug use in the year before RA}
    \begin{tabular}{|c|c|c|c|}
    \toprule
    $\#$ & description & $\#$ & description \\
    \toprule
        1 & did not use drugs  & 2 & \textbf{ marijuana $/$ hashish only} \\
        3  & drugs other than marijuana $/$ hashish  & 4 & both marijuana and other drugs \\
    \bottomrule
    \end{tabular}
    
    \label{tab:source}
\end{table}

\textbf{Frequency of marijuana use ($\text{FRQ\_POT}$) }. Applicants were asked to select one of $5$ possible answers best describing their marijuana $/$ hashish  use in the past year before RA.
\begin{table}[H]
    \centering
           \small
     \caption{Frequency of marijuana/hashish use in the year before RA}
    \begin{tabular}{|c|c|c|c|}
    \toprule
    $\#$ & description & $\#$ & description \\
    \toprule
        1 & daily & 2 & a few times each week \\
        3 & \textbf{  a few times each month } & 4 & less often  \\
        5 & missing & 6  & N/A \\
    \bottomrule
    \end{tabular}
   
    \label{tab:pot}
\end{table}

 \textbf{Applicant's welfare receipt history}. Applicants were asked whether they ever received food stamps ($\text{GOTFS}$), AFDC benefits ($\text{GOTAFDC}$) or other welfare ($\text{GOTOTHW}$) in the year prior to RA. In case of receipt, they asked about the duration of receipt in months ($\text{MOS\_ANYW},\text{MOS\_AFDC}$). For example, $\text{GOTAFDC}$=1 and $\text{MOS\_AFDC}$=8 describes an applicant who received AFDC benefits during 8 months before RA.

\textbf{Household welfare receipt history ($\text{WELF\_KID}$).} Applicants were asked about family welfare receipt history during childhood. 
\begin{table}[H]
    \centering
           \small
    \caption{Family was on welfare when growing up}
     \begin{tabular}{|c|c|c|c|}
    \toprule
    $\#$ & description & $\#$ & description \\
    \toprule
        1 & never & 2 & occasionally \\
        3 & half of the time & 4 & \textbf{ most or all time }  \\
    \bottomrule
    \end{tabular}
    
    \label{tab:welfare}
\end{table}

\textbf{Health status ($\text{HEALTH}$)}. Applicants were asked to rate their health at the moment of RA
\begin{table}[H]
    \centering
    \caption{Health status at RA}
     \begin{tabular}{|c|c|c|c|}
    \toprule
    $\#$ & description & $\#$ & description \\
    \toprule
        1 & excellent & 2 & good \\
        3 & \textbf{ fair } & 4 & poor  \\
    \bottomrule
    \end{tabular}
    
    \label{tab:welfare}
\end{table}

\textbf{Arrest experience}. $\text{CPAROLE21}$=1 is a binary indicator for being on probation or parole at the moment or RA. In addition, arrested applicants were asked about the time past since most recent arrest $\textbf{MARRCAT}$.  
\begin{table}[H]
    \centering
    \caption{Number of months since most recent arrest}
     \begin{tabular}{|c|c|c|c|}
    \toprule
    $\#$ & description & $\#$ & description \\
    \toprule
        1 & \textbf{ less than 12 } & 2 & 12 to 24 \\
        3 & 24 or more & 4 & N/A  \\
    \bottomrule
    \end{tabular}
    
    \label{tab:arrcat}
\end{table}

\begin{landscape}
\small
   \captionof{table}{Figure \ref{fig:Figure2} details: monotonicity test results}
 \label{tab:covgroups}
     \begin{tabularx}{1.4\textwidth}{@{}@{}@{\extracolsep{6pt}}YYY@{}}
			\toprule
 			 \toprule 
 \multicolumn{1}{p{3.0cm}}{\centering  Weeks  }    &  \multicolumn{1}{p{9cm}}{\centering  Cell with the largest $t$-statistic } & Average Test Statistic   \\
\multicolumn{1}{p{3.0cm}}{\centering  (1) }    &  \multicolumn{1}{p{9cm}}{\centering  (2) } & (3) \\
 
    \hline
 \multicolumn{1}{p{3.0cm}}{\centering  Weeks 60 -- 89 }  &  \multicolumn{1}{p{9cm}}{\centering MOS\_AFDC=8   or  \\
 PERS\_INC=3 and  EARN\_YR $\in [720, 3315]$    } & 2.390 \\
 \\
 \multicolumn{1}{p{3.0cm}}{\centering  Weeks 90 -- 116}   &   \multicolumn{1}{p{9cm}}{\centering  R\_HOME=1 and MARRCAT11=1 or \\
    WELF\_KID=4 and TYPEWORR=5  }  
    & 2.536   \\
    \\
     \multicolumn{1}{p{3.0cm}}{\centering  Weeks 117 -- 152 } & \multicolumn{1}{p{9cm}}{\centering R\_COMM=1 and IMP\_PRO=1 and FRQ\_POT=3 or \\
      DRG\_SUMP=2 and TYPEWORR=5 and IMP\_PRO=1 
     } & 2.690  \\
     \\
      \multicolumn{1}{p{3.0cm}}{\centering  Weeks 153 -- 186 }  &  \multicolumn{1}{p{9cm}}{\centering  IMP\_PRO=1  and MARRCAT11 or \\ 
      REASED\_R4 = 1 and R\_COMM=1 and DRG\_SUMP=2  }  & 3.303  \\
    \\
  
          \multicolumn{1}{p{3.0cm}}{\centering  Weeks 187 -- 208 }  &  \multicolumn{1}{p{9cm}}{\centering same as weeks 90--116 }   & 2.221 \\
                \bottomrule

\end{tabularx}

\caption*{Notes. This table shows the results for the monotonicity test in Figure \ref{fig:Figure2}. The test is conducted separately for each week using a week-specific test statistic and p-value. For each group of weeks, I partition $N=9,145$  subjects into $J=2$ cells $C_1, C_2$.  The cell with the largest $t$-statistic whose value is compared to the critical value, is sketched in Column (2). The cell is determined via a sequence of if/else statements, the two of which are presented in Column (2).   Column (3) shows the average test-statistic  across time period, defined in Column (1). The test statistic is $T= \max_{j \in \{1,2\}} \widehat{\mu}_j/\widehat{\sigma}_j$, where $\widehat{\mu}_j$ and $\widehat{\sigma}_j$ are sample average and  standard deviation of random variable $ \xi_j:= \E[ (2D-1) \cdot S |X \in C_j]$, weighted by design weights $\text{DSGN\_WGT}$.  The critical value  $c_{\alpha}$ is the self-normalized critical value of \cite{CCK}. For $\alpha=0.05$, $c_{\alpha}=1.960$. For $\alpha=0.01$, $c_{\alpha}=2.577$. Covariates are defined in Section \ref{sec:datadescription}.  }
\label{tab:monot:test}
\end{landscape}

 \bibliographystyle{apalike}
\bibliography{/Users/virasemenova/Desktop/my_new_bibtex}

\end{document}